\newcommand{\openone}{\leavevmode\hbox{\small1\normalsize\kern-.33em1}} 
\newcommand{\matriz}[1]{{\mathsf{#1}}}  
\newcommand{\Matriz}[1]{{\mathcal{#1}}} 
\newcommand{\coefr}{r}  
\newcommand{\coeft}{t}  
\newcommand{\re}{\mathop{\mathrm{Re}}\nolimits}
\newcommand{\im}{\mathop{\mathrm{Im}}\nolimits}
\newcommand{\Tr}{\mathop{\mathrm{Tr}}\nolimits} 

\documentclass[final,authoryear,1p]{elsarticle}
\usepackage{amsfonts,amssymb,amsbsy,bm}
\usepackage{graphicx,color,mathptmx,txfonts}

\journal{Physics Reports}

\begin{document}

\begin{frontmatter}

  \title{The transfer matrix: a geometrical perspective}

  \author[UCM]{Luis~L.~S\'{a}nchez-Soto}
  \author[UCM]{Juan~J.~Monz\'on} \author[UCM]{Alberto~G.~Barriuso}
  \author[UZ]{Jos\'{e} F. Cari\~{n}ena}

  \address[UCM]{Departamento de \'Optica, Facultad de F\'{\i}sica,
    Universidad Complutense, 28040~Madrid, Spain}

  \address[UZ]{Departamento de F\'{\i}sica Te\'{o}rica, Facultad de
    Ciencias, Universidad de Zaragoza, 50009~Zaragoza, Spain}

  \date{today}

  \begin{abstract}
    We present a comprehensive and self-contained discussion of the
    use of the transfer matrix to study propagation in one-dimensional
    lossless systems, including a variety of examples, such as
    superlattices, photonic crystals, and optical resonators. In all
    these cases, the transfer matrix has the same algebraic properties
    as the Lorentz group in a $(2+1)$-dimensional spacetime, as well
    as the group of unimodular real matrices underlying the structure
    of the $abcd$ law, which explains many subtle details.  We
    elaborate on the geometrical interpretation of the transfer-matrix
    action as a mapping on the unit disk and apply a simple
    trace criterion to classify the systems into three types with very
    different geometrical and physical properties.  This approach is
    applied to some practical examples and, in particular, an
    alternative framework to deal with periodic (and quasiperiodic)
    systems is proposed.
  \end{abstract}


\end{frontmatter}


\tableofcontents

\section{Introduction}

Quantum mechanical scattering in one dimension describes many actual
phenomena to a good approximation.  The advantage of this theory is
that it does not need special mathematical functions, while still
retaining sufficient complexity to illustrate the pertinent physical
concepts. It is therefore not surprising that there have been many
articles dealing with various aspects of such scattering at various
levels of detail, ranging from pedagogical issues~\citep{Eberly:1965,
James:1970, Formanek:1976,Kamal:1984,Dijk:1992,Nogami:1996,
Barlette:2000,Barlette:2001,Cattapan:2003,Sanchez-Soto:2005,Boonserm:2010} to
edge-cutting research~\citep{Peres:1983,Jaworski:1989,Trzeciakowski:1993,
Sassoli:1994,Rozman:1994a,Nockel:1994,Sassoli:1995,Chebotarev:1996,
Kiers:1996,Marinov:1996,Kerimov:1998,Visser:1999,Miyazawa:2000,Grossel:2002qy,
Boya:2008fj,Xuereb:2009yq,Boonserm:2009rt,Boonserm:2010fr}.

These papers emphasize notions such as partial-wave decomposition,
Lippmann-Schwinger integral equations, transition operator, or
parity-eigenstate representation, paralleling as much as possible
their analogues in two and three dimensions. In other words, these
approaches, like most of the standard textbooks on the
subject~\citep{Goldberger:1964,Newton:1966,Cohen:1977,Galindo:1990},
employ the $\matriz{S}$ matrix [note some significant
exceptions, such as,~e.g.~\citet{Mathews:1978},
\citet{Merzbacher:1997}, \citet{Ballentine:1998gf}, or
\citet{Singh:1997}].

The elegance and power of the $\matriz{S}$-matrix formulation is
beyond doubt. However, it is a ``black-box'' theory: the system under
study is isolated and is tested through asymptotic states. This is
well suited for experiments in elementary particle physics, but
becomes inadequate as soon as one couples several systems. The most
effective technique for studying such a coupling is the transfer
matrix, in which the amplitudes of two fundamental solutions on either
sides of a potential are connected by a matrix $\matriz{M}$.

The transfer matrix is a fruitful object widely used in the treatment
of layered systems, like
superlattices~\citep{Tsu:1973,Esaki:1986,Ram-Mohan:1988fe,
  Hauge:1989,Vinter:1991,Weber:1994,Sprung:2003} or photonic
crystals~\citep{Joannopoulos:1995,Bendickson:1996,Tsai:1998}.  Optics,
of course, is a field in which multilayers are important and the
method is time
honored~\citep{Brekovskikh:1960,Lekner:1987,Azzam:1987,Yeh:1988}.

An extensive and up-to-date review of the applications of the transfer
matrix to many problems can be found in the two excellent monographs
by \citet{Moliner:1992} and \citet{Perez:2004}. They are addressed to
anyone who wants to enter the field and provide a really professional
level of penetration into the basic issues.

A natural question thus arises: why yet another essay on the transfer
matrix? The answer is simple: a quick look at the literature
immediately reveals the different backgrounds and habits in which the
transfer matrix is used and the very little ``cross talk'' between
them. In fact, many scientists are usually not aware of the
mathematical basis behind the standard toolkits they are using in
their everyday research. The main goal of this review is precisely to
fill this gap.

When one thinks in a unifying mathematical scenario, geometry
immediately comes to mind. Although special relativity is the
archetypal example of the interplay between physics and geometry, one
cannot forget that geometrical ideas are essential in the
development of modern physics~\citep{Schutz:1997,Kauderer:1994}.

In recent years a number of geometrical concepts have been exploited to
gain further insights into the behavior of scattering in one
dimension~\citep{Yonte:2002,Monzon:2002,Barriuso:2003,Barriuso:2004,
  Sprung:2004,Martorell:2004kx,Sanchez-Soto:2005,Barriuso:2009}.  The
algebraic basis for these developments is the fact that the transfer
matrix is an element of the group SU(1,~1), which is locally
isomorphic to the $(2+1)$-dimensional Lorentz group SO(2,~1).  This
leads to a natural and complete identification between reflection and
transmission coefficients and the parameters of the corresponding
Lorentz transformation.

As soon as one realizes that SU(1,~1) is also the basic group of 
hyperbolic geometry~\citep{Coxeter:1968}, it is tempting to look for
an enriching geometrical interpretation. In fact, we propose to look
at the the action of the transfer matrix as a bilinear (or M\"{o}bius)
transformation on the unit disk, obtained by stereographic projection
of the unit hyperboloid associated with SO(2,~1).

Borrowing elementary techniques of hyperbolic geometry, we can
classify and reinterpret all the relevant features of these matrices
in a very elegant and concise way, largely independent of the model
considered.  We stress that this formulation does not offer any
inherent advantage in terms of efficiency in solving practical
problems; rather, we expect that it could supply a general and
unifying setting to analyze the transfer matrix in many fields of
physics, which, in our opinion, is more than a curiosity.

\section{Transfer matrix in quantum mechanics}

\subsection{Basic concepts on transfer matrix}

We consider the quantum scattering of a particle of mass $m$ in one
spatial dimension by a potential barrier $V(x)$. This is governed by
the time-independent Schr\"{o}dinger equation
\begin{equation}
  \label{eq:Schrind}
  \left [ - \frac{d^2}{dx^2} + U(x) \right ] \Psi (x) = \varepsilon \,
  \Psi (x) \, ,  
\end{equation}
where 
\begin{equation}
  \label{eq:varres}
  \varepsilon = \frac{2 m}{\hbar^2} E \, , 
  \qquad \qquad
  U(x) = \frac{2 m}{\hbar^2} V(x) \, ,
\end{equation}
$E$ being the energy of the particle.  We assume this potential to be
real (i. e., a Hermitian operator) but otherwise arbitrary in a finite
interval $(a, b)$. Outside this interval it is taken to be a constant
that we define to be the zero of energy. Complex potentials can be
used to model absorption, a situation which we shall not touch
upon~\citep{Monzon:2011fk}.

The treatment can be also adapted, with minor modifications, to deal with
potentials for which $V_{a} \neq V_{b}$, and also with the more subtle
case of a position-dependent effective mass $m(x)$, which usually
arise in superlattices~\citep{Leibler:1975vn,Bastard:1981,Perez-Alvarez:1988ly,
Thomsen:1989zr,Burt:1992ys}.

Since $E>0$, the spectrum is continuous and we have two linearly
independent solutions for a given value of
$E$~\citep{Galindo:1990}. Accordingly,  the general solution of the
time-independent Schr\"{o}dinger equation can be expressed as a
superposition of a right-mover $e^{+i k x}$ and a left-mover
$e^{-ikx}$:
\begin{equation}
  \label{eq:movers}
  \Psi(x) = \left \{
    \begin{array}{ll}
      A_{+} e^{+ i k (x-a)} + A_{-} e^{- i k (x-a)}
      \qquad &  x < a , \\
      & \\
      \Psi_{ab} (x) & a < x < b \\
      & \\
      B_{+} e^{+ i k (x-b)} + B_{-} e^{- i k (x-b)}
      & x > b , \\
    \end{array}
  \right . \, ,
\end{equation}
where $ k^2 = \varepsilon$ and the subscripts $+$ and $-$ indicate
that the waves propagate to the right and to the left, respectively
(see figure~1). The origins of the movers have been chosen to simplify
the subsequent calculations.

\begin{figure}
  \centering
  \includegraphics[height=3.5cm]{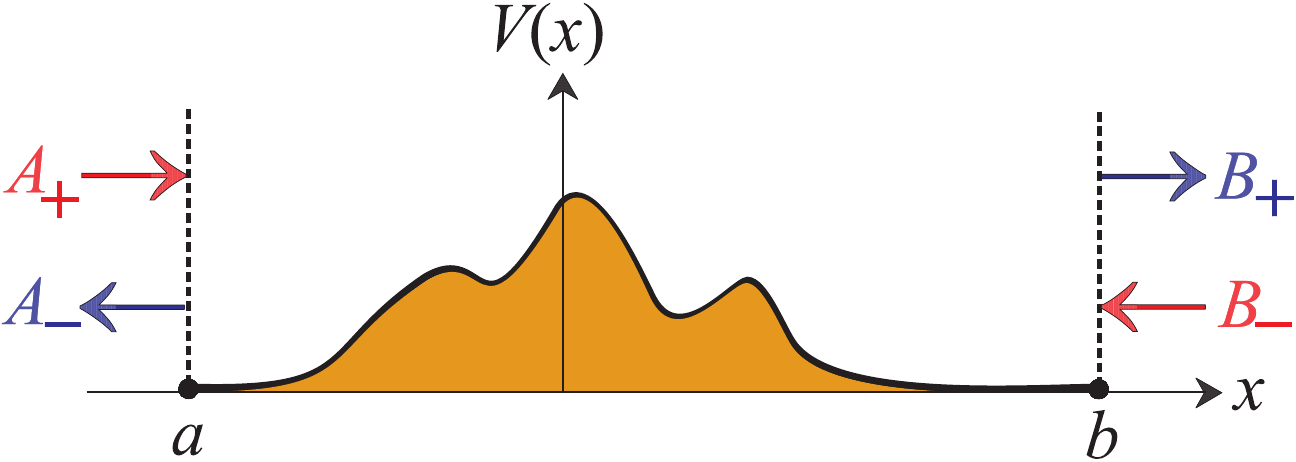}
  \caption{Illustration of the scattering from an arbitrary potential
    barrier, showing the input ($A_+$ and $B_-$) and output ($A_-$ and
    $B_+$) amplitudes.}
  \label{figure1}
\end{figure}

To solve the problem in a closed form one must work out the
Schr\"odinger equation in $(a, b)$ to compute $\Psi_{ab}(x)$ and
invoke the appropriate boundary conditions, involving not only the
continuity of $\Psi(x)$ itself, but also of its derivative. In this
way, one has two linear relations among the coefficients $A_{\pm}$
and $B_{\pm}$, which can be solved for any amplitude pair in terms of
the other two: the result can be expressed as a matrix equation, which
translates the linearity of the problem.  For our purposes, it is
more advantageous to express a linear relation between the wave
amplitudes on both sides of the scatterer, namely,
\begin{equation}
  \label{M} 
  \left (
    \begin{array}{c}
      A_{+} \\
      A_{-}
    \end{array}
  \right ) = 
  \matriz{M}_{ab} 
  \left (
    \begin{array}{c}
      B_{+} \\
      B_{-}
    \end{array}
  \right ) \, ,
\end{equation}
$\matriz{M}_{ab}$ being the transfer matrix for the potential.

The reflection and transmission coefficients are the ratio
of the amplitudes of the reflected and transmitted waves to the
amplitude of the incoming wave, respectively.  
Denoting the amplitudes for waves propagating from
the right as $\coefr_{ba}$ and $\coeft_{ba}$ and repeating the
procedure, one easily finds that time-reversal invariance imposes
\begin{equation}
  \label{Stokes}
  \coeft_{ab} = \coeft_{ba} \, ,  
  \qquad \qquad    
  \coefr_{ba}/\coeft_{ba} = -
  \coefr_{ab}^{\ast}/\coeft_{ab}^{\ast}  \, , 
\end{equation}
while the conservation of the flux (\ref{fluxcon}) gives
\begin{equation}
  \label{norm}
  | \coefr_{ab} |^{2} + | \coeft_{ab} |^{2} = 1 \, . 
\end{equation}

In conclusion, the  form of the transfer matrix is
\begin{equation}
  \label{TranM} 
  \matriz{M}_{ab} = 
  \left (
    \begin{array}{cc}
      1/\coeft_{ab} & \coefr_{ab}^{\ast}/\coeft_{ab}^{\ast} \\
      \coefr_{ab}/\coeft_{ab}  & 1/\coeft_{ab}^{\ast}
    \end{array}
  \right )  \, .
\end{equation}
In the particular case of a symmetric potential [i.e., $V(x) =
V(-x)$], it is clear that $r_{ab} = r_{ba}$ and therefore the matrix
element $\beta$ is an imaginary number.

Thus far, we have related the amplitudes $A_{\pm}$ to the $B_{\pm}$,
as in equation (\ref{M}). This choice is by no means essential and we
could relate the amplitudes taken in the reverse order.  The
corresponding transfer matrix, represented by $\matriz{M}_{ba}$, can be
expressed as
\begin{equation}
  \label{eq:MReverse} 
  \matriz{M}_{ba} = 
  \left (
    \begin{array}{cc}
      1/\coeft_{ab} & - \coefr_{ab}/\coeft_{ab} \\
      -  \coefr_{ab}^{\ast}/\coeft_{ab}^{\ast}   & 1/\coeft_{ab}^{\ast}
    \end{array}
  \right )  \, ,
\end{equation}
where we have used (\ref{Stokes}).

We will now bring up the paradigmatic example when the
potential $V(x)$ reduces to a rectangular potential barrier of width
$L$ and height $V_0$. The calculations can be easily carried out, so
we skip the details~\citep{Cohen:1977} and simply quote the results
for $\coefr_{ab}$ and $\coeft_{ab}$ (with the choice of movers in
figure~\ref{figure1})
\begin{eqnarray}
  \label{Barellip}
  \coefr_{ab} & = &
  \frac{(k^2 - \kappa^2) \sin(\kappa L)}
  {(k^2+ \kappa^2) \sin (\kappa L)
    + 2 i k \kappa \cos (\kappa L)}  \,  , 
  \nonumber \\
  & & \\
  \coeft_{ab} & = &
  \frac{2 i k \kappa}
  {(k^2 + \kappa^2) \sin (\kappa L)
    + 2 i k \kappa \cosh (\kappa L)}   \, ,
  \nonumber
\end{eqnarray}
with $\kappa^{2} = 2m (E - V_0)/\hbar^2$. These coefficients
correspond to $E > V_0$. When $E < V_0$ the expressions are
\begin{eqnarray}
  \label{Barhyper}
  \coefr_{ab} & = & 
  \frac{(k^2 + \bar{\kappa}^2) \sinh( \bar{\kappa} L)}
  {(k^2- \bar{\kappa}^2) \sinh ( \bar{\kappa} L)
    + 2 i k \bar{\kappa} \cosh ( \bar{\kappa} L)}   \, , 
  \nonumber \\
  & & \\
  \coeft_{ab} & = &    
  \frac{2 i k \bar{\kappa}}
  {(k^2 - \bar{\kappa}^2) \sinh ( \bar{\kappa} L)
    + 2 i k \bar{\kappa} \cosh ( \bar{\kappa} L)}   \,  ,
  \nonumber
\end{eqnarray}
where now $\bar{\kappa}^{2} = 2 m (V_{0} - E)/\hbar^{2}$. This can be
obtained from the previous case with the formal substitution
$\bar{\kappa} \rightarrow i \kappa$. Finally, when $E = V_0$ a
limiting procedure yields
\begin{equation}
  \label{Barpar}
  \coefr_{ab} =    \frac{kL}{kL + 2i} \, ,
  \qquad \qquad
  \coeft_{ab} =   \frac{2i}{kL + 2i} \, .
\end{equation}
A detailed discussion of the significance of these three situations
can be found in \citet{Bohm:1989mz}.

It is worth stressing that one could also relate outgoing amplitudes
in terms of the incoming amplitudes (which are the magnitudes one can
externally control). This is precisely the scattering matrix, which
can be concisely written as
\begin{equation}
  \label{S} 
  \left (
    \begin{array}{c}
      B_{+} \\
      A_{-}
    \end{array}
  \right ) = 
  \matriz{S}_{ab} 
  \left (
    \begin{array}{c}
      A_{+} \\
      B_{-}
    \end{array}
  \right ) \, ,
\end{equation}
and $\matriz{S}_{ab}$ reads
\begin{equation}
  \label{SM} 
  \matriz{S}_{ab} = 
  \left(
    \begin{array}{cc}
      \coeft_{ab} & \coefr_{ba} \\
      \coefr_{ab}  & \coeft_{ba}
    \end{array}
  \right) \, .
\end{equation}
Due to the properties (\ref{Stokes}) and (\ref{norm}),
$\matriz{S}_{ab}$ is unitary and with unit determinant, that is, an
element of the group SU(2).
 
Note carefully that the transfer matrix depends on the choice of basis
vectors~\citep{Perez:2001,Perez:2004} and special care must be paid
when comparing results from different sources. For example, instead of
specifying the amplitudes of the right and left-moving waves, we could
also write a linear relation between the values of the wave function
and its derivative at the points $a$ and
$b$~\citep{Sprung:1993}. prefer to employ the adimensional
variables
\begin{equation}
  \label{canvar}
  \Psi(x) + \frac{1}{k} \Psi^{\prime} (x) \, , \qquad \qquad
  \Psi(x) - \frac{1}{k} \Psi^{\prime} (x) \, .  
\end{equation}
This looks very much as passing from position-momentum to
creation-annihilation operators~\citep{Ballentine:1998gf}. The
amplitudes $\mathcal{A}_{\pm}$ associated to these variables are
related to $A_{\pm}$ by
\begin{equation}
  \label{eq:StoA}
  \left (
    \begin{array}{c}
      \mathcal{A}_{+} \\
      \mathcal{A}_{-}
    \end{array}
  \right ) = 
  \Matriz{U} 
  \,
  \left (
    \begin{array}{c}
      A_{+} \\
      A_{-}
    \end{array}
  \right ) \, ,
\end{equation}
with
\begin{equation}
  \label{Cayley}
  \Matriz{U} = \frac{1}{\sqrt{2}}
  \left(
    \begin{array}{cc}
      1 & i \\
      i & 1
    \end{array}
  \right) ,
\end{equation}
and analogously at the point $b$. The transfer matrix in this
equivalent representation is
\begin{equation}
  \label{MconjU}
  \Matriz{M}_{ab} = \Matriz{U} \, \matriz{M}_{ab} \, \Matriz{U}^{\dagger} = 
  \left (
    \begin{array}{cc}
      \mathfrak{a} & \mathfrak{b} \\
      \mathfrak{c} & \mathfrak{d}
    \end{array}
  \right ) \, ,
\end{equation}
where $\dagger$ stands for the Hermitian conjugate and
\begin{eqnarray}
  & \mathfrak{a} = \re \alpha + \im \beta ,
  \qquad  \qquad
  \mathfrak{b} =  \im \alpha + \re \beta  , & \nonumber \\
  & & \\
  & \mathfrak{c} = - \im  \alpha + \re \beta , 
  \quad \qquad 
  \mathfrak{d} = \re \alpha - \im \beta \, , & \nonumber
\end{eqnarray}
are real numbers. The role that the transformation $\Matriz{U}$ will
play in what follows justifies our choice in (\ref{canvar}).

Since the determinant are preserved by matrix conjugation, we have
that $\det \Matriz{M}_{ab}= +1$.  In other words, the matrices
$\Matriz{M}_{ab}$ belong to the group SL(2, $\mathbb{R}$) of
unimodular $2 \times 2$ matrices with real elements. The
transformation by $\Matriz{U}$ establishes in fact a one-to-one map
between the group SL(2, $\mathbb{R}$) of matrices $\Matriz{M}_{ab}$
and the group SU(1,~1) of matrices $\matriz{M}_{ab}$, which allows for
a direct translation of the properties from one to the other, as we
will have occasion to check.

Irrespective of the representation used, transfer matrices are very
convenient mathematical objects.  Suppose we know how the wave
functions ``propagate'' from point $b$ to point $a$, with a transfer
matrix we symbolically write as $\matriz{M}_{ab}$, and also from $c$
to $b$, with $\matriz{M}_{bc}$.  The crucial point is that the
propagation from $c$ to $a$ is described by the product
\begin{equation}
  \label{propag}
  \matriz{M}_{ac} = \matriz{M}_{ab}  \, \matriz{M}_{bc} \, .
\end{equation}
This property is rather helpful: we can connect simple scatterers to
create an intricate potential landscape and determine its transfer
matrix by simple multiplication~\citep{Jonsson:1990,Kalotas:1991,Walker:1994,
Rozman:1994a,Yuan:2010}. However, this important property does not
seem to carry over into the scattering matrix in any simple
way~\citep{Aktosun:1992kx,Aktosun:1996fj,Giust:2009ly}, because the
incoming amplitudes for the overall system cannot be obtained in terms
of the incoming amplitudes for every subsystem. While this is not a
difficulty for a single scatterer (the typical situation arising in
particle physics and for which the $\matriz{S}$-approach was specially
tailored), it constitutes a drawback in applications where a number of
cascaded systems are present.

\subsection{Building the transfer matrix}

The complete determination of the transfer matrix for an arbitrary
potential $V(x)$ amounts to solving the Schr\"{o}dinger equation and,
in consequence, it is not, in general, a simple exercise. Very
accurate approximation schemes are available, among which we cite the
WKB~\citep{Chebotarev:1995,Chebotarev:1997ly}, the
variational~\citep{Bastard:1983vn,Ahn:1986ys,Gould:1995,Ando:2003mz},
the Monte-Carlo~\citep{Singh:1986qf,Kalos:2007}, or the finite-element
methods~\citep{Hayata:1988cr,Nakamura:1989pd,Ram:2002,Liu:2004dq}.

\begin{figure}
  \centering
  \includegraphics[height=4cm]{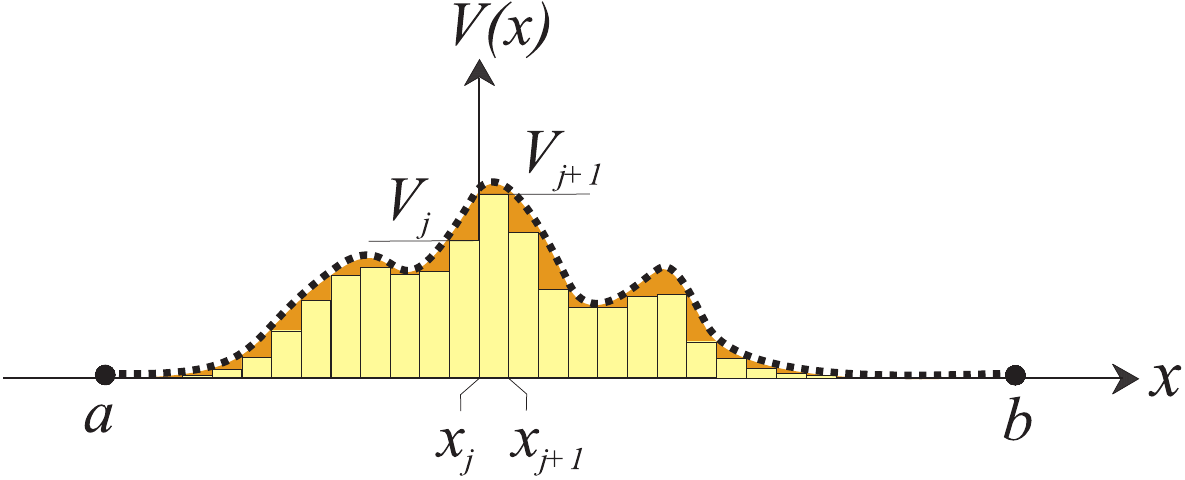}
  \caption{Decomposition of the potential $V(x)$ in elementary
    rectangular barriers.}
  \label{figure2}
\end{figure}

Here, we favor a method inspired by the long experience in dealing with
layered systems~\citep{Kennett:1983fu,Perez-Alvarez:1988xr,
Rodriguez-Coppola:1990qa,Moliner:1992,Perez:2004}. Roughly
speaking, the idea is that one can consider $V(x)$ as made of
successive constant barriers, as schematized in figure~\ref{figure2}
\citep{Kalotas:1991,Rozman:1994,Grossel:1994zr,Cao:2001eu,Rakityansky:2004tg,
He:2005qe,Monsoriu:2005ij,Su:2008dp,Hutem:2008th,Wen:2010ul}.
The $j$th barrier, of height $V_{j}$ and width $d_j$, is situated
between the points $x_{j-1}$ and $x_{j}$ (we take $x_{0} = a$ and
$x_{N+1} = b$). In this way, we express $\matriz{M}_{ab}$ as a product
of matrices that characterize the effects of the individual
discontinuities and propagations of the entire discretized structure,
taken in the proper order, as follows~\citep{Yeh:1988}:
\begin{equation}
  \label{matrizS2} 
  \matriz{M}_{ab} = 
  \matriz{I}_{01} \matriz{P}_1 \matriz{I}_{12} \matriz{P}_2 \matriz{I}_{23} 
  \ldots \matriz{I}_{(j-1) j} \matriz{P}_j  \matriz{I}_{j (j+1)} \ldots   
  \matriz{I}_{(N-1)N} \matriz{P}_N \matriz{I}_{N (N+1)} \, .
\end{equation}
When the number of barriers is large enough, the method should provide
satisfactory results for any potential~\citep{Jirauschek:2009}.  Here
$\matriz{I}_{ij}$ accounts for the discontinuity at the interface
between $V_{i}$ and $V_{j}$, and has the form~\citep{Landau:2001}
\begin{equation}
  \label{eq:interface}
  \matriz{I}_{ij}= 
  \frac{1}{\coeft_{ij}} 
  \left (
    \begin{array}{cc}
      1 & \coefr_{ij} \\
      \coefr _{ij} & 1
    \end{array}
  \right ) \, ,
\end{equation}
where $\coefr_{ij}$ and $\coeft_{ij}$ are the reflection and
transmission coefficients at the interface $ij$ and are given by
\begin{equation}
  \label{es}
  \coefr_{ij}= \frac{\kappa_{i} - \kappa_{j}}{\kappa_{i} + \kappa_{j}} \,,
  \qquad \qquad
  \coeft_{ij}= \frac{2 \kappa_{i}}{\kappa_{i} + \kappa_{j}} \, .
\end{equation}
The wave number is $\kappa_j^2 = 2m (E - V_{j})/\hbar^2$ and,
for simplicity, we have assumed $E > V_{j}$.  They verify
\begin{equation}
  \label{determinante1} 
  \det  \matriz{I}_{ij} =  \frac{\kappa_{j}}{\kappa_{i}} \, ,
\end{equation}
and the outstanding composition law
\begin{equation}
  \label{compI} 
  \matriz{I}_{ij} \; \matriz{I}_{j(j+1)} =  \matriz{I}_{i(j+1)} \, .
\end{equation}

The matrix $\matriz{P}_j$ describes the effect of propagation through
the barrier $j$ alone, and reads
\begin{equation}
  \label{eq:Prop}
  \matriz{P}_j = \left (
    \begin{array}{cc}
      \exp ( i  \delta_{j}) & 0 \\
      0 & \exp(-i \delta_{j} )
    \end{array}
  \right ) \, ,
\end{equation}
where the phase shift is $\delta_{j} = \kappa_{j} d_{j}$.  Now, we
have
\begin{equation}
  \label{detP} 
  \det \matriz{P}_j = 1 \,  .
\end{equation}

By taking the determinant in equation~(\ref{matrizS2}) and using
(\ref{determinante1}) and (\ref{detP}) we get a simple but relevant
result:
\begin{equation}
  \label{determinante2} 
  \det \matriz{M}_{ab}  =  \frac{\kappa_{N+1}}{\kappa_{0}}= 
  \frac{\coeft_{ba}}{\coeft_{ab}} \, .
\end{equation}
Therefore, when $V_a$ and $V_b$ are the same (as we have assumed until
now), the determinant of $\matriz{M}_{ab}$ is +1.  When these
potentials are different, this result also holds by renormalizing
conveniently the amplitudes~\citep{Monzon:1999}.

If we denote
\begin{equation}
  \label{eq:isbar}
  \matriz{M}_{j}  = \matriz{I}_{0j} \,   \matriz{P}_{j} \,
  \matriz{I}_{j0} \, ,
\end{equation}
which corresponds to the $j$th barrier sandwiched between two
identical constant potentials (that for simplicity we take as 0), we
can also rewrite equation~(\ref{matrizS2}) as
\begin{equation}
  \label{eq:matrizSlam}
  \matriz{M}_{ab} = \prod_{j=1}^{N} \matriz{M}_{j} \, .
\end{equation}

Sometimes, it is more convenient to work with unimodular matrices in
SL(2, $\mathbb{R}$). The counterparts of interface and
propagation can be obtained by conjugating (\ref{eq:interface}) and
(\ref{eq:Prop}) with $\Matriz{U}$; the final result is
\begin{equation}
  \label{eq:IPslR}
  \Matriz{I}_{ij}= 
  \frac{1}{\coeft_{ij}} 
  \left (
    \begin{array}{cc}
      1 & \coefr_{ij} \\
      \coefr _{ij} & 1
    \end{array}
  \right) ,
  \qquad \qquad
  \Matriz{P}_j = \left (
    \begin{array}{cc}
      \cos \delta_{j} &   \sin  \delta_{j} \\
      - \sin \delta_{j} & \cos \delta_{j} 
    \end{array}
  \right ) \, .
\end{equation}

\subsection{Hyperbolic Stokes parameters}

To move ahead  let us construct the matrices
\begin{equation}
  \matriz{J} =
  \left (
    \begin{array}{c}
      X_{+}  \\
      X_{-}
    \end{array}
  \right ) 
  \otimes
  \left (
    \begin{array}{cc}
      X_{+}^{\ast}  &      X_{-}^{\ast}
    \end{array}
  \right ) =
  \left (
    \begin{array}{cc}
      |X_{+}|^2 &  X_{+} X_{-}^{\ast}  \\
      X_{+}^\ast X_{-} & |X_{-}|^2
    \end{array}
  \right ) \, , 
\end{equation}
where $X= A$ or $B$ are the amplitudes that determine the behavior at the
points $a$ and $b$, respectively. They are quite reminiscent of the
coherence matrix in optics or the density matrix in quantum
mechanics~\citep{Mandel:1995}. Observe that $\matriz{J}$ is Hermitian
and $\det \matriz{J} = 0$. In addition, one can readily verify that
\begin{equation}
  \label{congruence}
  \matriz{J}_{a} = \matriz{M}_{ab} \, \matriz{J}_{b} \,
  \matriz{M}_{ab}^\dagger \, ,
\end{equation}
so they transform under $\matriz{M}_{ab} $ by congruence.

Let now $\sigma_{\mu}$ (the Greek indices run from 0 to 3) be the set
of four Hermitian matrices $\sigma_{0} = \openone$ (the identity) and
$(\sigma_{1}, \sigma_{2}, \sigma_{3} )$ (the standard Pauli
matrices). They constitute a natural basis of the vector space of $2
\times 2$ complex matrices, so the coordinates $s^{\mu}$ with respect
to that basis are
\begin{equation}
  s^{\mu} = \frac{1}{2} \Tr (\matriz{J}  \sigma_{\mu} )  \, ,
\end{equation}
so that
\begin{eqnarray}
  \label{equivalencia} 
  s^0 & = & \frac{1}{2} ( | X_{+} |^2 + | X_{-}|^2 ) \, ,
  \nonumber  \\
  s^1 & = & \re ( X_{+}^{\ast} X_{-} ) \, ,
  \nonumber \\
  s^2 & = & \im ( X_{+}^{\ast} X_{-} ) \, ,
  \nonumber \\
  s^3 & = &  \frac{1}{2} ( | X_{+}|^2 - | X_{-}|^2 ) \, .
\end{eqnarray}
The congruence (\ref{congruence}) induces in this manner a
transformation on the variables $s^\mu$ of the form
\begin{equation}
  \label{var} 
  s_{a}^{\mu}= \Lambda^{\mu}_{\ \nu} \, s_{b}^{\nu} ,
\end{equation}
where $\Lambda^{\mu}_{\ \nu}$ can be found to be
\begin{equation}
  \label{relation} 
  \Lambda^{\mu}_{\ \nu} (\matriz{M}_{ab}) =
  \frac{1}{2} \Tr \left ( \sigma_{\mu} \matriz{M}_{ab} \sigma_{\nu}
    \matriz{M}_{ab}^\dagger \right ) ,
\end{equation}
and it turns out to be a Lorentz transformation.  This equation can be
solved to obtain $\matriz{M}_{ab}$ from $\Lambda$.  The matrices
$\matriz{M}_{ab}$ and $- \matriz{M}_{ab}$ generate the same $\Lambda$,
so this homomorphism is two-to-one~\citep{Barut:1980}.

\begin{figure}[t]
  \centerline{\includegraphics[height=6cm]{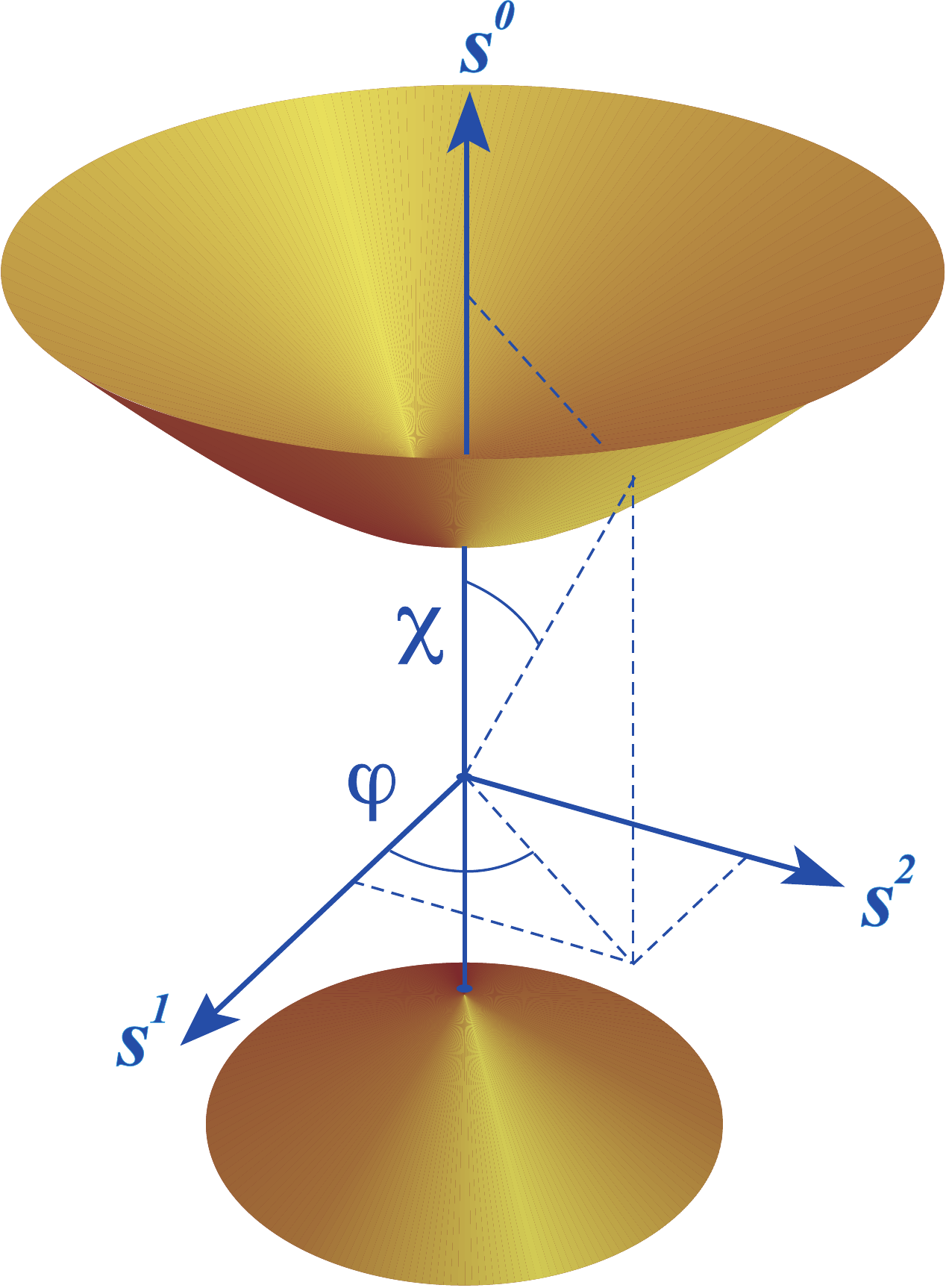}}
  \caption{Pseudospherical coordinates of the unit two-sheeted
    hyperboloid associated with a transfer matrix $\matriz{M}_{ab}$.}
  \label{figure3}
\end{figure}

The variables $s^{\mu}$ are coordinates in a Minkovskian space. Since
$\det \matriz{J} = 0$, the value of the interval (in both points $a$
and $b$) is
\begin{equation}
  \label{interval}
  (s^0)^2 - (s^1)^2 - (s^2)^2 - (s^3)^2  = 0 \, ,
\end{equation}
so it is lightlike. Moreover, the conservation of the probability
current expressed in equation (\ref{fluxcon}) means that the
coordinate $s^3$, defined in (\ref{equivalencia}), remains invariant
and (\ref{interval}) reduces to
\begin{equation}
  \label{twosheet}
  (s^0)^2 - (s^1)^2 - (s^2)^2 = (s^3)^2 = \mathrm{constant}  \, ,
\end{equation}
that is, a two-sheeted hyperboloid of radius $s^3$, which without loss
of generality will be taken henceforth as unity (see
figure~\ref{figure3}).  All this shows that the group SU(1,1) of
transfer matrices is locally isomorphic to the (2+1)-dimensional
Lorentz group SO(2,1). In fact, in more technical terms, the matrices
$\matriz{M}_{ab}$ form a two-dimensional spinor representation of the
restricted Lorentz group SO(2,1): $\Lambda_{1} \Lambda_{2}
\longleftrightarrow \matriz{M}(\Lambda_{1}) \matriz{M}(\Lambda_{2} ) =
\pm \matriz{M}(\Lambda_{1} \Lambda_{2} )$.

Let us now rewrite the complex amplitudes $X_{\pm}$ in polar form
\begin{equation}
  X_{\pm} = |X_{\pm}|  \exp ( i \varphi_{\pm} ) \, .
\end{equation}
Denoting $ \varphi = \varphi_{+} - \varphi_{-}$ and introducing the
angle $ \chi$ in pseudospherical coordinates (shown in
figure~\ref{figure3}), we have
\begin{eqnarray}
  \label{pseudosp}
  s^0 & = &   \cosh  \chi \,  ,  \nonumber \\
  s^1 & = &   \sinh \chi \, \cos \varphi \,  , \\
  s^2 & = &   \sinh \chi \, \sin  \varphi \, . \nonumber
\end{eqnarray}
This parametrization is very evocative of the standard one for the
Stokes parameters in the unit Poincar\'e sphere~\citep{Born:1999},
except for the fact that now the angle $\chi$ appears as the argument
of hyperbolic functions. We refer to these parameters as hyperbolic
Stokes parameters.  One can interpret the rotations $\varphi$ and the
hyperbolic rotations $\chi$ much in the same way as it is done for the
Poincar\'e sphere~\citep{Giust:2002fu}: a rotation of angle $\varphi$
about the axis $s^0$ corresponds to a dephasing between left- and
right-traveling waves without changing their relative amplitudes; as
it happens, for example, in the propagation inside a barrier. On the
contrary, a hyperbolic rotation of angle $\chi$ corresponds to a
change in amplitude between left- and right-traveling waves, as it
occurs at the discontinuity between two barriers.

\subsection{Lorentz transformation associated to a transfer matrix}

To better appreciate the physical meaning of the Lorentz
transformation induced by a transfer matrix, we recall that a Lorentz
transformation $\Lambda$ can be always decomposed into a product of a
spatial rotation $\matriz{R}$ and a boost $\matriz{L}$ along an
arbitrary direction~\citep{Moretti:2006}
\begin{equation}
  \label{LR} 
  \Lambda = \matriz{L} \,  \matriz{R}  \, .
\end{equation}
The analogous factorization for a transfer matrix is the polar
decomposition, which ensures that any matrix $\matriz{M} \in$ SU(1,1)
(to simplify the notation, we drop the subscript $ab$ in the rest of
this section) can be expressed in a unique way as
\begin{equation}
  \matriz{M} = \matriz{H} \, \matriz{U} \, ,
\end{equation}
where $\matriz{H}$ is positive definite Hermitian (``modulus'') and
$\matriz{U}$ is unitary (``argument''). Under the homomorphism
discussed previously, $\matriz{H}$ generates a boost and $\matriz{U}$
a rotation, in agreement with equation~(\ref{LR}).

To find their explicit form we note that polar decomposition for the
matrix $\matriz{M}$ in equation~(\ref{TranM}) reads
\begin{eqnarray}
  \label{polar} 
  \matriz{M} & = &  \matriz{H} \, \matriz{U} = 
  \frac{1}{|\coeft|} 
  \left (
    \begin{array}{cc}
      \ 1 \ & \ \coefr^\ast  \\
      \ \coefr \ & \ 1 \
    \end{array}
  \right )  
  \left (
    \begin{array}{cc}
      \exp (- i \tau ) & 0 \\
      0 & \exp (i \tau )
    \end{array}
  \right ) \, ,
\end{eqnarray}
where we have expressed the reflection and transmission coefficients
in the form
\begin{equation}
  \label{eq:frt}
  \coefr = |\coefr| \exp (i \rho) \, ,
  \qquad \qquad
  \coeft = |\coeft| \exp (i \tau) \,  .
\end{equation}
Using (\ref{relation}), the unitary component $\matriz{U}$ generates
the rotation [in $(2+1)$ dimensions]
\begin{equation}
  \matriz{R} (\matriz{U}) = 
  \left (
    \begin{array}{cccc}
      1 & 0 & 0 \\
      0 & \; \;   \cos (2 \tau) \; \; &
      \; \;  - \sin (2 \tau)  \; \; \\
      0 & \; \;  \sin (2 \tau) \; \; &
      \; \;  \cos (2 \tau) \; \; 
    \end{array}
  \right ) \,  ,
\end{equation}
that is, a spatial rotation in the plane 1-2 of angle twice the phase
of the transmission coefficient.

The Hermitian component $\matriz{H}$ generates the boost
\begin{eqnarray}
  \matriz{L} (\matriz{H}) = 
  \left (
    \begin{array}{cccc}
      \gamma &  - \gamma  \varv \cos \rho
      &  -\gamma  \varv \sin \rho  \\
      -\gamma  \varv \cos \rho  &  1 + (\gamma -1) \cos^2 \rho  &
      (\gamma - 1) \cos \rho \sin \rho   \\
      -\gamma  \varv \sin \rho &  (\gamma -1) \cos \rho \sin \rho &
      1 + (\gamma- 1)\sin^2 \rho  
    \end{array}
  \right ) \, .
\end{eqnarray}
The modulus of the velocity $\varv$ (we take $c=1$ everywhere) and the
relativistic factor $\gamma = 1/\sqrt{1 - \varv^2}$ of this boost are
\begin{equation}
  \varv  = \frac{2 |\coefr|}{1 + |\coefr|^2} , 
  \qquad \qquad
  \gamma  =   \frac{1 + |\coefr|^2} {1 - |\coefr|^2} . 
\end{equation}
The matrix $\matriz{L} (\matriz{H})$ is then a boost to a reference
frame moving with a constant velocity $\varv $ in the plane $1-2$, in
a direction forming a counterclockwise angle $\rho$ with the axis $1$.

If, as it is usual, we introduce the rapidity $\zeta$ from~\citep{Jackson:1975}
\begin{equation}
  \varv  =   \tanh \zeta \,  ,   
\end{equation}
we have the following appealing identification of the reflection and
transmission coefficients with the parameters of the Lorentz
transformation:
\begin{equation}
  \label{rtLor}
  \coefr =  \tanh (\zeta/2) \exp(i  \rho ) ,
  \qquad \qquad 
  \coeft =  \mathrm{sech} (\zeta/2) \exp( i \tau) . 
\end{equation}
Therefore, $|\coefr| = \tanh(\zeta/2)$, behaves as a velocity, while
$|\coeft|$ behaves as $1/\gamma$.

The convenient properties of the hyperbolic tangent have been
exploited in dealing with layered
systems~\citep{Khashan:1979la,Corzine:1991kl}, and explain why the
reflection coefficients are examined in greater detail than the
transmission ones.

\section{Transfer matrix in other contexts}

One-dimensional continuous models provide a detailed account of the
behavior of a variety of systems~\citep{Lieb:1966,Albeverio:2004}.
The nature of the actual particles, or states, or elementary
excitations, as they may be variously called, is irrelevant for many
purposes: there is always two input and two output channels related by
a $2 \times 2$ transfer matrix. In fact, this matrix can be viewed as
a compact way of setting out the integration of the differential
equations involved in the model with the pertinent boundary
conditions; this is what makes the method so effective.

From this perspective, one can construct a general theory of the
transfer matrix for second-order differential
equations~\citep{Khorasani:2003,Khorasani:2003a}. However, we prefer to explore some
selected examples~\citep{Perez:2004}; this restricted choice reflects
the authors personal bias, but it is illustrative enough to grasp what
the method is about.

\subsection{Mechanical waves}

Transverse waves on weighted strings, longitudinal waves on loaded
rods, acoustic waves in corrugated tubes, and water waves crossing
sandbars, among other examples~\citep{Griffiths:2001}, are governed by
the classical wave equation
\begin{equation}
  \label{eq:cwe}
  \frac{\partial^{2} \psi}{\partial t^{2}} = \varv^{2}  \frac{\partial^{2}
    \psi}{\partial x^{2}} \, .
\end{equation}
Here $\psi (x,t)$ is the amplitude of the considered phenomena (in the
above-mentioned examples $\psi (x,t)$ stands for the transverse
displacement of the string, the displacement of a point whose
equilibrium position is $x$, the pressure above ambient, or the height
of the surface above its equilibrium level, respectively), and
$\varv(x)$ is the local propagation speed of the perturbation.  We
continue to use complex notation, with the understanding that the
physical wave is the real part. For a monochromatic perturbation of
angular frequency $\omega$ [$\psi (x, t) = \Psi (x) e^{ - i \omega
  t}$] equation (\ref{eq:cwe}) reduces to
\begin{equation}
  \label{eq:Helm}
  \left [ \frac{d^2}{dx^2} + k^{2} (x) \right ] \Psi (x) = 0 \,  , 
\end{equation}
and the local wave number is
\begin{equation}
  k(x) = \frac{\omega}{\varv(x)} \, .
\end{equation}
The expansion in left- and right-movers in section 2 can be
transplanted here without modifications. In addition, as we did in
2.2, we can build the transfer matrix by assuming that the material
parameters of the medium are piecewise constant and vary in a stepwise
manner, so each constituent slab is a homogeneous material by itself.
The mismatched impedances generate the reflected and transmitted
waves~\citep{Crawford:1968}, while the application of the proper
boundary conditions at the discontinuity points (which depend on the
particular model under consideration) provide the corresponding
amplitude coefficients.

\subsection{Electromagnetic waves}

We next consider the propagation of plane electromagnetic waves in a
(nonmagnetic) stratified medium, whose optical properties are
contained in the dielectric function $\epsilon (x) = n^2 (x)$ [$n (x)$
is the local value of the refractive index]. For a monochromatic
component of frequency $\omega$ we write the field components
as~\citep{Monsivais:1995}
\begin{equation}
  \label{eq:Abel}
  \mathbf{E} (\mathbf{r}, t)  =   \bm{\mathcal{E}} (x)  
  \exp[ - i ( \omega t - \mathbf{K} \cdot \mathbf{r} )] \, , 
  \qquad \qquad
  \mathbf{B} (\mathbf{r}, t)  =   \bm{\mathcal{B}} (x) 
  \exp[ - i (\omega t - \mathbf{K} \cdot \mathbf{r} )] \, , 
\end{equation}
where $\mathbf{K}$ is the component of the wave vector in the plane
perpendicular to the $x$ axis. By eliminating, e.g., the magnetic
field $\mathbf{B}$ from Maxwell equations, it turns out
that~\citep{Born:1999}
\begin{equation}
  \label{eq:Abel2}
  \left [ \frac{d^2 }{dx^2} + k^2(x) \right ] \bm{\mathcal{E}} (x) = 0 \, ,
\end{equation}
and $k(x)$ is the local value of the normal component of the wave
vector
\begin{equation}
  \label{eq:norm}
  k (x) = \sqrt{\epsilon (x) \frac{\omega^2}{c^2} - K^2} \, .
\end{equation}
A completely analogous equation can be written for the magnetic field
by eliminating $\mathbf{E}$.

Apparently, equation (\ref{eq:Abel2}) is identical to (\ref{eq:Helm}),
and the theory can be immediately extended here, expressing the
solution as a superposition of a left- and right-mover fields. But
this requires some extra care because the amplitude in
(\ref{eq:Abel2}) is a vector.

For linear isotropic media, any plane wave can be written as a
superposition of an $s$ (or TE) wave and a $p$ (or TM) wave. The $s$
wave has its electric vector perpendicular to the plane of incidence,
and the $p$ wave has its electric vector in the plane of incidence
(and its magnetic vector perpendicular to the plane of incidence;
hence its designation as a TM, or transverse magnetic, wave).  If we
further take the plane of incidence to be the $(x, z)$ plane, the
vectors are~\citep{Azzam:1987,Lekner:1987,Yeh:1988}
\begin{equation}
  \label{eq:2}
  \bm{\mathcal{E}} = 
  \left (
    \begin{array}{c}
      0 \\
      \mathcal{E}_{y} \\
      0 
    \end{array}
  \right ) 
  \quad 
  (s \; \mathrm{polarization} )  
  \qquad
  \qquad
  \bm{\mathcal{B}} = 
  \left (
    \begin{array}{c}
      0 \\
      \mathcal{B}_{y} \\
      0 
    \end{array}
  \right ) 
  \quad 
  (p \; \mathrm{polarization}) \, . 
\end{equation}
This has to be taken into account when matching the boundary
conditions between two media. For these two basic polarizations the
problem reduces to a scalar one, and the transfer matrix can be
applied as before.

\begin{figure}
  \centering
  \includegraphics[height=3.5cm]{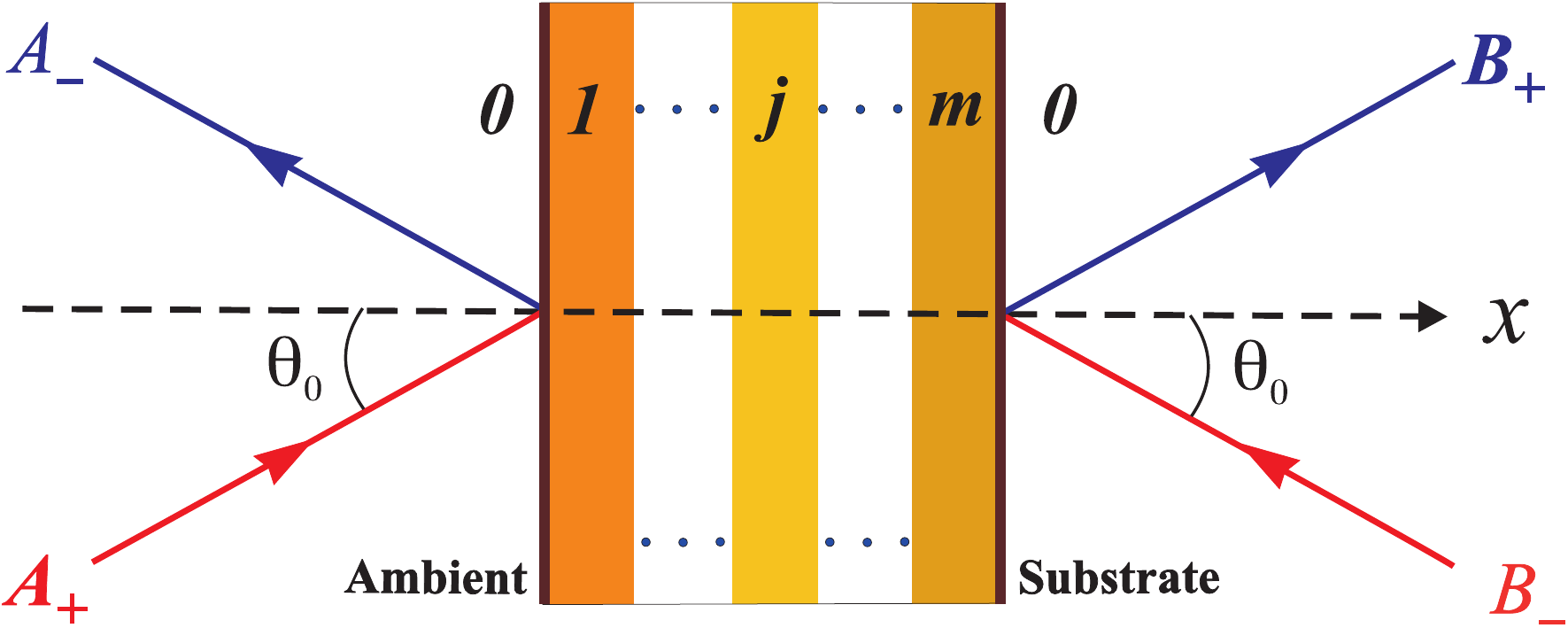}
  \caption{Amplitudes of the input [$A_{+}$ and $B_{-}$] and output
    [$A_{-}$ and $B_{+}$] fields in a multilayer sandwiched between
    two semi-infinite ambient and substrate identical media. The angle
    of refraction in the $j$th medium is denoted $\theta_{j}$.}
  \label{figure4}
\end{figure}

If the medium extends from $x = a$ to $x = b$, bounded by the same
homogeneous media (ambient and substrate) of index $n_{0}$, the formal
solution can be written in full analogy with
equation~(\ref{eq:movers}) and one can use a transfer matrix that
relates the field amplitudes $A_{\pm}$ and $B_{\pm}$.

Again, one can take the medium as consisting of a stack of $1, 2,
\ldots, j, \ldots, m$ plane-parallel layers, as sketched in
figure~\ref{figure4}. We denote by $n_{j}$, $d_{j}$, and $\theta_{j}$,
respectively, the refractive index, the thickness, and the angle of
refraction of the $j$th medium, which can be obtained by a repeated
application of Snell's law (which is a consequence of the conservation
of the modulus of $\mathbf{K}$)
\begin{equation}
  n_{0} \sin \theta_{0} =  \cdots   n_{j} \sin \theta_{j} = \cdots = 
  n_{m} \sin \theta_{m} \, .
\end{equation}

The transfer matrix is given by the ordered product in
equation~(\ref{matrizS2}) and the interface matrix has the same
expression as in equation~(\ref{eq:interface}), but with
\begin{equation}
  \begin{array}{lll}
    r_{ij}^{p}  = \displaystyle 
    \frac{n_j \cos \theta_i - n_i \cos\theta_j}
    {n_j \cos\theta_i + n_i \cos\theta_j},  &
    \qquad \qquad &
    t _{ij}^{p}  = 
    \displaystyle
    \frac{2n_i \cos \theta_i}
    {n_j \cos \theta_i + n_i \cos \theta_j} ,   \\
    &  \\
    r_{ij}^{s}  =  
    \displaystyle
    \frac{n_i \cos \theta_i - n_j \cos \theta_j}
    {n_i \cos \theta_i  + n_j \cos \theta_j},   &
    \qquad \qquad &  
    t_{ij}^{s} = 
    \displaystyle  
    \frac{2n_i \cos \theta_i}
    {n_i \cos \theta_i + n_j \cos \theta_j} , 
  \end{array}
\end{equation}
for each one of the basic polarizations. The propagation matrix is
also as in (\ref{eq:Prop}), but now the phase shift is
\begin{equation}
  \label{eq:shift}
  \delta_{j} = \frac{2 \pi}{\lambda} n_j d_j \cos \theta_j \, ,
\end{equation}
$\lambda$ being the wavelength in vacuum. All this gives the formalism
developed by Hayfield and White in terms of movers~\citep{Azzam:1987}.

It is also possible to develop an equivalent formalism by employing
the amplitudes in equation~(\ref{eq:StoA}), which, roughly speaking,
are the electric field and its derivative at each point. This is the
idea behind the pioneering work of \citet{Abeles:1948}.

\subsection{Geometrical optics}

\begin{figure}[b]
  \centering
  \includegraphics[height=3.5cm]{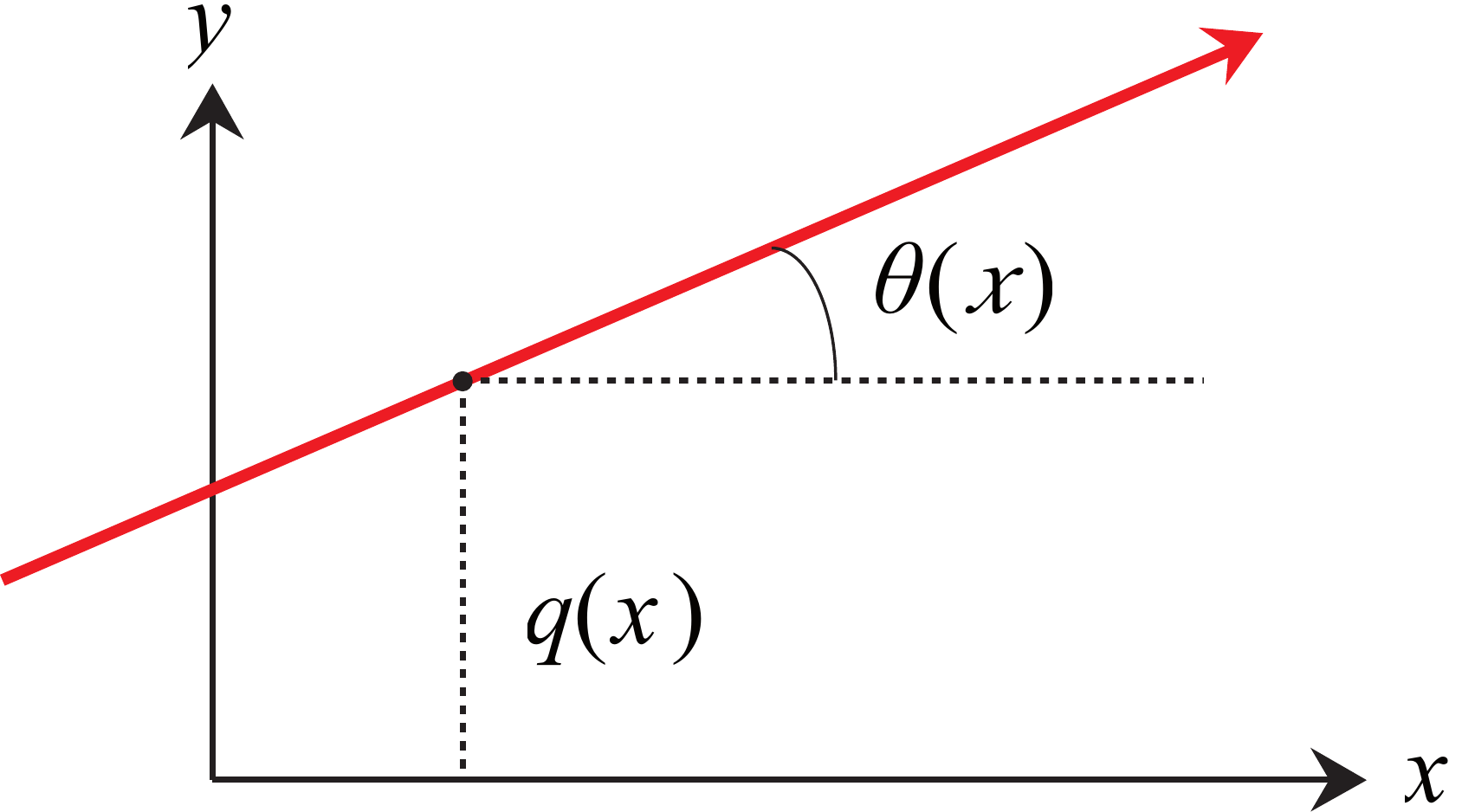}
  \caption{Notation for the ray vector. The optical axis is $x$,
    $q(x)$ is the transverse position at a point $x$ reached by a ray
    and $\theta$ is the ray inclination at that point.}
  \label{figure5}
\end{figure}

Finally, we look at the paraxial propagation of light through axially
symmetric systems, containing no tilted or misaligned
elements~\citep{Wolf:2004}. We take a Cartesian coordinate system
whose $x$ axis is along the axis of the optical system (see
figure~\ref{figure5}) and represent a ray at a plane $x$ by the
transverse position vector $q (x)$ (which can be chosen in the
meridional plane) and by the momentum $ p (x) = n \theta (x)$, which
in the paraxial limit is $ p (x) = n d q(x) /dx$, where $n$ is the
refractive index of the medium. These are canonical coordinates and
satisfy all the mathematical requirements for a consistent
description~\citep{Guillemin:1984}.

In homogeneous media, to fully specify the ray behavior three basic
matrices are needed, namely
\begin{equation}
  \label{eq:3bm}
  \left ( 
    \begin{array}{cc}
      1      &  d/n \\
      0 & 1
    \end{array}
  \right ) \, ,
  \qquad \qquad
  \left ( 
    \begin{array}{cc}
      1      &  0 \\
      (n^{\prime} - n)/R  & 1
    \end{array}
  \right ) \, ,
  \qquad \qquad
  \left ( 
    \begin{array}{cc}
      1      &  0 \\
      - 2 n/R  & 1
    \end{array}
  \right ) \, .
\end{equation}
The first one gives the propagation through a distance $d$, the second
gives the changes in the ray parameters for a refraction in a dioptre
of radius $R$ separating two homogeneous media of refractive indices
$n$ and $n^{\prime}$, and the third one is the reflection in a mirror
of radius $R$.

\begin{figure}
  \centering
  \resizebox{0.60\columnwidth}{!}{\includegraphics{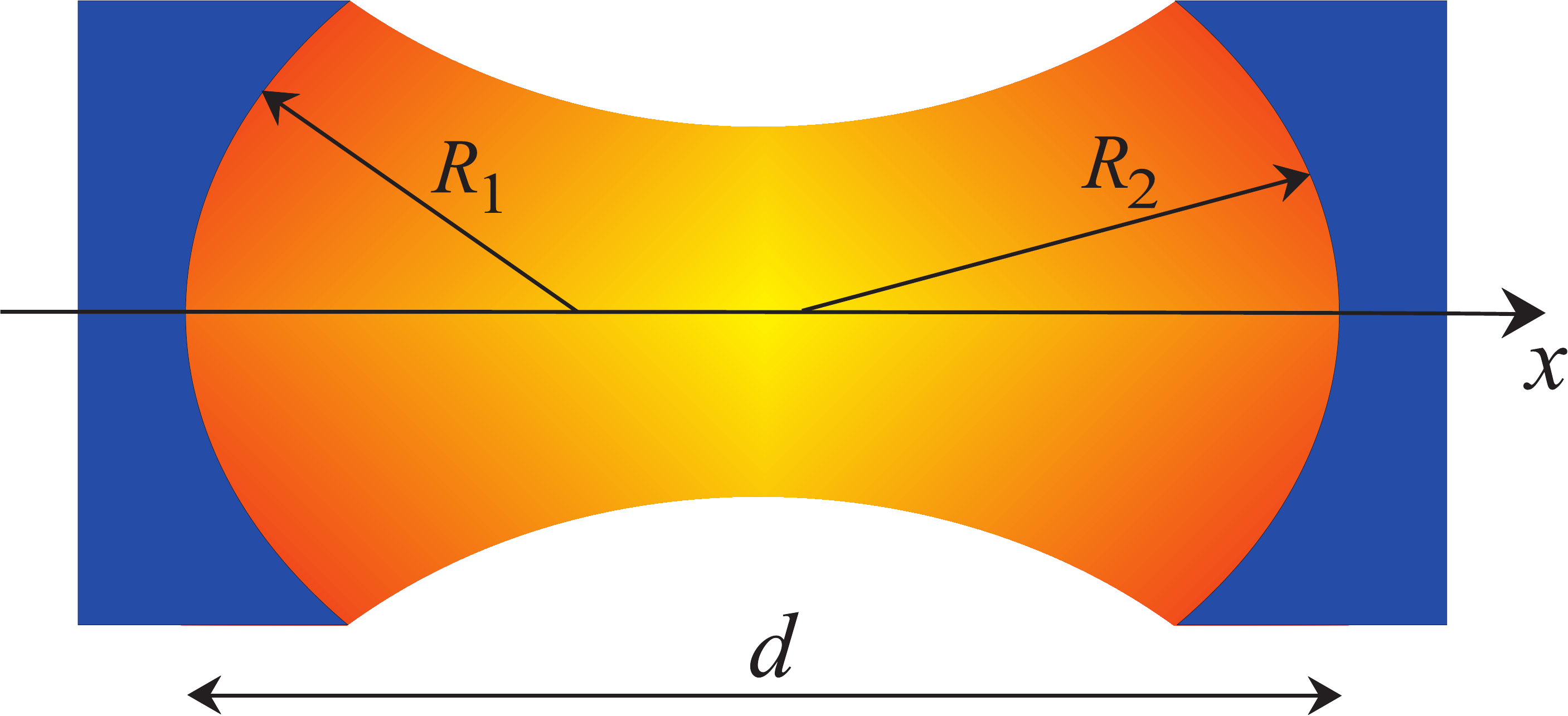}}
  \caption{Optical cavity consisting of two spherical mirrors of radii
    $R_{1}$ and $R_{2}$ separated a distance $d$.}
  \label{figure6}
\end{figure}

Let us apply these matrices to the illustrative example of an optical
cavity consisting of two spherical mirrors of radii $R_1$ and $R_2$,
separated a distance $d$, which will be examined in more detail in
Section 6.3 (see figure~\ref{figure6}) .  The ray-transfer matrix
corresponding to a round trip can be routinely computed using
(\ref{eq:3bm})~\citep{Gerrard:1975}:
\begin{equation}
  \label{CAV}
  \Matriz{M} =
  \left (
    \begin{array}{cc}
      2 g_1 g_2 - g_1 + g_2 -1 &
      \displaystyle
      \frac{d}{2} (2g_1 g_2 + g_1 + g_2 ) \\
      \displaystyle
      \frac{2}{d} (2g_1 g_2 - g_1 - g_2 ) &
      2 g_1 g_2 + g_1 - g_2 - 1
    \end{array}
  \right ) \,  ,
\end{equation}
where we have introduced the parameters ($i= 1, 2$)
\begin{equation}
  g_i = 1- \frac{d}{R_i} \, .
\end{equation}

In the same vein, a general first-order system can be built as a
cascaded application of these three basic elements and the ray
parameters change according to the simple
transformation~\citep{Simon:2000,Wolf:2004,Baskal:2004}
\begin{equation}
  \label{Mgeo}
  \left (
    \begin{array}{c}
      q_{a} \\
      p_{a}
    \end{array}
  \right ) =
  \Matriz{M}_{ab}
  \left (
    \begin{array}{c}
      q_{b} \\
      p_{b}
    \end{array}
  \right ) \, ,
\end{equation}
and $\Matriz{M}_{ab}$ is the ray-transfer matrix
\begin{equation}
  \label{defMgeo}
  \Matriz{M}_{ab} =
  \left (
    \begin{array}{cc}
      \mathfrak{a}  &    \mathfrak{b} \\
      \mathfrak{c}  &    \mathfrak{d}
    \end{array}
  \right )  \, .
\end{equation}
Since the three basic matrices in (\ref{eq:3bm}) have
unit determinant, this means that $\det \Matriz{M} = +1$, so that they
belong to the group SL(2, $\mathbb{R}$) of real unimodular $2 \times
2$ matrices. This is the essence of the celebrated $abcd$ law in
geometrical optics.

\section{The geometry of the transfer matrix}

\subsection{Transfer function in the unit disc}

\begin{figure}
  \centering
  \includegraphics[height=5.5cm]{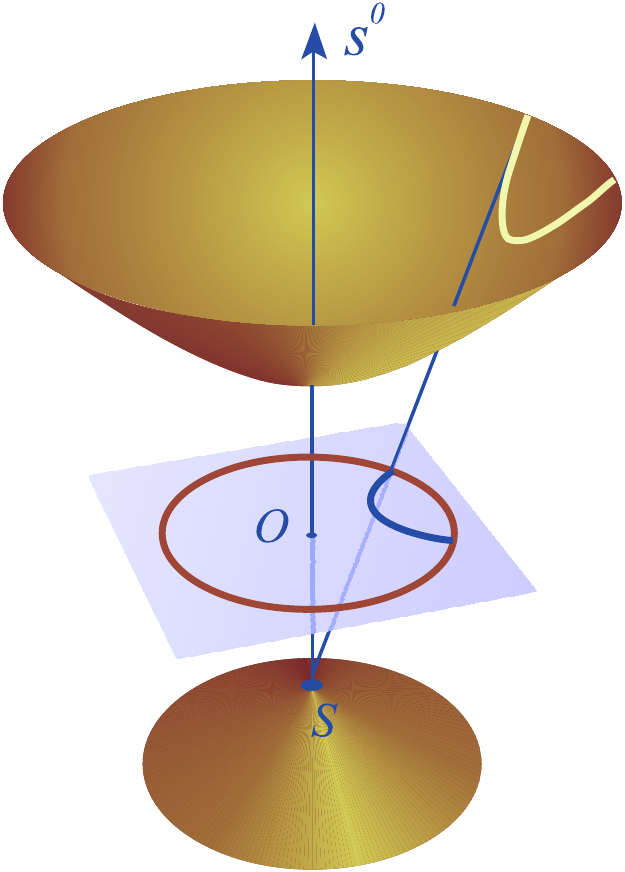}
  \caption{Outline of the unit hyperboloid and a geodesic on it.  We
    also show how a hyperbolic line is obtained in the unit disk by
    stereographic projection, taking the south pole as projection
    center.}
  \label{figure7}
\end{figure}

Let us go back to the unit two-sheeted hyperboloid (\ref{twosheet})
that is the phase space for our problem. If one uses stereographic
projection taking the south pole $S = (-1, 0, 0)$ as projection center
(see figure~\ref{figure7}), the projection of the point $(s^0, s^1,
s^2)$ becomes in the complex plane
\begin{equation}
  \label{defz}
  z = \frac{s^{1} + i s^{2}}{1 + s^{0}}=
  \frac {X_{-}}{X_{+}} \, ,
\end{equation}
for $X= A$ or $B$. This confirms that what matters here are the
transformation properties of amplitude quotients rather than the amplitudes
themselves. In terms of the pseudospherical coordinates
(\ref{pseudosp}), this point can be written as
\begin{equation}
  z = \frac{\sinh \chi}{1 + \cosh \chi} \exp(i \varphi) =
  \tanh(\chi/2) \exp(i \varphi)  \, ,
\end{equation}
which allows to interpret $\chi$ as a rapidity and $\varphi$ as a
phase shift. The upper sheet of the unit hyperboloid is projected into
the unit disc, we shall denote $\mathbb{D}$, the lower sheet into the
external region, while the infinity goes to the boundary of the unit
disc. We mention in passing that stereographic projection is
conformal, meaning that it preserves the angles at which curves cross
each other on the two-sheeted hyperboloid.

Through stereographic projection, the standard Minkowski distance in
the unit hyperboloid becomes in $\mathbb{D}$~\citep{Anderson:1999}
\begin{equation}
  \label{metricH}
  ds^{2} = \frac{dz \, dz^{\ast}}{( 1 - |z|^{2} )^{2} }  \, .
\end{equation}
The geodesics in the hyperboloid are intersections with the
hyperboloid of planes passing through the origin. Consequently,
hyperbolic lines are obtained from these by stereographic projection
and they correspond to circle arcs that orthogonally cut the boundary
of the unit disk (diameters are a particular instance of these
geodesics), as equation~(\ref{metricH}) confirms after some
calculations~\citep{Mischenko:1988}.

It seems natural to consider the complex variables in
equation~(\ref{defz}) for both points $a$ and $b$. The basic linear
relation expressed in equation~(\ref{M}) settles a transformation on
the complex plane $\mathbb{C}$, mapping the point $z_{b}$ into the
point $z_{a}$ according to
\begin{equation}
  \label{accion}
  z_{a} = \Phi [\matriz{M}_{ab} , z_{b}] =
  \frac{\alpha^\ast z_{b} + \beta^\ast}
  {\beta z_{b} + \alpha } \, ,
\end{equation}
which is a bilinear or M\"{o}bius transformation.  The action of the
transfer matrix appears then as a function $z_{a} = f(z_{b})$ that can
be appropriately called the transfer function~\citep{Yonte:2002}.  One
can check that the unit disk $\mathbb{D}$, the external region and the
boundary remain also invariant under this action.

We shall need the concept of hyperbolic distance in the unit disc. To this end,
it is customary to define the cross ratio of four distinct points
$z_A$, $z_B$, $z_C$, and $z_D$ as the number
\begin{equation}
  (z_A, z_B|z_C, z_D ) =
  \frac{(z_A -z_C )/(z_B -z_C)}
  {(z_A -z_D )/(z_B -z_D)} \, ,
\end{equation}
which is real only when the four points lie on a circle or a straight
line. In fact, bilinear transformations preserve this cross
ratio~\citep{Pedoe:1970}.

Let now $z$ and $z^\prime$ be two points that are joined by the
hyperbolic line whose endpoints on the unit circle are $e$ and
$e^\prime$.  The hyperbolic distance between $z$ and $z^\prime$ is
\begin{equation}
  \label{hdis}
  d_\mathrm{H} (z, z^\prime )= \frac{1}{2} | 
  \ln (e, e^\prime|z,  z^\prime ) | \, .
\end{equation}
The fundamental point for us is that bilinear transformations are
isometries; i.e., they preserve this distance.

The visual import of the disk with this metric is that a pair of
points with a given distance between will appear to be closer and
closer as their location approaches the boundary circle. Or,
equivalently, a pair of points near the unit circle are actually
farther apart (via the metric) that a pear near the center of the
disc, which appear to be the same distance apart.

An important tool for the classification of the transfer-matrix action
are the fixed points, which correspond to the wave functions
such that $z_{a} = z_{b}$ in equation~(\ref{accion}). If we denote
them by $z_{f}$ we have that
\begin{equation}
  z_{f} = \Phi [\matriz{M}_{ab} , z_{f}]  \, ,
\end{equation}
whose solutions are
\begin{equation}
  z_{f \pm} = \frac{1}{2 \beta}
  \left \{  
    -2 i \im (\alpha) \pm   \sqrt{[\Tr ( \matriz{M}_{ab} )]^2 -4}
  \right \} \, .
\end{equation}

When $ |\Tr ( \matriz{M}_{ab} ) | < 2$ the action is said elliptic and
it has only one fixed point inside $\mathbb{D}$, while the other lies
outside. Since in the Euclidean geometry a rotation is characterized
for having only one invariant point, this action can be appropriately
called a hyperbolic rotation.

When $ |\Tr ( \matriz{M}_{ab} ) | > 2$ the action is hyperbolic and it
has two fixed points both on the boundary of $\mathbb{D}$. The
hyperbolic line joining these two fixed points remains invariant and
thus, by analogy with the Euclidean case, this action will be named a
hyperbolic translation.

Finally, when $ | \Tr (\matriz{M}_{ab} ) | = 2$ the action is
parabolic and it has only one (double) fixed point on the boundary of
$\mathbb{D}$. This action has no Euclidean analogy and will be called
a parallel displacement for reasons that will become clear soon.

In is worth mentioning that for the example of the rectangular barrier
discussed in equations~(\ref{Barellip})-(\ref{Barpar}), the associated
actions are elliptic, hyperbolic, or parabolic according to whether
$E$ is greater than, less than, or equal to $V_{0}$, respectively.

To proceed further, let us note that by taking the conjugate of
$\matriz{M}_{ab}$ with any matrix $\matriz{C} \in $ SU(1,~1); i.e.,
\begin{equation}
  \label{conjC}
  \hat{\matriz{M}}_{ab} = 
  \matriz{C} \,  \matriz{M}_{ab} \, \matriz{C}^{-1} \,  ,
\end{equation}
we get another matrix of the same type, forasmuch as $\Tr (
\hat{\matriz{M}}_{ab} ) = \Tr (\matriz{M}_{ab})$. Conversely, if two
transfer matrices have the same trace, one can always find a matrix
$\matriz{C}$ satisfying equation~(\ref{conjC}). The fixed points of
$\hat{\matriz{M}}_{ab}$ are the image by $\matriz{C}$ of the fixed
points of $\matriz{M}_{ab}$. In fact, if we write the matrix
$\matriz{C}$ as
\begin{equation}
  \label{matC}
  \matriz{C} =
  \left (
    \begin{array}{cc}
      \ \mathfrak{c}_1  \ & \ \mathfrak{c}_2 \ \\
      \ \mathfrak{c}^\ast_2 \  & \ \mathfrak{c}^\ast_1 \
    \end{array}
  \right ) \,  ,
\end{equation}
the matrix elements of $\hat{\matriz{M}}_{ab}$ (marked by carets) and
those of $\matriz{M}_{ab}$ are related by
\begin{eqnarray}
  \hat{\alpha} & = &  \alpha |\mathfrak{c}_1|^2 -
  \alpha^\ast |\mathfrak{c}_2|^2 -
  2 i \im (\beta \mathfrak{c}_1
  \mathfrak{c}^\ast_2 ) \, , \nonumber \\
  & & \\
  \hat{\beta} & = &  \beta \mathfrak{c}_1^2 -
  \beta^\ast \mathfrak{c}_2^2 -
  2 i  \mathfrak{c}_1 \mathfrak{c}_2
  \im (\alpha) \, . \nonumber
\end{eqnarray}
In consequence, given any transfer matrix $\matriz{M}_{ab}$ one can
always reduce it to a $\hat{\matriz{M}}_{ab}$ with one of the
following canonical forms
\begin{eqnarray}
  \label{canonical}
  \hat{\matriz{K}} (\phi) & = &
  \left (
    \begin{array}{cc}
      \exp (i\phi/2) & 0 \\
      0 & \exp (-i\phi/2)
    \end{array}
  \right ) \,  ,
  \nonumber \\
  \hat{\matriz{A}}(\xi) & = &
  \left (
    \begin{array}{cc}
      \cosh (\xi/2) & i \, \sinh(\xi/2) \\
      -i\, \sinh(\xi/2) & \cosh (\xi/2)
    \end{array}
  \right ) \, , \\
  \hat{\matriz{N}}( \nu) & = &
  \left (
    \begin{array}{cc}
      1 - i\, \nu/2 & \nu/2 \\
      \nu/2 & 1+ i\, \nu/2
    \end{array}
  \right ) \,  ,
  \nonumber
\end{eqnarray}
where $ 0 \le \phi \le 4 \pi$ and $\xi, \nu \in \mathbb{R}$.  They
have as fixed points the origin (elliptic), $+i$ and $-i$ (hyperbolic)
and $+i$ (parabolic).  All these SU(1, 1) matrices leave invariant
$|X_{+}|^2 - |X_{-}|^2$ at each side of the potential, in agreement
with equation~(\ref{fluxcon}).  In addition, $\hat{\matriz{K}} (\phi)$
preserves the product $X_{+} X_{-}$, $\hat{\matriz{A}} (\xi) $
preserves the quadratic form $X_{+}^{2} + X_{-}^{2}$, and
$\hat{\matriz{N}} (\nu)$ preserves the sum $X_{+} + i \,
X_{-}$~\citep{Yonte:2002}.

The matrix $\hat{\matriz{K}} (\phi)$ represents the free propagation
in a constant potential barrier with a dephasing of $\phi/2$.
Obviously, this reduces to a mere shift of the origin of phases.  The
second matrix $\hat{\matriz{A}}(\xi)$ represents a symmetric system
with reflection and transmission phase shifts of
$\tau_{\hat{\matriz{A}}} = 0$ and $\rho_{\hat{\matriz{A}}} = \pm
\pi/2$, and a transmission coefficient $\coeft_{\hat{\matriz{A}}} =
\mathrm{sech} (\xi/2)$. Finally, the third matrix,
$\hat{\matriz{N}}(\nu)$, represents a system having
$\coeft_{\hat{\matriz{N}}} = \cos( \tau_{\hat{\matriz{N}}}) \exp ( i
\tau_{\hat{\matriz{N}}})$ and $\coefr_{\hat{\matriz{N}}} =
\sin(\tau_{\hat{\matriz{N}}}) \exp(i \tau_{\hat{\matriz{N}}})$, with $
\tan (\tau_{\hat{\matriz{N}}}) = \nu/2$. There are many ways to
implement these elementary actions depending on the physical system
under consideration~\citep{Simon:1998}.

The explicit construction of the family of matrices $\matriz{C}$ is
easy: it suffices to impose that $\matriz{C}$ transforms the fixed
points of $\matriz{M}_{ab}$ into the ones of $\hat{\matriz{K}}(\phi)$,
$\hat{\matriz{A}}(\xi)$, or $\hat{\matriz{N}}(\nu)$.  By way of
example, let us examine the case when $\matriz{M}_{ab}$ is elliptic
and its fixed point inside the unit disk is $z_{f}$. One should
have
\begin{equation}
  \Phi[\matriz{C} \matriz{M}_{ab} \matriz{C}^{-1}, 0]
  = \Phi[\matriz{C} \matriz{M}_{ab}, z_f] =
  \Phi[\matriz{C}, z_f] = 0 \, .
\end{equation}
Solving this equation one gets directly
\begin{equation}
  \mathfrak{c}_1  =  \frac{1}{\sqrt{1 - |z_f|^2}}
  \exp(i \vartheta ) \, , 
  \qquad \qquad
  \mathfrak{c}_2  =  - \mathfrak{c}_1 z^\ast_f \, ,
\end{equation}
and $\vartheta$ is a real free parameter. The same procedure applies
to matrices $\hat{\matriz{A}}(\xi)$ and $\hat{\matriz{N}}(\nu)$.

\begin{figure}
  \centering
  \resizebox{0.80\columnwidth}{!}{\includegraphics{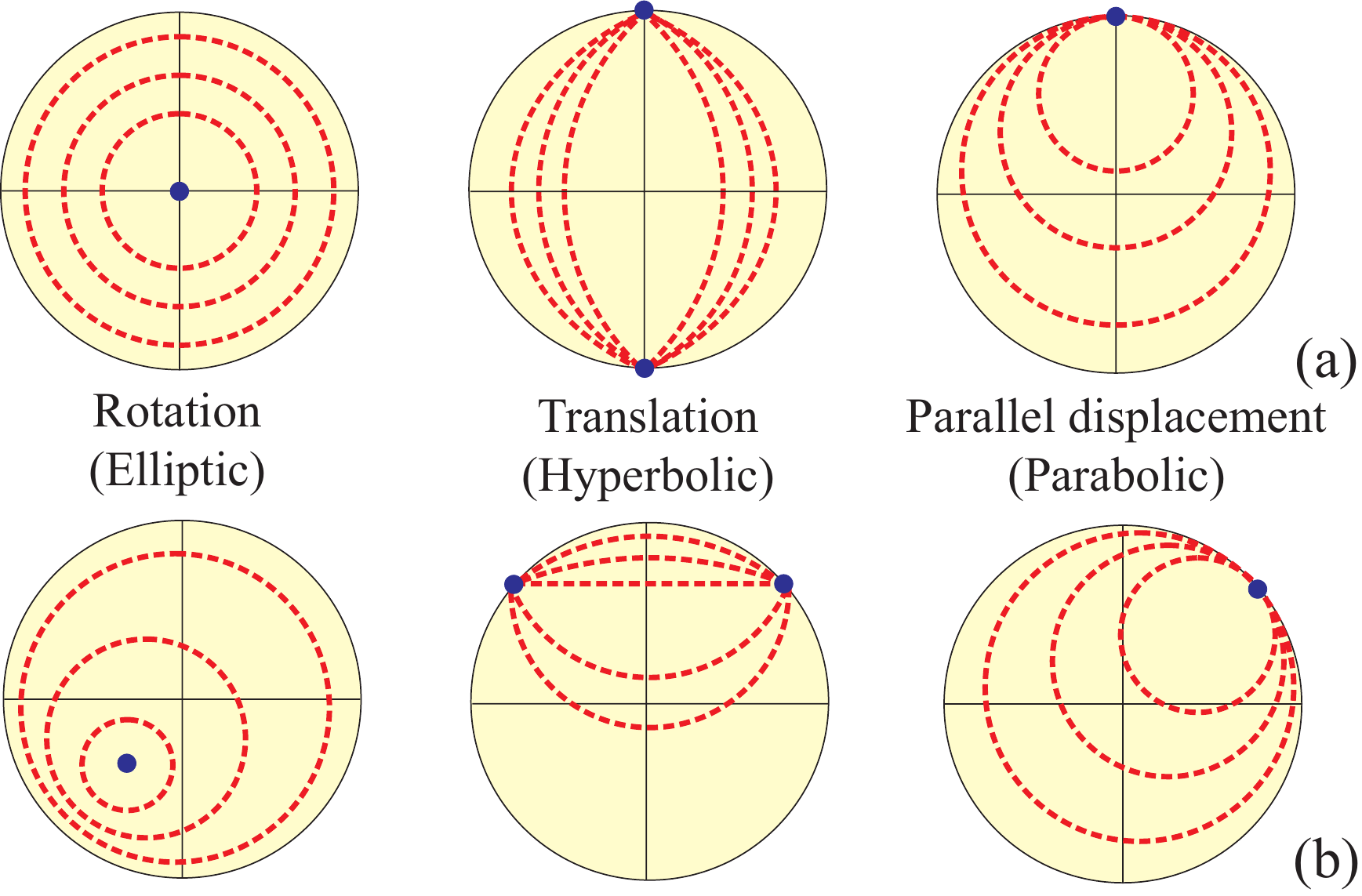}}
  \caption{Plot of orbits in the unit disk for: (a) canonical transfer
    matrices in equation~(\ref{canonical}) and (b) arbitrary
    transfer matrices.}
  \label{figure8}
\end{figure}

The concept of orbit is especially appropriate for getting an
intuitive picture of these actions. Given a point $z$, its orbit is
the set of points $z^\prime$ obtained from $z$ by the action of all
the elements of the group. In figure~\ref{figure8}.a we have plotted
typical orbits for each one of the canonical forms
$\hat{\matriz{K}}(\phi)$, $\hat{\matriz{A}} (\xi)$, and
$\hat{\matriz{N}} (\nu)$. They are
\begin{eqnarray}
  z^\prime & = &  \Phi [\hat{\matriz{K}} (\phi), z ] = 
  z \,  \exp (-i\phi) \, , \nonumber \\
  z^\prime & = & \Phi [\hat{\matriz{A}} (\xi), z ] =  
  \frac{z - i\, \tanh(\xi/2)}{1 + i\, z  \tanh(\xi/2)} \,  , \\
  z^\prime & =  & \Phi [ \hat{\matriz{N}} (\nu), z ] = 
  \frac{z + (1 + iz) \nu/2}{1 + (z - i) \nu/2} \,  . 
  \nonumber
\end{eqnarray}
For matrices $\hat{\matriz{K}}(\phi)$ the orbits are circumferences
centered at the origin and there are no invariant hyperbolic
lines. For $\hat{\matriz{A}} (\xi)$, they are arcs of circumference
going from the point $ +i$ to the point $-i$ through $z$ and they are
known as hypercicles. Every hypercicle is equidistant [in the sense of
the distance (\ref{hdis})] from the imaginary axis, which remains
invariant (in the Euclidean plane the locus of a point at a constant
distance from a fixed line is a pair of parallel lines). Finally, for
$\hat{\matriz{N}} (\nu)$ the orbits are circumferences passing through
the point $+ i$ and joining the points $z$ and $-z^\ast$ and they are
denominated horocycles: they can be viewed as the locus of a point that
is derived from the point $+ i$ by a continuous parallel
displacement~\citep{Coxeter:1968}.

For a general $\matriz{M}_{ab}$ the corresponding orbits can be
obtained by transforming with the appropriate matrix $\matriz{C}$ the
orbits delineated before.  In figure~\ref{figure8}.b we have plotted
examples of such orbits for elliptic, hyperbolic, and parabolic
actions. We stress that once the fixed points of the transfer matrix
are known, one can ensure that $z_a$ will lie in the orbit associated
to $z_b$.

\begin{figure}
  \centering
  \resizebox{0.95\columnwidth}{!}{\includegraphics{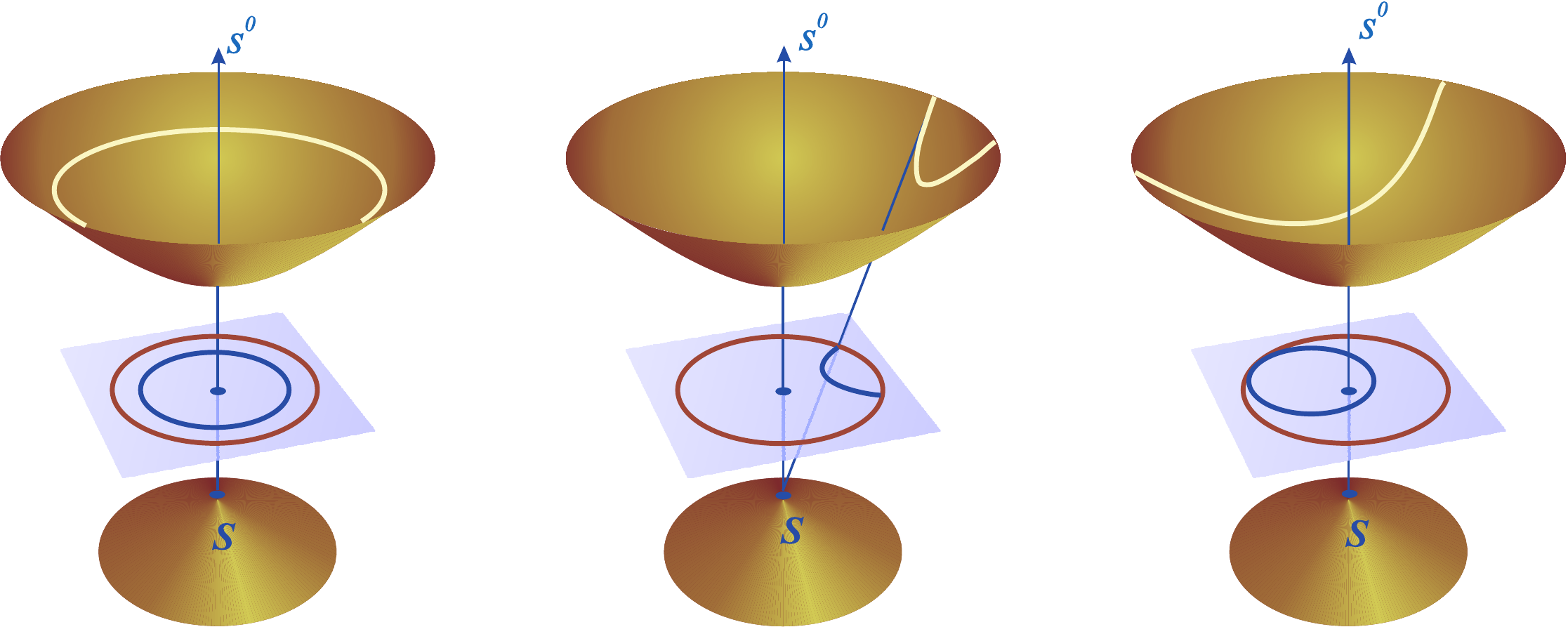}}
  \caption{Unit hyperboloids defined in equation~(\ref{twosheet}),
    representing the space of states for SO(2,1). In each one of them
    we have plotted a typical orbit for the matrices
    $\Lambda_{\hat{\matriz{K}}}$,   $\Lambda_{\hat{\matriz{A}}}$, and
    $\Lambda_{\hat{\matriz{N}}}$ (from left to right). In all the figures we have
    performed stereographic projection from the south pole $S$ of the
    hyperboloid, to obtain the unit disk in the plane $s^0 =0$ and the
    corresponding orbits, which are the actions of the SU(1,~1)
    matrices.}
  \label{figure9}
\end{figure}

An alternative way to understand these results is to look at the
canonical matrices in the Lorentz group SO(2, 1). In fact, using
(\ref{relation}) one finds that
\begin{eqnarray}
  \label{Iwarel}
  \Lambda_{\hat{\matriz{K}}} (\phi ) & = & 
  \left (
    \begin{array}{ccc}
      1 &  0 & 0 \\
      0 &   \cos \phi &   \sin \phi \\
      0 & -\sin \phi &  \cos \phi
    \end{array}
  \right ) \, , 
  \nonumber \\
  \Lambda_{\hat{\matriz{A}}} (\xi) & = & 
  \left (
    \begin{array}{ccc}
      \cosh \xi &  0 & -\sinh \xi \\
      0 &   1 & 0 \\
      -\sinh \xi & 0 &  \cosh \xi
    \end{array}
  \right ) \, ,  \\
  \Lambda_{\hat{\matriz{N}}} (\nu) & = & 
  \left (
    \begin{array}{ccc}
      1 + (\nu^2/2) &  \nu & -\nu^2/2 \\
      \nu &   1 & -\nu \\
      \nu^2/2 & \nu &  1 - (\nu^2/2)
    \end{array}
  \right ) \, .  
  \nonumber
\end{eqnarray}
The action of these matrices in SO(2,1) is clear:
$\Lambda_{\hat{\matriz{K}}} (\phi )$ is a space rotation of angle
$\phi$ in the $1-2$ plane, $\Lambda_{\hat{\matriz{A}}} (\xi)$ is a
boost in the direction of the axis $2$ with velocity $\varv = \tanh
\xi$; and, finally, $\Lambda_{\hat{\matriz{N}}} (\nu)$ is a space
rotation of angle $\tau_{\hat{\matriz{N}}}$ [such that $\tan
(\tau_{\hat{\matriz{N}}}) = \nu/2$] followed by a boost of angle
$\tau_{\hat{\matriz{N}}}$ and velocity $v = \tanh (\nu/2)$, both in
the $1-2$ plane. In figure~\ref{figure9} we have plotted examples of
the orbits for each one of the subgroups in (\ref{Iwarel}).  For
$\Lambda_{\hat{\matriz{K}}} (\phi)$ the orbits are the intersection of
the hyperboloid with planes $s^0= \mathrm{constant}$, for
$\Lambda_{\hat{\matriz{A}}} (\xi)$ with planes $s^1=
\mathrm{constant}$, and for $\Lambda_{\hat{\matriz{N}}}(\nu)$ with
planes $s^0 - s^2 = \mathrm{constant}$. Through stereographic
projection we get the corresponding orbits for the matrices
(\ref{canonical}) in the unit disc.

\subsection{Transfer function in the half-plane}

The unitary matrix (\ref{Cayley}) plays an important role in the
intertwining between the two basic vector bases used for the
transfer-matrix description. In mathematical terms, $\Matriz{U}$
establishes a one-to-one map between the groups SU(1,~1) and SL(2,
$\mathbb{R}$). To investigate the meaning of this map we observe that
if the point $w \in \mathbb{C}$ is defined in terms of $z$ by
\begin{equation}
  \label{DtoH}
  w = \Phi [\Matriz{U}, z ]  =  \frac{z+i}{1 + i\,z} \,  ,
\end{equation}
then the interior of $\mathbb{D}$ is mapped onto the upper half-plane
of the complex plane $w$, the boundary maps onto the real axis, while
the exterior of $\mathbb{D}$ becomes the lower half-plane. This
remarkable map is known as the Cayley transform.

The metric now reads as
\begin{equation}
  ds^{2} = \frac{dw \, dw^{\ast}}{( \im w )^{2}} \, ,
\end{equation}
and the geodesic lines are the open semicircles
orthogonal to the real axis.  The M\"{o}bius transformations are
\begin{equation}
  \label{bilinearH} 
  w_{b} =  \Phi [\Matriz{M}_{ab}, w_{a} ] = 
  \frac{\mathfrak{d}  w + \mathfrak{c} }
  {\mathfrak{b} w + \mathfrak{a} } \, ,
\end{equation}
with $\Matriz{M}_{ab}$ obtained from $\matriz{M}_{ab}$ through
conjugation with $\Matriz{U}$ as in equation~(\ref{MconjU}). They are
also isometries.

The points $w$ in the upper half-plane constitute the Poincar\'e model
of the hyperbolic plane $\mathbb{H}$~\citep{Stahl:1993}. In this way,
one can transport all the geometrical properties of the unit disc
$\mathbb{D}$ to the upper half-plane $\mathbb{H}$.

Since the matrix conjugation does not change the trace, the same
geometrical classification in three basic actions still holds. In
fact, by conjugating with $\Matriz{U}$ the canonical forms
(\ref{canonical}), the corresponding ones for SL(2, $\mathbb{R}$) are
\begin{eqnarray}
  \label{Iwasa2}
  \hat{\Matriz{K}} (\phi ) & = & 
  \left (
    \begin{array}{cc}
      \cos ( \phi/2) & \sin ( \phi/2) \\
      - \sin (\phi/2) & \cos ( \phi/2)
    \end{array}
  \right ) \,  ,
  \nonumber \\
  \hat{\Matriz{A}} (\xi) & = & 
  \left (
    \begin{array}{cc}
      e^{\xi/2}  & 0 \\
      0  & e^{-\xi/2}
    \end{array}
  \right ) \,  , \\
  \hat{\Matriz{N}} ( \nu ) & = &
  \left (
    \begin{array}{cc}
      1 & 0 \\
      \nu & 1
    \end{array}
  \right ) \, . 
  \nonumber
\end{eqnarray}
These matrices have as fixed points $+i$ (elliptic), 0 and $ \infty$
(hyperbolic), and $ \infty$ (parabolic), respectively.  Clearly,
$\hat{\Matriz{K}} (\phi )$ is a rotation in phase space, also
termed a fractional Fourier transform, while $\hat{\Matriz{A}}$ is
sometimes called a squeezer or hyperbolic magnifier: it scales
the positive amplitude $+$ up by the factor $e^{\xi/2}$ and the
negative one $-$ down by the same factor. Finally, $\hat{\Matriz{N}} (
\nu )$ represents the action of a thin lens of power $\nu$ (i.e.,
focal length $1/\nu$) in geometrical optics~\citep{Wolf:2004}.

For the canonical forms (\ref{Iwasa2}), the orbits for a point $w$ are
\begin{eqnarray}
  \label{QactH} 
  w^\prime & = & \frac{\cos (\phi/2) w - \sin
    (\phi/2)} {\sin (\phi/2) w  +
    \cos (\phi /2)} \, , 
  \nonumber \\
  &  & \nonumber \\
  w^\prime & = & e^{- \xi} w \, , \\
  & & \nonumber \\
  w^\prime & = & w + \nu \, . 
  \nonumber
\end{eqnarray}

\begin{figure}
  \centering
  \resizebox{0.80\columnwidth}{!}{\includegraphics{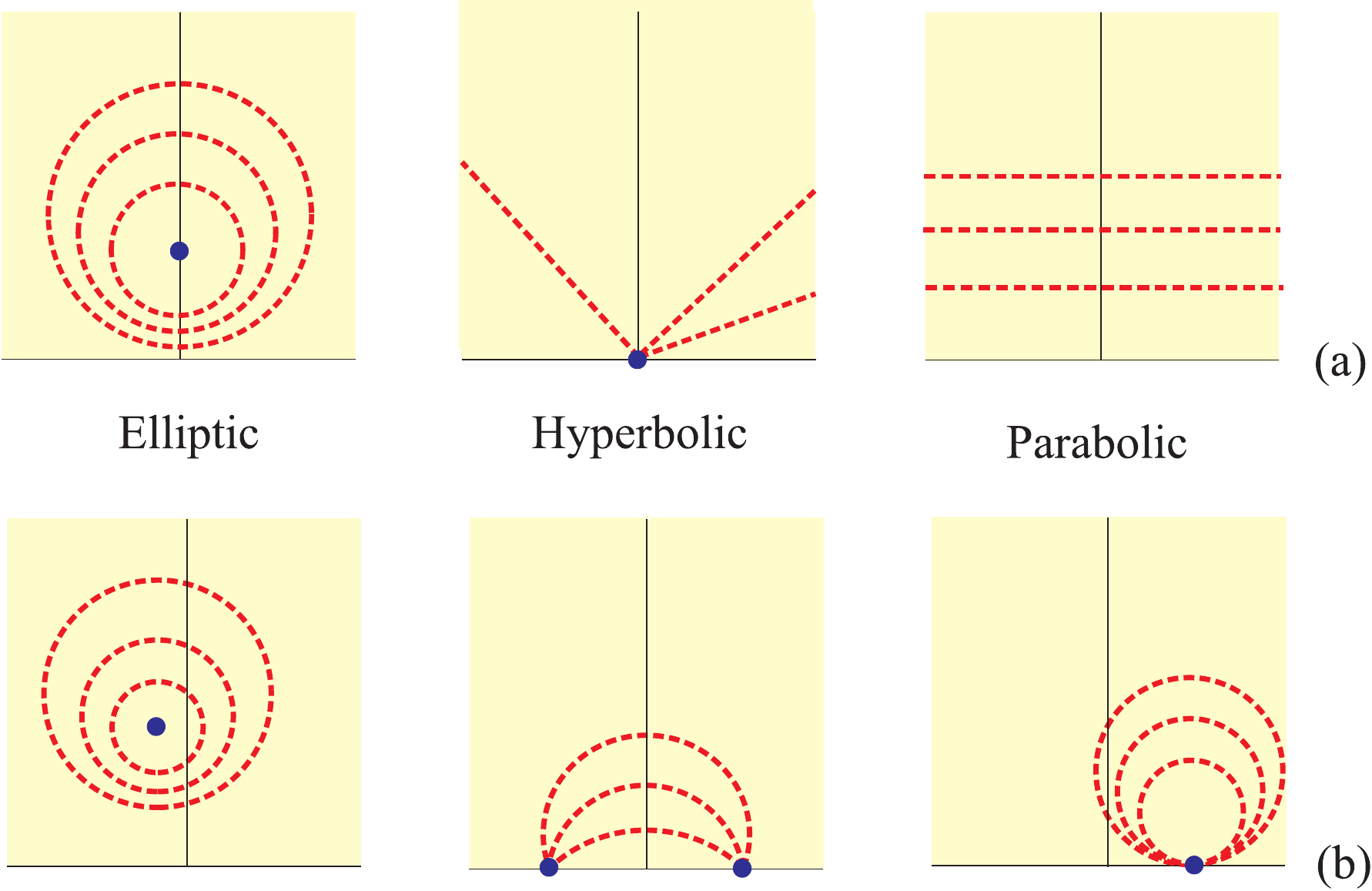}}
  \caption{Plot of typical orbits in the hyperbolic plane
    $\mathbb{H}$: (a) canonical transfer matrices as in
    equation~(\ref{Iwasa2}) and (b) arbitrary transfer matrices
    obtained by matrix conjugation.}
  \label{figure10}
\end{figure}

In figure~\ref{figure10}.a we have plotted these orbits. For matrices
$\hat{\Matriz{K}} (\phi)$ they are circumferences centered at the
invariant point $+ i$ and passing through $w$ and $-1/w$. For
$\hat{\Matriz{K}} (\xi)$, they are lines going from 0 to the $\infty$
through $w$ and they corresponds to hypercicles. Finally, for matrices
$\hat{\Matriz{N}} (\nu)$ the orbits are lines parallel to the real
axis passing through $w$ and they are the
horocycles~\citep{Coxeter:1969}. It is in the plane $\mathbb{H}$ where
the denomination of parallel displacements becomes manifest.

As we did before, for a general matrix $\Matriz{M}_{ab}$ the
corresponding orbits can be obtained by transforming with the
appropriate matrix $\Matriz{C}$ that transforms the fixed points of
$\Matriz{M}_{ab} $ into the ones of $\hat{\Matriz{K}} (\phi)$,
$\hat{\Matriz{A}} (\xi)$, or $\hat{\Matriz{N}} (\nu)$, respectively.
In figure~\ref{figure10}.b we have plotted typical examples of such
orbits.

\subsection{Factoring the transfer matrix}

Many types of factorizations have been considered in the
literature~\citep{Arsenault:1983,Abe:1994,Shamir:1995}, all of them
decomposing the matrix as a unique product of other matrices of
simpler interpretation. Particularly, given the relevant role played
by the Iwasawa decomposition, both in fundamental studies and in
applications to several fields, one is tempted to investigate also its
role in the transfer-matrix formalism.

Without embarking us in mathematical subtleties, the Iwasawa
decomposition is established as
follows~\citep{Barut:1977,Helgason:1978}: any element of a (noncompact
semi-simple) Lie group can be written as an ordered product of three
elements, taken one each from a maximal compact subgroup $K$, a
maximal Abelian subgroup $A$, and a maximal nilpotent subgroup
$N$. Furthermore, such a decomposition is global and essentially
unique.

For a matrix $\matriz{M}_{ab} \in$ SU(1,~1), the decomposition reads
as
\begin{equation}
  \label{Iwa1} 
  \matriz{M}_{ab} = \hat{\matriz{K}}(\phi^\prime) \,
  \hat{\matriz{A}} (\xi^\prime ) \, \hat{\matriz{N}} (\nu^\prime) \,  ,
\end{equation}
where the matrices appearing here are of the form of the canonical
ones in equation~(\ref{canonical}), but the parameters $\phi^\prime,
\xi^\prime$, and $\nu^\prime$ are given in terms of the elements
$\alpha$ and $\beta$ of the transfer matrix by
\begin{eqnarray}
  \label{param}
  \phi^\prime /2 & = & \arg (\alpha + i \beta)\ ,  \nonumber  \\
  \xi^\prime /2  &  =  & \ln  (1/|\alpha +i \beta | )\ , \\
  \nu^\prime /2 & = & \re (\alpha  \beta^\ast)/ |\alpha + i \beta|^2 \, .  
  \nonumber
\end{eqnarray}

\begin{figure}
  \centering
  \includegraphics[height=5cm]{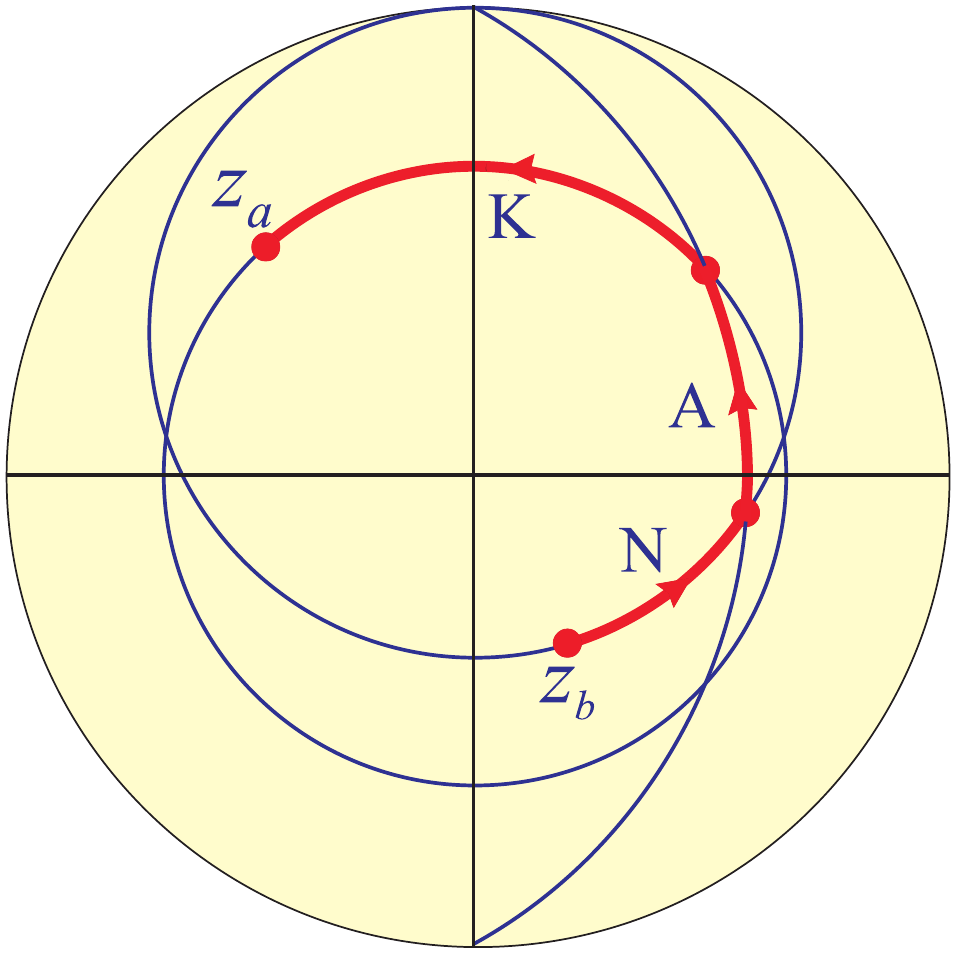}
  \caption{Representation in the unit disk of a transformation with
    the parameters indicated in the text. The point $z_{b}$ is
    transformed by the system into $z_a$. The three orbits of the
    Iwasawa decomposition are indicated.}
  \label{figure11}
\end{figure}

The importance of the Iwasawa decomposition reflects at the
geometrical level: no matter how complicated a system is, its action
can always be viewed in terms of these three basic actions with a
patent meaning. Let us show this with a practical example: we consider
a system that transforms the point $z_{b} = 0.4 \exp(- i \pi/3)$ into
$z_a = -0.44 + 0.49 \, i $ [see \citet{Monzon:2002} for a realistic
implementation]. In figure~\ref{figure11} we have plotted these points
$z_{b}$ and $z_a$ in $\mathbb{D}$. Obviously, from these data alone
we cannot infer at all the path for this discrete transformation.

The Iwasawa decomposition remedies this drawback: once we know the
values of $\phi^\prime$, $\xi^\prime$, and $\nu^\prime$ [that are
easily computed from equation~(\ref{param})] we get the intermediate
values of $z^\prime$ for the ordered application of the matrices
$\hat{\matriz{K}}(\phi^\prime)$, $\hat{\matriz{A}}(\xi^\prime)$, and
$\hat{\matriz{N}}(\nu^\prime)$, which, in fact, ensures that the
trajectory from $z_{b}$ to $z_a$ is defined through the corresponding
orbits, as shown in figure~\ref{figure11}.

\subsection{Geometrical reflections as building blocks}

In the Euclidean plane any isometry is either a rotation or a
translation. In any case, reflections are the ultimate building
blocks, since any isometry can be expressed as a composition of
reflections. In this Euclidean plane, two distinct lines are either
intersecting or parallel. Accordingly, the composition of two
reflections in two intersecting lines forming an angle $\phi$ is a
rotation of angle $2 \phi$ around the intersection point, while the
composition of two reflections in two parallel lines separated a
distance $d$ is a translation of value $2 d$.

On the other hand, in hyperbolic geometry any two distinct lines are either
intersecting (they cross in a point inside the unit disc), parallel
(they meet at infinity; i.e., at a point on the boundary of the unit
disc), or ultraparallel (they have no common points).  A natural
question arises: what is the composition of reflections in these three
different kind of lines? To some extent, the answer could be expected:
the composition is a rotation, a translation, or a parallel
displacement, respectively. However, to gain further insights one
needs to know how to deal with reflections in the unit disc.

In the Euclidean plane, given any straight line and a point $P$ which
does not lie on the line, its reflected image $P^\prime$ is such that
the line is equidistant from $P$ and $P^\prime$. In other words, a
reflection is a special kind of isometry in which the invariant points
consist of all the points on the line.

The concept of hyperbolic reflection is completely analogous: given
the hyperbolic line $\ell$ and a point $P$, to obtain its reflected
image $P^\prime$ in $\ell$ we must drop a hyperbolic line
$\mathcal{L}$ from $P$ perpendicular to $\ell$ (such a hyperbolic line
exists and it is unique) and extending an equal hyperbolic distance
[according to (\ref{hdis})] on the opposite side of $\mathcal{L}$ from
$P$. In the unit disc, this corresponds precisely to the notion of an
inversion.

We recall some facts about inversion~\citep{Coxeter:1969}.  Let $C$ be
a circle with center $\Omega$ and radius $R$. An inversion on the
circle $C$ maps the point $z$ into the point $z^\prime$ along the same
radius in such a way that the product of distances from the center
$\Omega$ satisfies
\begin{equation}
  | z^\prime - \Omega | \ | z - \Omega | = R^{2}  \, ,
\end{equation}
and hence
\begin{equation}
  z^\prime = \Omega + \frac{R^2}{z^\ast - \Omega^\ast}
  = \frac{R^2 + \Omega z^\ast - \Omega^\ast \Omega}
  {z^\ast - \Omega^\ast}  \, .
\end{equation}
If the circle $C$ is a hyperbolic line, it is orthogonal to the
boundary of the unit disk and fulfills $\Omega \, \Omega^\ast = R^2 +
1$.  In consequence
\begin{equation}
  \label{verinv}
  z^\prime = \frac{\Omega \, z^\ast -1}
  {z^\ast - \Omega^\ast}  \, .
\end{equation}
One can check~\citep{Pedoe:1970} that inversion maps circles and lines
into circles and lines, and transforms angles into equal angles
(although reversing the orientation). If a circle $C^\prime$ passes
through the points $P$ and $P^\prime$, inverse of $P$ in the circle
$C$, then $C$ and $C^\prime$ are perpendicular.  Moreover, the
hyperbolic distance (\ref{hdis}) is invariant under inversions. This
confirms that inversions are indeed reflections and so they appear as
the most basic isometries of the unit disc.

It will probe useful to introduce the conjugate bilinear
transformation associated with a matrix $\matriz{M}_{ab}$ as [compare
with equation~(\ref{accion})]
\begin{equation}
  \label{accionC}
  z_a = \Phi^\ast [\matriz{M}_{ab}, z_{b}] =
  \frac{\alpha^\ast \, z_{b}^\ast + \beta^\ast}
  {\beta \, z_{b}^\ast + \alpha}  \, .
\end{equation}
With this notation we can recast equation~(\ref{verinv}) as
\begin{equation}
  \label{accionI}
  z^\prime = \Phi^\ast [\matriz{I}_{\Omega}, z]  \, ,
\end{equation}
where the matrix $\matriz{I}_\Omega \in$ SU(1,~1) associated to the
inversion is~\citep{Barriuso:2003fk}
\begin{equation}
  \matriz{I}_{\Omega} =
  \left (
    \begin{array}{cc}
      - i \ \Omega^\ast/R & i/R \\
      - i/R & i \ \Omega/R
    \end{array}
  \right )  \, .
\end{equation}
The composition law for inversions can be stated as follows: if
$z^\prime = \Phi^\ast [\matriz{I}_\Omega, z]$ and $z^{\prime \prime} =
\Phi^\ast [\matriz{I}_{\Omega^\prime}, z^\prime]$ then
\begin{equation}
  z^{\prime \prime} =
  \Phi [\matriz{I}_{\Omega^\prime} \matriz{I}_\Omega^\ast , z]  \, .
\end{equation}

To appreciate the physical meaning of the inversion, assume that
incoming and outgoing amplitudes are interchanged in the
configuration shown in figure~1. This is tantamount to reversing the
time arrow. It is  known that for a right-traveling mover
$X_{+}$, the conjugate amplitude $[X_{+}]^\ast$ is a left
phase-conjugate wave of the original one~\citep{Zeldovich:1985}. In
other words, the time-reversal operation can be viewed in this context
as the transformation
\begin{equation}
  z \mapsto \frac{1}{z^\ast} \, ,
\end{equation}
for both $a$ and $b$; that is, it can be depicted by an inversion
in the unit circle. The transformed points lie outside the unit circle
because time reversal transforms the upper sheet into the lower sheet
of the hyperboloid.

It is easy to convince oneself that the matrix relating these
time-reversed amplitudes is precisely $\matriz{M}_{ab}^\ast$ and so
the action can be put as
\begin{equation}
  (1/z_a)^\ast = \frac{\alpha^\ast  (1/z_{b})^\ast +\beta^\ast}
  {\beta (1/z_{b})^\ast + \alpha} \, ,
\end{equation}
which expresses a general property of the time-reversal invariance.

\begin{figure}
  \centering
  \resizebox{0.80\columnwidth}{!}{\includegraphics{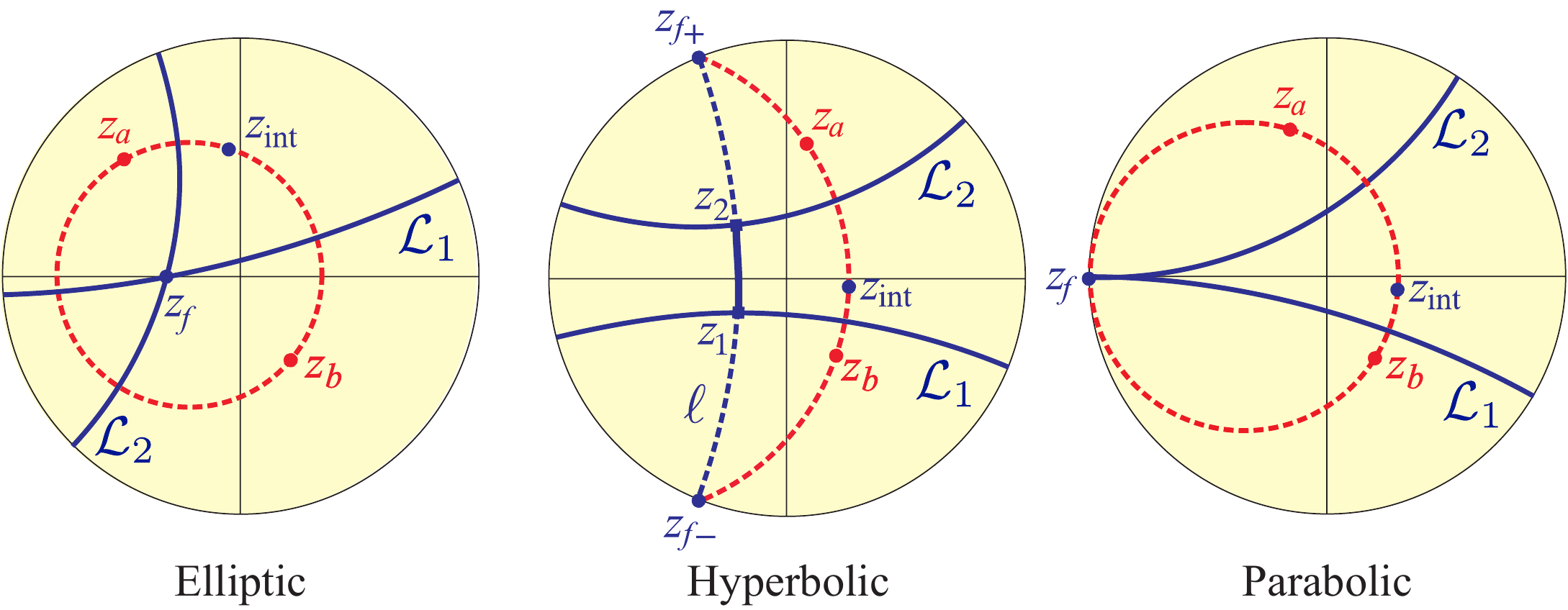}}
  \caption{Decomposition of the transfer-matrix action in terms of two
    reflections for the elliptic, hyperbolic and parabolic cases.}
  \label{figure12}
\end{figure}

In figure~\ref{figure12}, we have drawn the three basic actions as a
product of two reflections in two hyperbolic lines.  For the elliptic
case, the two hyperbolic lines $\mathcal{L}_1$ and $\mathcal{L}_2$
intersect at the fixed point $z_f$ and form an angle $\phi$, which is
just one half of the rotation angle.  The first inversion maps $z_b$
into the intermediate point $z_{\mathrm{int}}$, which is mapped into
$z_a$ by the second inversion. Note that there are infinity pairs of
lines satisfying these conditions, but chosen arbitrarily one of them,
the other is uniquely determined.  Once these lines are known, they
delimit automatically the associated inversions.

For the hyperbolic case, there are no invariant points in the unit
disc, but the hyperbolic line $\ell$ joining the fixed points $z_{f-}$ and
$z_{f+}$ is the axis of the hyperbolic translation. We have also
plotted the hypercicle passing through $z_{b}$ and $z_{a}$. The action
can be now interpreted as the composition of two reflections in two
ultraparallel hyperbolic lines $\mathcal{L}_1$ and $\mathcal{L}_2$
orthogonal to the translation axis. If $\mathcal{L}_1$ and
$\mathcal{L}_2$ intersect the axis $\ell$ at the points $z_1$ and
$z_2$, they must fulfill
\begin{equation}
  \label{tras}
  d_{\mathrm{H}} (z_b, z_a) = 2   d_{\mathrm{H}} (z_1, z_2) \, ,
\end{equation}
in complete analogy with what happens in the Euclidean plane. Once
again, there are infinity pairs of lines fulfilling this condition.

Finally, in the parabolic case, we have plotted the horocyle
connecting $z_b$ and $z_a$ and the fixed point. Now, we have the
composition of two reflections in two parallel lines $\mathcal{L}_1$
and $\mathcal{L}_2$ that intersect at the fixed point $z_f$, with
the same constraints as before.

\section{A closer look at the composition of transfer matrices}

\subsection{Setting up the inverse system}

The property (\ref{propag}) has allowed us to characterize a compound
system, as expressed more explicitly in (\ref{eq:matrizSlam}).  In its
simplest form, it states that given two potentials $V_{1}$ and
$V_{2}$, described by the transfer matrices $\matriz{M}_{1}$ and
$\matriz{M}_{2}$, with scattering amplitudes $(\coefr_1, \coeft_1)$
and $(\coefr_2, \coeft_2)$, respectively (once more we drop the
subscript $ab$ to simplify the notation), the action of the compound
system is
\begin{equation}
  \matriz{M}_{12} = \matriz{M}_1 \matriz{M}_2 \,  ,
\end{equation}
and the reflection and transmission amplitudes associated to
$\matriz{M}_{12}$ are
\begin{equation}
  \label{12}
  \coefr_{12} =  \frac{\coefr_1 + \coefr_2
    \exp (i 2 \tau_1)}
  {1 + \coefr_1^\ast \coefr_2 \exp (i 2 \tau_1)} \, ,
  \qquad \qquad
  \coeft_{12} =  \frac{\coeft_1 \coeft_2}
  {1 + \coefr_1^\ast \coefr_2 \exp (i 2 \tau_1)} \, ,
\end{equation}
with the same notation as in equation~(\ref{eq:frt}). With a bit of
effort, one can derive valuable bounds on these
coefficients~\citep{Visser:1999,Boonserm:2010fr}.

According to the general form (\ref{TranM}), the identity matrix has
unit transmission and zero reflection coefficients. In other words, it
represents an antireflection system (without transmission phase
shift). In consequence, two systems that are inverse, when composed
give an antireflection system. 

Let us investigate the outstanding example of a single potential
barrier of width $L_{1}$ and height $V_{0}$. For the time being, we
take $E > V_{0}$. Consequently, we look for another barrier of the
same height such that, when put together with the original, gives the
identity transfer matrix.

A simple glance at the coefficients in (\ref{Barellip}), reveals two
possibilities. The first is to couple other barrier with the same wave
number $\kappa$ and width $L_{2}$ such that
\begin{equation}
  \label{eq:4}
  \sin [ \kappa (L_{1} + L_{2})] = 0 \, , 
\end{equation}
in such a way that the resulting barrier presents a transmission
resonance.  The second solution is to use a ``complementary'' barrier;
that is, one having the same length and height, but opposite wave
number $- \kappa$, so it acts as canceling the effects of the first.

\begin{figure}
  \centering
  \includegraphics[height=5cm]{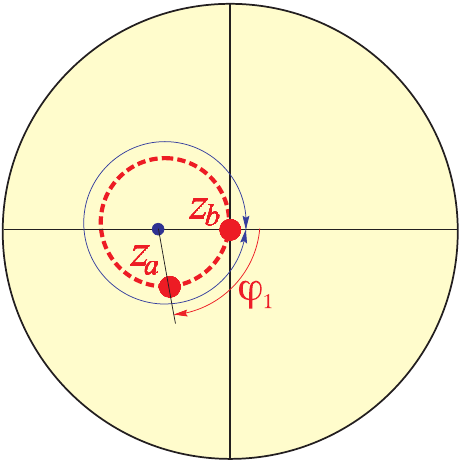}
  \caption{Rotation in the unit disk associated with a potential
    barrier described by the transfer matrix $\matriz{M}_{1}$
    transforming the point $z_{b} = 0$ into $z_{a} = r_{1} $ with the
    scattering amplitudes $(r_{1} = -0.3804 - i 0.3339i, t_{1} = 0.5689
    - i 0.6482).$ The orbits associated with the two solutions in the
    text can be either clockwise (of angle $2 \pi - \varphi_{1}$) or
    counterclockwise (of angle $\varphi_{1}$) to give the identity.}
  \label{figure13}
\end{figure}

Our geometrical picture leads to an appealing interpretation of these
facts. In figure~\ref{figure13} we have schematized the action of the
original barrier as a rotation of angle $\varphi_{1}$ around the fixed
points. The two previous solutions are a rotation of angle $2 \pi -
\varphi_{1}$ and a rotation of angle $- \varphi_{1}$, respectively,
getting thus the identity in two different ways.
 
The barrier can be seen as a quantum analog of a layer in classical
optics. The condition $E > V_{0}$ ensures the classical regime in
which light striking the layer is partly reflected and partly
transmitted. On the contrary, $E < V_{0}$ corresponds to an imaginary
refractive index, producing total internal
reflection~\citep{Bohm:1989mz}. So, the condition $E > V_{0}$ prevents
the appearance of total reflection.

\begin{figure}
  \centering
  \includegraphics[height=4.5cm]{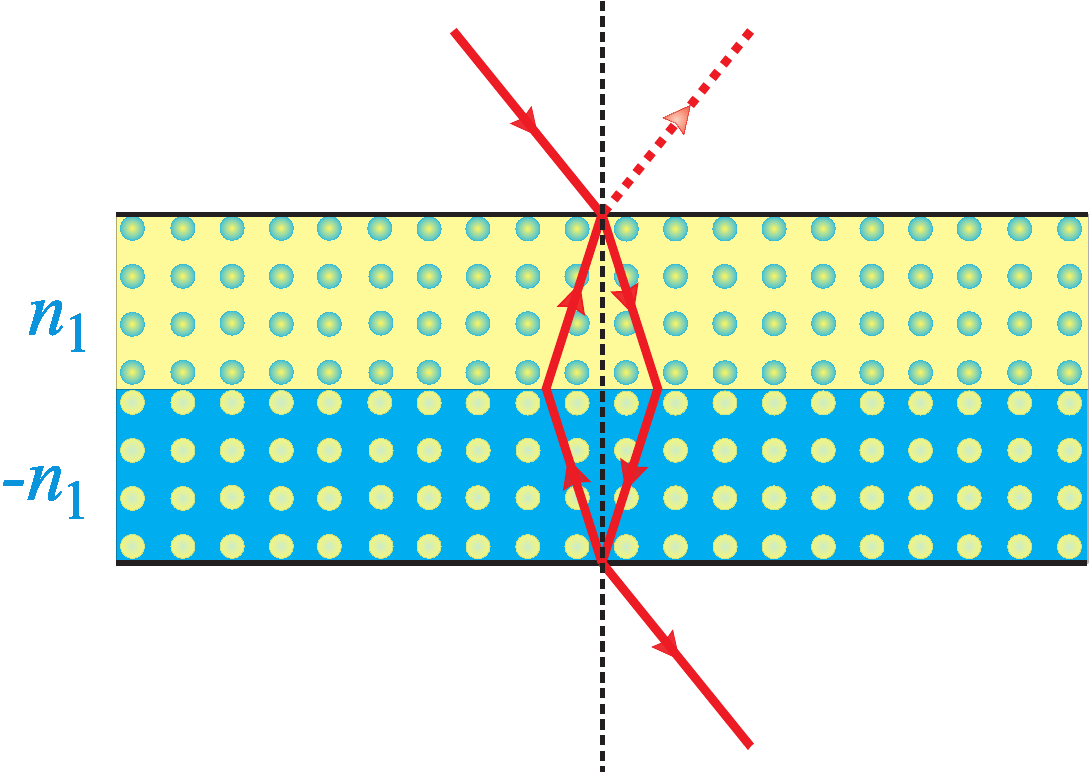}
  \caption{Scheme of the energy flow for the system resulting by
    putting together two identical slab, one made of RH and other of
    LH material. Both constitute a pair of complementary media, each
    canceling the effect of the other. Consequently, no reflection
    occurs, as indicated by the dotted line.}
  \label{figure14}
\end{figure}

Negative values of the wave vector $\kappa$ can be realized in terms
of negative effective particle mass
$m$~\citep{Kobayashi:2006ve,Dragoman:2007pd}. The analog phenomenon in
optics is a medium with both negative electrical permittivity
$\epsilon$ and magnetic permeability $\mu$. This is at the center of a
lively and sometimes heated debate~\citep{Cai:2009lr}. This idea dates
back to 1968, when \citet{Veselago:1968qy} theoretically predicted
that these remarkable materials would exhibit a number of unusual
effects derived from the fact that in them the vectors ($\mathbf{k},
\mathbf{E}, \mathbf{B}$) of a plane wave form a left-handed (LH)
rather than a right-handed (RH) set. For this reason, he called them
LH media.  One of the most interesting properties of these LH
materials is a negative refraction at the interface with a RH
medium. Our solution for complementary barriers can be interpreted as
putting together two RH and LH
slabs~\citep{Monzon:2006uq,Monzon:2008yq}. The scheme of the energy
flow in the resulting system appears in figure~\ref{figure14}.

The discussion so far admits a straightforward generalization for any
arbitrary potential. Indeed, let $\matriz{M}_{ab}$ be the transfer
matrix of a potential that can be decomposed in an arbitrary number of
barriers (some having positive and some negative values of the wave
vector $\kappa$), which can be constructed by a direct extension of
(\ref{eq:matrizSlam}).  Now, we take the potential in the reverse
order, which is represented by $\matriz{M}_{ba}$ in
(\ref{eq:MReverse}).  Next, either we complete every barrier as in
equation~(\ref{eq:4}) or we switch every barrier with positive
$\kappa_{j}$ to an identical one with negative $\kappa_{j}$ and viceversa.  In
both cases, this new system is represented by $\mathsf{M}^\ast_{ba}$.
Since one can check that
\begin{equation}
  \matriz{M}^\ast_{ba} = \matriz{M}^{-1}_{ab} \, ,
\end{equation}
when both systems are put together they give the identity. 

This substitution $\kappa_{j} \mapsto - \kappa_{j}$ formalizes in a
different framework the notion of ``complementary'' media introduced
by \citet{Pendry:2003fr}: any medium can be optically cancelled by an
equal thickness of material constructed to be an inverted mirror image
of the medium, with $\epsilon$ and $\mu$ reversed in sign. That is,
complementary media cancel one another and become invisible (i.e., a
perfect antireflector).

\subsection{The Wigner angle}

In special relativity there is an intriguing phenomenon that emerges
in the composition of two noncollinear pure boosts: the combination of
two such successive boosts cannot result in a pure boost, but renders
an additional rotation, usually known as the Wigner
rotation~\citep{Wyk:1984mz,Ben-Menahem:1985,Strandberg:1986fk,Aravind:1997,
  Ungar:2001,Malykin:2006,ODonnell:2011} [sometimes the name of Thomas
rotation~\citep{Jackson:1975,Ungar:1989,Muller:1992uq,Hamilton:1996kx}
is also used]. In other words, boosts are not a subgroup.

\begin{figure}
  \centering
  \resizebox{0.80\columnwidth}{!}{\includegraphics{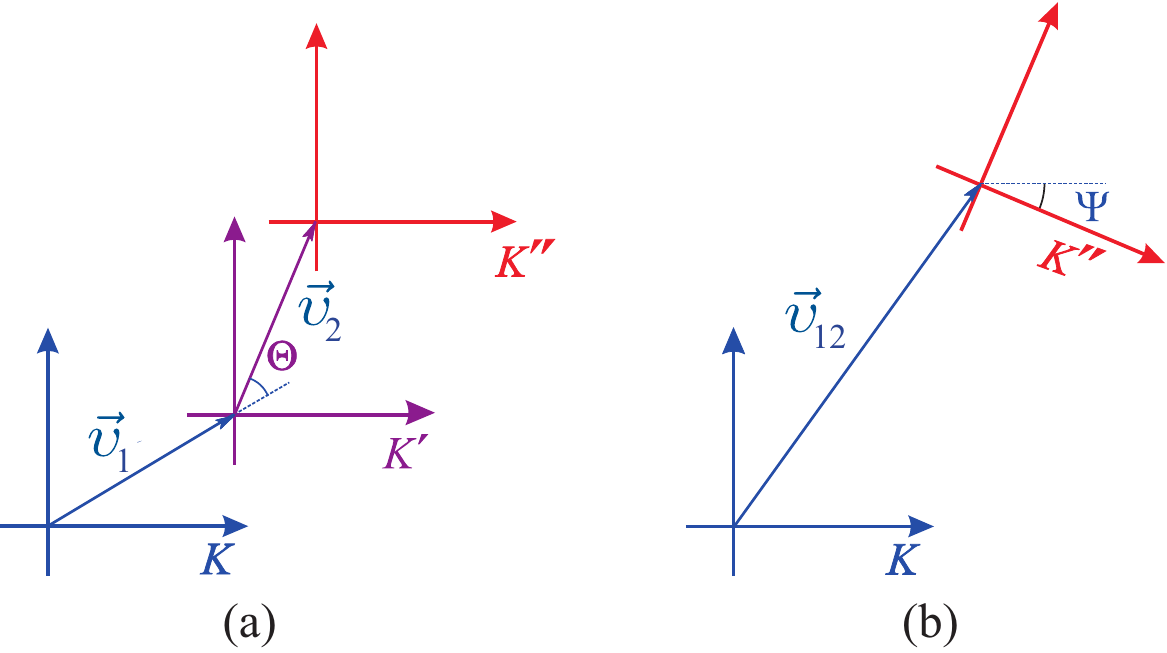}}
  \caption{(a) Reference frame $K^{\prime \prime}$ moves with velocity
    $\vec{\varv}_2$ relative to frame $K^\prime$ while frame
    $K^\prime$ moves with velocity $\vec{\varv}_1$ relative to frame
    $K$.  (b) The axes of $K^{\prime \prime}$ appears rotated relative
    to $K$ by the Wigner angle $\Theta$. Time and one space dimension
    are suppressed for clarity.}
  \label{figure15}
\end{figure}

To fix the physical background, consider three reference frames $K$,
$K^\prime$ and $K^{\prime \prime}$ (see figure~\ref{figure15}). Frames
$K$-$K^\prime$ and $K^\prime$-$K^{\prime \prime}$ have parallel
respective axes. Frame $K^{\prime \prime}$ moves with uniform velocity
$\vec{\varv}_2$ with respect to $K^\prime$, which in turn moves with
velocity $\vec{\varv}_1$ relative to $K$. The Lorentz transformation $\Lambda_{12}$
that connects $K$ with $K^{\prime \prime}$ is given by the product
$\matriz{L}_{1} \matriz{L}_{2} $, which can be decomposed as
\begin{equation}
  \label{bcom} 
  \matriz{L}_{1} \matriz{L}_{2} = \Lambda_{12} = 
  \matriz{R}(\psi) \, \matriz{L}_{(12)} \, .
\end{equation}
An equivalent decomposition in terms of a boost with the same modulus
of $\vec{\varv}_{12}$ but with a different direction postmultiplied by
the same rotation is also possible~\citep{Ungar:1989,Aravind:1997}.

In words, this means that an observer in $K$ sees the axes of
$K^{\prime \prime}$ rotated relative to the observer's own axes by a
Wigner rotation $\matriz{R}(\psi)$. More explicitly, it
is possible to show that the axis $\hat{\mathbf{n}}$ and angle $\psi$
of this rotation are~\citep{Ben-Menahem:1985,Malykin:2006,Ritus:2008sp}
\begin{equation}
  \label{eq:exWan}
  \hat{\mathbf{n}}  = \frac{\vec{\varv}_2 \times \vec{\varv}_1} 
  {| \vec{\varv}_2 \times \vec{\varv}_1  |} \, ,
  \qquad \qquad
  \tan  ( \psi/2  )  =  \frac{\sin \Theta}{K+ \cos \Theta} \,  ,  
\end{equation}
where $\Theta$ is the angle between $\vec{\varv}_1$ and
$\vec{\varv}_2$, and
\begin{equation}
  K^2 = \frac{\gamma_1 + 1}{\gamma_1 -1} \frac{\gamma_2 + 1}{\gamma_2 -1} = 
  \frac{1}{\tanh^2(\zeta_1/2) \tanh^2(\zeta_2/2)}  \,  ,
\end{equation}
$\gamma_1$ and $\gamma_2$ being the corresponding factors for
$\vec{\varv}_1$ and $\vec{\varv}_2$, while $\zeta_1$ and $\zeta_2$ are
the rapidities. This means that $\tan(\psi/2)$ depends on the
velocities as $\varv_1 \varv_2$, so the Wigner rotation is a
second-order effect and is absent in the non-relativistic limit.

On the other hand, the resulting boost $\matriz{L}_{(12)}$ has a
velocity fulfilling
\begin{equation}
  \label{Gamma12} 
  \gamma_{12} = \gamma_1 \gamma_2  ( 1 +  \vec{\varv}_1 \cdot \vec{\varv}_2  ) 
  =  \gamma_1 \gamma_2 ( 1 + \varv_1  \varv_2 \cos \Theta  ) \, ,
\end{equation}
while the direction of $\vec{\varv}_{12}$ has a complicated expression
of little interest here.

The set of boosts could be regarded as a hyperbolic space provided the
velocity $\varv$ is replaced by the hyperbolic parameter $\zeta=
\tanh^{-1} ( \varv )$, which constitutes the usual rapidity space and
whose line element has a Lobachevskian metric, as known from long
times ago~\citep{Landau:2000,Rhodes:2004rt}. A triangle in this
rapidity space obeys a non-Euclidean geometry and, in our context,
this results in the fact that the parameters (\ref{Gamma12}) of the
compound boost $\matriz{L}_{(12)}$ can be recast as
\begin{equation}
  \label{hcos}
  \cosh \zeta_{12} = \cosh \zeta_{1}  \cosh \zeta_{2}  + 
  \sinh \zeta_{1} \sinh \zeta_{2}  \cos \Theta \,  ,
\end{equation}
which is nothing but the hyperbolic law of cosines for the triangle
induced by the boosts $\matriz{L}_{1}$, $\matriz{L}_{2}$, and
$\matriz{L}_{(12)}$.  Therefore, given two sides and the included angle
of the triangle (corresponding to the two non-collinear boosts we wish
to combine) one can determine the third side and its angle by a simple
use of hyperbolic trigonometry.

Moreover, a standard calculation shows that the expression
(\ref{eq:exWan}) for the Wigner angle $\psi$ gives precisely the area
of this triangle \citep{Chen:1998}. We recall that for a hyperbolic
(spherical) triangle the sum of the angles is less (greater) than
$\pi$ with the angular defect (excess) being the area.

This suggests to look at the Wigner angle as a geometric phase.
Roughly speaking, geometric phases are associated with the cyclic
evolution of a system and the crucial concept to their understanding
is anholonomy.  Anholonomy~\citep{Shapere:1989} is a phenomenon in
which non-integrability causes some variables to fail to return to
their original values when others, which drive them, are altered round
a cycle (the simplest example occurs in the parallel transport of
vectors). This behaviour was anticipated by Pancharatnam when
discussing the phase shift that appears in the coherent addition of
two polarized beams on the Poincar\'e sphere.

It is worth mentioning that geometric phases associated with the group
SU(1,1) have been previously
identified~\citep{Simon:1993gf,Monzon:1999fr,Monzon:2001mz}.  The idea
is to view the rapidity triangle as imbedded in the unit hyperboloid,
which is a manifold of constant negative curvature (of value
$-1$). The analogous triangle for rotations instead of boosts is
traced on the unit sphere (of curvature $+1$) and the geometric phase
appears as the area enclosed by the triangle on the
sphere~\citep{Levi:1994}. Thus, it is tempting to infer that the
Wigner angle is just the area of the triangle on the hyperboloid, with
the opposite sign to that of rotations: in fact, this is true as
proved by \citet{Aravind:1997}  and others~\citep{Jordan:1988,Urbantke:1990}.

After our discussion in the previous section, it is clear that the
problem should arise in the context of transfer matrices.  First of
all, it is worth emphasizing that the two combining boosts and the
resulting one are in the same plane, usually assumed for simplicity to
be the 1-2 plane.

In consequence, we restrict our attention to the composition of two
Hermitian matrices $\matriz{H}_1$ and $\matriz{H}_2$ for, as
explained before, they are the equivalent to pure boosts. In complete
analogy with equation~(\ref{bcom}) we have now
\begin{equation}
  \label{Htot} 
  \matriz{H}_1 \matriz{H}_2  = \matriz{M}_{12}  = \matriz{U} \matriz{H}_{(12)} = 
  \left (
    \begin{array}{cc}
      \exp (- i \psi /2 ) & 0 \\
      0 & \exp ( i \psi /2  )
    \end{array}
  \right )   \times  
  \frac{1}{|\coeft_{12}|}
  \left (
    \begin{array}{cc}
      1 & \coefr_{12}^\ast \exp(i \psi) \\
      \coefr_{12} \exp(- i \psi)&  1
    \end{array}
  \right ) \, ,
\end{equation}
where $r_{12}$ and $t_{12}$ are given by (\ref{12}) and
\begin{equation}
  \psi =  2 \arg \coeft_{12} = \arg(1+ \coefr_1 \coefr_2^\ast) \, . 
\end{equation}
The appearance of an extra unitary matrix is the signature of a Wigner
rotation and, accordingly, the Wigner angle $\psi$ viewed in SO(2,1)
is just twice the phase of the transmission coefficient of the
compound system.  Obviously, when $\rho_1 = \rho_2$, the Wigner
rotation is absent, since then we are dealing with two parallel
boosts, whose composition leads to the famous Einstein addition law of
velocities \citep{Vigoureux:1992,Vigoureux:1993}.

\begin{figure}
  \centering
  \includegraphics[height=5cm]{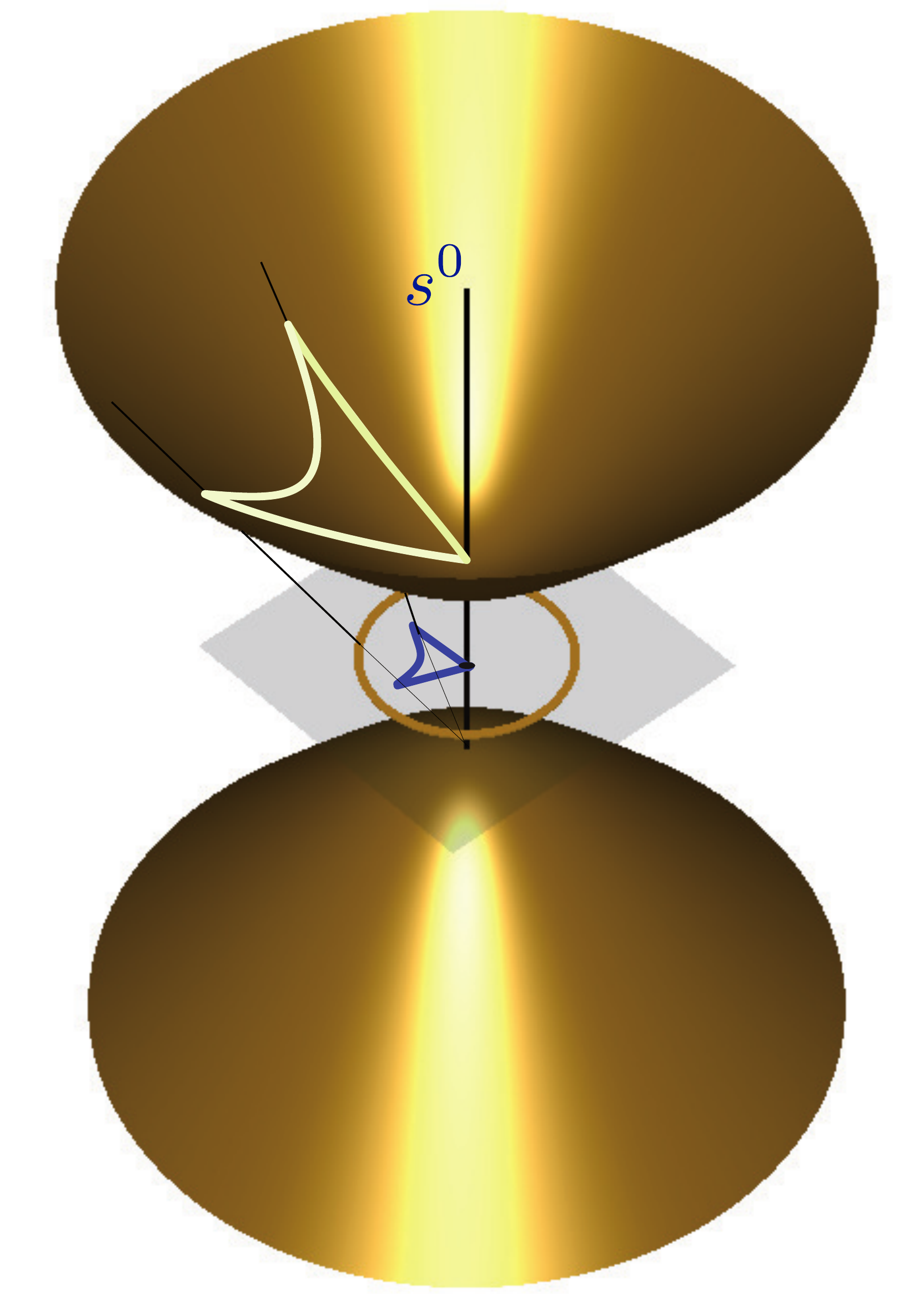}
  \caption{Geodesic triangle in the unit disk for the
    potentials indicated in the text. By stereographic projection the
    associated triangle is also plotted in the unit hyperboloid.}
  \label{figure16}
\end{figure}

To show an explicit implementation of this
phenomenon~\citep{Monzon:2001mz}, we take two potentials with
Hermitian transfer matrices $\matriz{H}_{1}$ and $\matriz{H}_{2}$, and
scattering coefficients $(r_{1} = 0.3736 -i 0.2014 , t_{1} =0.9055)$
and $(r_{2} = 0.3413 i, t_{2} =0.9399)$. Equation (\ref{Htot}) fixes a
third potential $\matriz{H}_{(12)}^{-1}$ such that when put together the
compound system is an antireflection system with phase in transmission
equal to the Wigner angle. This is shown in figure~\ref{figure16},
where the triangle is also plotted in the unit hyperboloid.

\subsection{Hyperbolic turns}

According to \citet{Hamilton:1853}, the turn associated with a
rotation of axis $\hat{\mathbf{n}}$ and angle $\vartheta$ is a
directed arc of length $\vartheta/2$ on the great circle orthogonal to
$\hat{\mathbf{n}}$ on the unit sphere. By means of these objects, the
composition of rotations is described through a parallelogram-like
law: if these turns are translated on the great circles until the head
of the arc of the first rotation coincides with the tail of the arc of
the second one, then the turn between the free tail and the head is
associated with the resultant rotation
\citep{Biedenharn:1981}. Hamilton turns are thus analogous for
spherical geometry to sliding vectors in Euclidean geometry. It is
unfortunate that this elegant idea of Hamilton is not as widely known
as it rightly deserves.

The purpose of this section is to show how the use of turns affords an
intuitive and visual image of all problems involved in quantum
scattering and reveals the emergence of hyperbolic geometry in the
composition law of transfer matrices.

Let us focus on the case of $ | \Tr ( \matriz{M} ) | > 2$. This is not
a serious restriction, since any matrix of SU(1,~1) can be written (in
many ways) as the product of two hyperbolic
translations~\citep{Juarez:1982,Simon:1989,Sanchez-Soto:2005,Simon:2006}. The
axis of the hyperbolic translation is the geodesic line joining the
two fixed points.

\begin{figure}
  \centering
  \includegraphics[height=5cm]{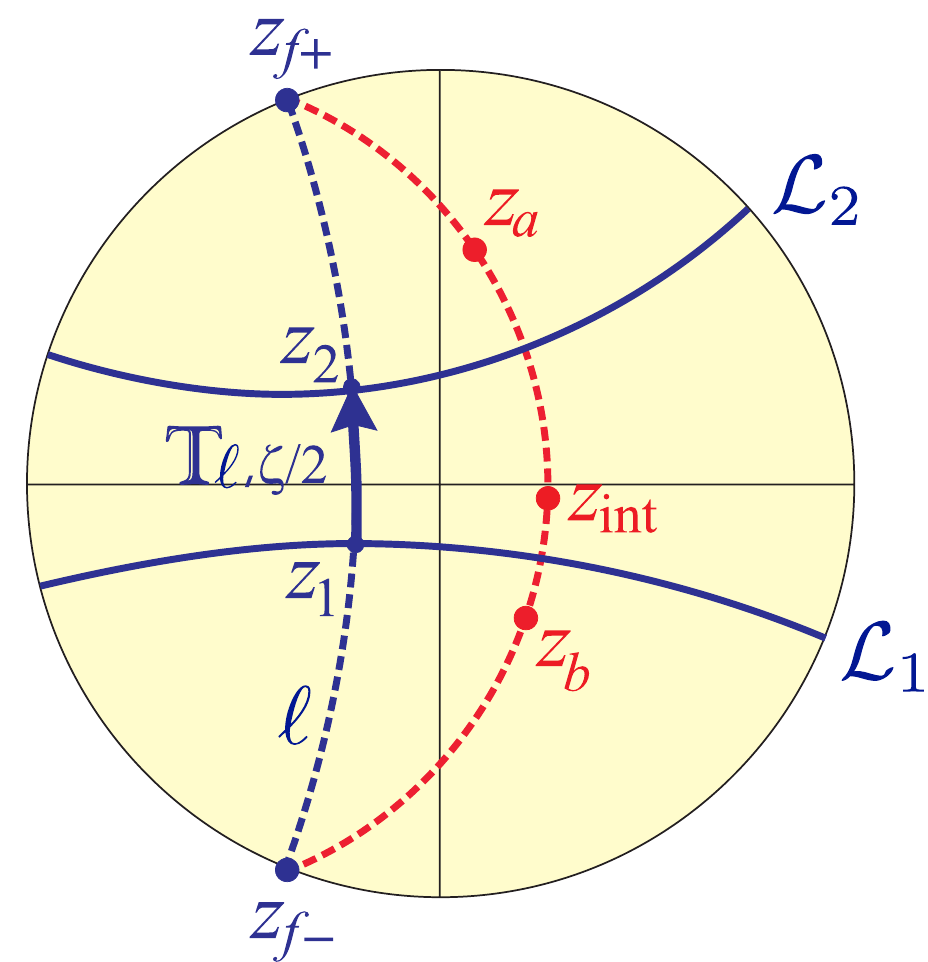}
  \caption{Representation of the sliding turn $\mathbb{T}_{\ell,
      \zeta/2}$ in terms of two reflections in two lines
    $\mathcal{L}_1$ and $\mathcal{L}_2$ orthogonal to the axis of the
    translation $\ell$, which has two fixed points $z_{f-}$ and
    $z_{f+}$. The transformation of a typical off axis point $z_b$ is
    also shown.}
  \label{figure17}
\end{figure}

As explained in section 4.4, any pair of points $z_1$ and $z_2$ on the axis of
the translation $\ell$ at a distance $\zeta/2$ can be chosen as
intersections of $\mathcal{L}_1$ and $\mathcal{L}_2$ (orthogonal lines
to $\ell$) with $\ell$. It is natural to associate to the
translation an oriented segment of length $\zeta/2$, with
\begin{equation}
  \label{eq:defzeta}
  \zeta = d_{\mathrm{H}} (z_{b}, z_{a}) \, ,
\end{equation}
but otherwise free to slide on $\ell$ (see
figure~\ref{figure17}). This is analogous to Hamilton's turns, and
will be called a hyperbolic turn $\mathbb{T}_{\ell, \zeta/2}$ \citep{Simon:1989a}.

Using this construction, an off-axis point such as $z_{b}$
will be mapped by these two reflections (through an intermediate point
$z_{\mathrm{int}}$) to another point $z_a$ along a curve equidistant
to the axis. These other curves, unlike the axis of translation, are
not hyperbolic lines. What matters is that once the turn is
known, the transformation of every point in the unit disk is
automatically established.

Alternatively, we can formulate the concept of turn as follows.  Let
$\matriz{M}$ be a hyperbolic translation with $\Tr (\matriz{M} ) $
positive [equivalently, $\re (\alpha ) > 1$].  Then, $\matriz{M}$ is
positive definite and one can ensure that its positive square root
exists and reads as \citep{Barriuso:2004}
\begin{equation}
  \sqrt{\matriz{M}} = \frac{1}{\sqrt{ \Tr (\matriz{M} ) + 2 }} 
  \left (
    \begin{array}{cc}
      \alpha + 1 &  \beta \\
      \beta^\ast & \alpha^\ast + 1
    \end{array}
  \right ) \, .
\end{equation}
This matrix has the same fixed points as $\matriz{M}$, but the
translated distance is just half the induced by $\matriz{M}$; i.e., we
set
\begin{equation}
  \zeta ( \matriz{M} ) = 2 \zeta  (\sqrt{\matriz{M}} ) \, .
\end{equation}
This suggests that the matrix $ \sqrt{\matriz{M}} $ can be
appropriately associated to the turn $\mathbb{T}_{\ell, \zeta/2}$ that
represents the translation induced by $\matriz{M}$. Therefore, we
symbolically write
\begin{equation}
  \label{mapsto} 
  \mathbb{T}_{\ell, \zeta/2} \mapsto \sqrt{\matriz{M}} \, .
\end{equation}

\begin{figure}[b]
  \centering
  \includegraphics[height=4.5cm]{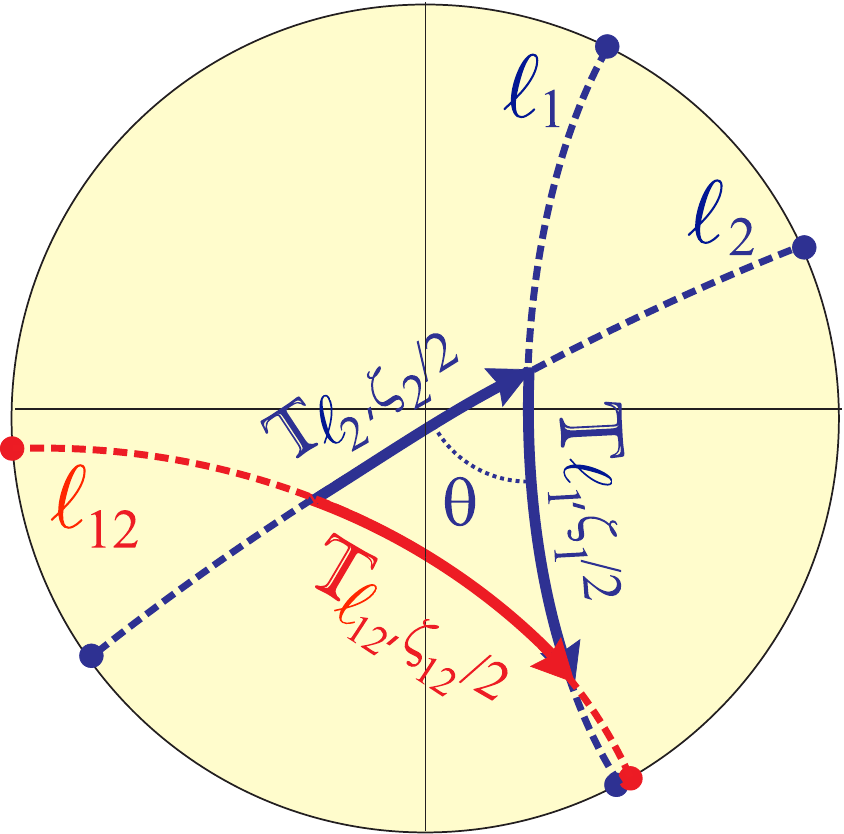}
  \caption{Composition of two hyperbolic turns $\mathbb{T}_{\ell_1,
      \zeta_1/2}$ and $\mathbb{T}_{\ell_2, \zeta_2/2}$ by using a
    parallelogramlike law when the axes $\ell_1$ and $\ell_2$ of the
    translations intersect.}
  \label{figure18}
\end{figure}

Let $\zeta_1$ and $\zeta_2$ be the corresponding translated distances
along intersecting axes $\ell_1$ and $\ell_2$, respectively. Take now
the associated turns $\mathbb{T}_{\ell_1, \zeta_1/2}$ and
$\mathbb{T}_{\ell_2, \zeta_2 /2}$ and slide them along $\ell_1$ and
$\ell_2$ until they are ``head to tail''. Afterwards, the turn
determined by the free tail and head is the turn associated to the
resultant, which can be interpreted as a translation of parameter
$\zeta_{12}$ along the line $\ell_{12}$.

This construction is illustrated in figure~\ref{figure18}, where the
pertinent parameters are $(\coefr_1 = 0.3103 - i 0.8274, \coeft_1 =
0.4383 + i 0.1644)$ and $(\coefr_2 = 0.6820 + i 0.3079, \coeft_2 =
0.6601 - i 0.0659)$. The application of (\ref{12}) gives $(\coefr_{12} =
0.5210 - i 0.7331, \coeft_{12} = 0.3915 - i 0.1947)$.  The
noncommutative character is evident, and can also be inferred from the
obvious fact that $\matriz{M}_{12} \neq \matriz{M}_{21}$.

In Euclidean geometry, the resultant of this parallelogram law can be
quantitatively determined by a direct application of the cosine
theorem. For any hyperbolic triangle with sides of lengths $\zeta_1$
and $\zeta_2$ that make an angle $\Theta$, the expression is precisely
given in equation~(\ref{hcos}).

\section{Periodic systems}

\subsection{Finite periodic structures}

Periodic potentials are those whose shape is repeated indefinitely
with period $d$; i.e., $V(x) = V(x+d)$. The distinctive feature of
these potentials is that the frequencies fall into continuous bands,
separated by forbidden gaps. In the quantum context this was first
noted by \citet{Kronig:1931} in the classic paper that laid the
foundation for the modern theory of solids. This band structure also
occurs, in principle, for mechanical, acoustical, and electromagnetic
waves \citep{Griffiths:2001}.

There is a recurring interest in the related instance where the
potential $V(x)$ consists of a finite number (say $N$) of identical
cells. We shall call that situation a finite periodic structure (the
term locally periodic is also employed); such potentials are produced
by any finite lattice and they are of great importance for a number of
applications, such as superlattices, photonic crystals, multilayers,
etc, where the finite size must unavoidably be taken into
account~\citep{Felbacq:1998ly,Busch:2007uq}.

From a theoretical standpoint finite periodic systems are more
difficult to analyze because Bloch theorem~\citep{Ashcroft:1976},
which so dramatically simplifies the periodic problem, does not
apply. It is amazing that the finite periodic case can be solved
analytically for arbitrary $N$. This was first discovered by
\citet{Abeles:1948} and rediscovered in the quantum context by
\citet{Kiang:1974ve} and \citet{Cvetic:1981} and later by several
others~\citep{Vezzetti:1986,Lee:1989bh,Kalotas:1991,Griffiths:1992,
 Sprung:1993,Wu:1993,Rozman:1994,Liviotti:1994,Chuprikov:1996,
  Erdos:1997,Barra:1999,Sprung:1999fv,Morozov:2002oq,Pereyra:2002}.

\begin{figure}
  \centering
  \includegraphics[height=4cm]{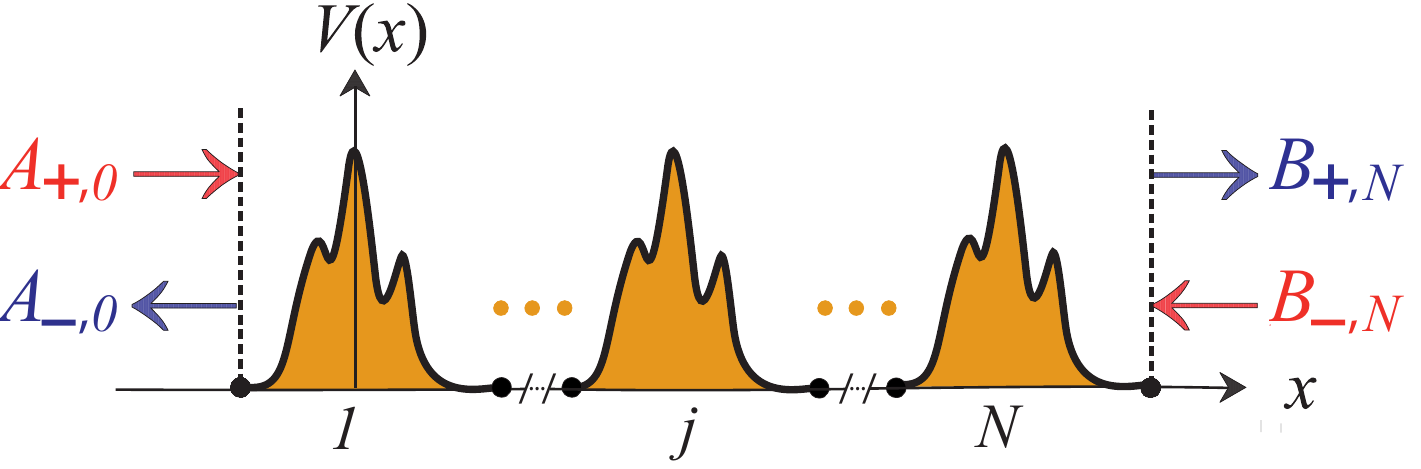}
  \caption{A finite periodic potential constructed from the basic cell
    $V(x)$.}
  \label{figure19}
\end{figure}

Let us suppose that the arbitrary potential $V(x)$ (the basic unit
cell) is replicated $N$ times at regular intervals, as schematized in
figure~\ref{figure19}. Our problem is to construct the transfer matrix
for the whole array, given the transfer matrix $\matriz{M}$ for the
single cell. The amplitudes at the $j$th cell are
\begin{equation}
  \label{Mnsites} 
  \left (
    \begin{array}{c}
      A_{+,j} \\
      A_{-,j}
    \end{array}
  \right ) = 
  \matriz{M} \, 
  \left (
    \begin{array}{c}
      B_{+,j+1} \\
      B_{-,j+1}
    \end{array}
  \right ) \,  .
\end{equation}
Using this equation recursively we have that
\begin{equation}
  \label{Mnsitesfin} 
  \left (
    \begin{array}{c}
      A_{+,0} \\
      A_{-,0}
    \end{array}
  \right ) = 
  \matriz{M}^{N} 
  \left (
    \begin{array}{c}
      B_{+,N} \\
      B_{-,N}
    \end{array}
  \right ) \, ,
\end{equation}
so the whole problem reduces to the evaluation of $\matriz{M}^{N}$.
There are several elegant ways of calculating this power. Perhaps, the
most efficient is to use the Cayley-Hamilton theorem, which states
that any square matrix satisfies its own characteristic
equation~\citep{Gantmacher:2000}. This means that
\begin{equation}
  \label{eq:CHam}
  \matriz{M}^2 - 2 u  \, \matriz{M}    + \openone = 0 \, .
\end{equation}
where $u = [ \Tr (\matriz{M}) ]/2$. Consequently, any higher power of
$\matriz{M}$ can be reduced to a linear combination of $\matriz{M}$
and the identity $\openone$. By induction, we obtain the
expression
\begin{equation}
  \label{eq:C}
  \matriz{M}^N = U_{N - 1} (u) \, \matriz{M} - U_{N - 2} (u) \,
  \openone \, .
\end{equation}
Here
\begin{equation}
  U_{N} (\theta) = \frac{\sin [ (N+1) \theta]}{\sin \theta} \, ,
\end{equation}
with $\cos \theta = u$, are the Chebyshev polynomials of the second
kind satisfying the recursion relation \citep{Abram:1996ul}
\begin{equation}
  U_{N+1} = 2 u \, U_{N} + U_{N-1} , \qquad \qquad N \ge 1  \, ,
\end{equation}
and $U_{0} (u) = 1$, $U_{1} (u) = 2 u$. This provides a closed
solution to the problem; in particular, for incidence from the left
the reflectance $\mathcal{R} = | \coefr |^2$ of the array is
\begin{equation}
  \label{eq:refTsch}
  \mathcal{R}^{(N)} = \frac{[ | \beta \, | U_{N-1} (u)]^{2}}{1 + [ | \beta|
    U_{N-1} (u)]^{2}} \, ,
\end{equation}
which requires to know the transfer matrix for the unit cell.

The Chebyshev polynomials play, for finite systems, a similar role to
the one played by Bloch functions in the description of infinite
periodic systems. In addition, they are very useful to perform
numerical calculations. However, it is not easy to separate the
different behaviors according to the value of the trace, as it happens
for infinite periodic media.  For this reason, we will follow an
alternative route based on geometrical properties.

\subsection{Bandgaps in the unit disc}

First, we note that the reflectance associated to each one of the
canonical matrices (\ref{canonical}) is
\begin{eqnarray}
  \label{RefKAN}
  \mathcal{R}_{\hat{\matriz{K}}} & = & 0 \, , \nonumber \\
  \mathcal{R}_{\hat{\matriz{A}}}  & = & \tanh^2 (\xi/2) \, , \\
  \mathcal{R}_{\hat{\matriz{N}}} & = &  (\nu/2)^2/[1 + (\nu/2)^2] \, .
  \nonumber
\end{eqnarray}
While $\mathcal{R}_{\hat{\matriz{K}}}$ is identically zero,
$\mathcal{R}_{\hat{\matriz{A}}}$ and $\mathcal{R}_{\hat{\matriz{N}}}$
tend to unity when $\xi$ and $\nu$, respectively, increase. Nevertheless,
they have distinct growths: $\mathcal{R}_{\hat{\matriz{A}}}$ goes to
unity exponentially, while $\mathcal{R}_{\hat{\matriz{N}}}$ goes as
$O(\nu^{-2})$.

Let $\matriz{C}$ be the matrix (\ref{matC}) that goes by conjugation
from an arbitrary $\matriz{M}$ to its canonical version.  All the
subgroups generated by $\hat{\matriz{K}}(\phi)$, $\hat{\matriz{A}}
(\xi)$, or $\hat{\matriz{N}}(\nu) $ are one-parametric  and
therefore Abelian, so we have that
\begin{equation}
  \hat{\matriz{M}} (\mu_1)   \hat{\matriz{M}} (\mu_2)
  = \hat{\matriz{M}} (\mu_1 + \mu_2) \,  ,
\end{equation}
where $\mu$ is the appropriate parameter $\phi$, $\xi$, or
$\nu$. For an $N$-cell system the overall transfer matrix is
\begin{equation}
  \label{Npow}
  \matriz{C}^{-1} \   [ \hat{\matriz{M}} (\mu) ]^N \   \matriz{C} =
  \matriz{C}^{-1} \   \hat{\matriz{M}} (N \mu) \   \matriz{C}  \,   ,
\end{equation}
From this equation, one must expect three universal behaviors of the
reflectance according the transfer matrix for the basic cell is
elliptic, hyperbolic, or parabolic. We shall work in what follows the
detailed structure of these three basic laws.

Since the stop bands are given by the condition $|\Tr (\matriz{M})|
> 2$, we first consider the case when $\matriz{M}$ is
hyperbolic. We can rewrite equation~(\ref{conjC}) as
\begin{equation}
  \label{MCA}
  \matriz{M} = \matriz{C}^{-1} \   \matriz{A}(\xi) \  \matriz{C}  \, ,
\end{equation}
and $\xi$ is given by $\Tr ( \matriz{M}) = 2 \, \cosh \xi  > 2$,
because we are taking into account only positive values of $\Tr (
\matriz{M})$. One solution of equation~(\ref{MCA}) is
\citep{Monzon:2003bh}
\begin{equation}
  \mathfrak{c}_1 = F (\beta^\ast + i \ \sinh \xi )   \,  ,
  \qquad \qquad
  \mathfrak{c}_2 = - i  F  \ \mathrm{Im}(\alpha)  \,  ,
\end{equation}
with $F = 1/\sqrt{2 \sinh \xi [ \sinh \xi - \mathrm{Im}(\beta)]}$.

Carrying out the matrix multiplications in~(\ref{Npow}) it is easy to
compute the reflectance:
\begin{equation}
  \label{RasA}
  \mathcal{R}_{\matriz{A}}^{(N)} =   \frac{ | \beta|^2}
  {| \beta|^2 +   [\sinh ( \xi) / \sinh (N \xi) ]^2 }  \,  .
\end{equation}
This is an exact expression for any value of $N$.  As $N$ grows,
$\mathcal{R}_{\matriz{A}}^{(N)}$ approaches unity exponentially, as
expected from a stop band.

The band edges are determined by $| \Tr(\matriz{M}
) | = 2$; that is, when $\matriz{M}$ is parabolic. A calculation very
similar to the previous one shows that 
\begin{equation}
  \label{RasN}
  \mathcal{R}_{\matriz{N}}^{(N)}  = \frac{ | \beta|^2}
  {| \beta|^2 + ( 1/ N ) ^2 }  \,  ,
\end{equation}
with a typical behavior $\mathcal{R}_{\matriz{N}}^{(N)} \sim 1 -
O(N^{-2})$ that is universal in the physics of reflection. The general
results (\ref{RasA}) and (\ref{RasN}) have been obtained in a
different framework by \citet{Yeh:1988} and \citet{Lekner:1994}.

Finally, in the allowed bands we have $| \Tr (\matriz{M} ) | < 2$,
$\matriz{M}$ is elliptic, and
\begin{equation}
  \label{RasK}
  \mathcal{R}_{\matriz{K}}^{(N)}  =   \frac{ \mathcal{Q}^2 - 2  \mathcal{Q}
    \cos (2 N \Xi )}{1 + \mathcal{Q}^2 - 2  \mathcal{Q} \cos (2 N \Xi )}  \, ,
\end{equation}
where
\begin{equation}
  \mathcal{Q} = \frac{ |\beta |^2}{ |\beta |^2 - | \alpha - e^{i \Xi} |^2} \, ,
  \qquad \qquad
  e^{i \Xi} = \re ( \alpha )   + i \sqrt{1 - [ \re  ( \alpha )] ^2}  \,  .
\end{equation}
Now the reflectance oscillates with $N$ between the values
$(\mathcal{Q}^2 - 2\mathcal{Q})/(\mathcal{Q} - 1)^2$ and
$(\mathcal{Q}^2 + 2\mathcal{Q})/ (\mathcal{Q} + 1)^2$.

It seems quite pertinent to picture these behaviors in the unit
disc. Note that if we have only an incident wave from the left ($B_{-}
= 0$; that is, $z_b = 0$) and simultaneously $|z_a| = 1$ the system
behaves as a perfect mirror. Therefore, a mirror maps the
origin into a point on the unit circle.

\begin{figure}
  \centering
  \includegraphics[height=4.5cm]{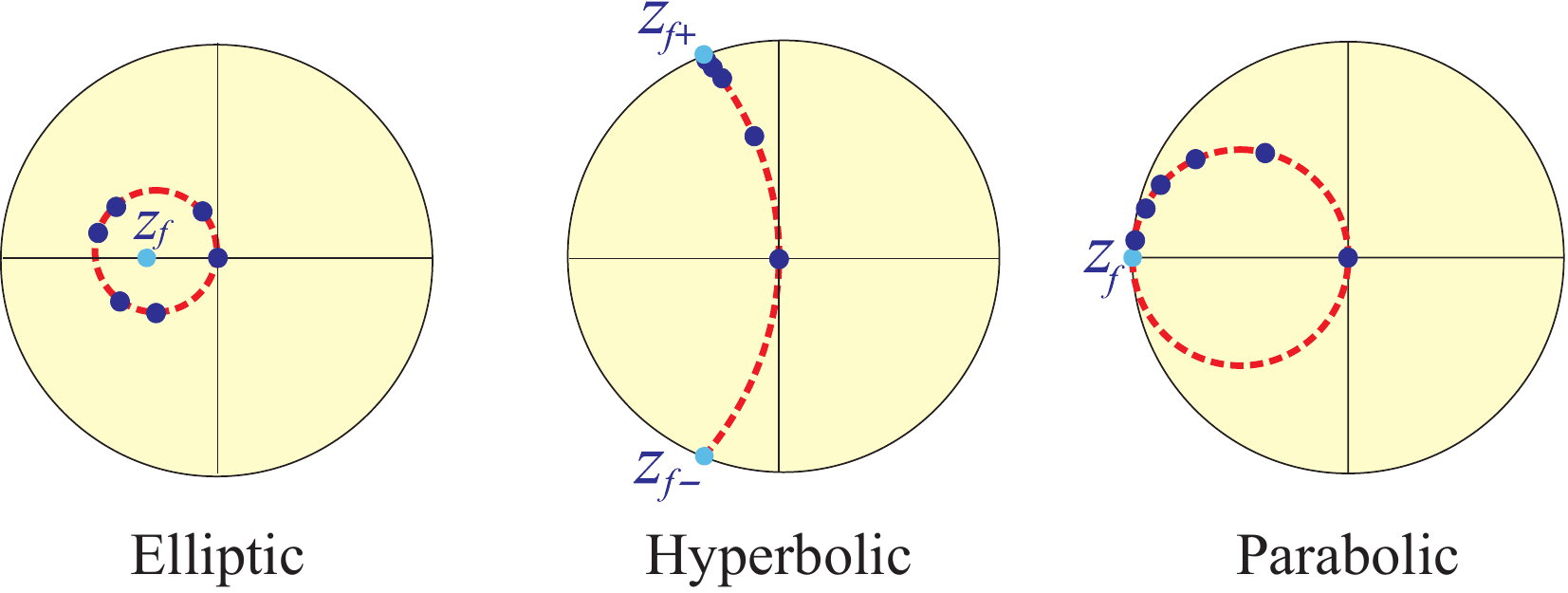}
  \caption{Successive iterates ($N = 1, \ldots, 5$) for an elliptic,
    hyperbolic, and parabolic action starting from the origin as the
    initial point. The fixed points are marked in cyan. Only
    hyperbolic and parabolic actions tend to the unit circle.}
  \label{figure20}
\end{figure}

Henceforth, we shall take $z_0 \equiv z_b= 0$. As mentioned before,
all the points $z_N$ obtained by iteration from $z_{0}$ lie in the
orbit associated to the initial point $z_0$ by the single cell, which
is determined by its fixed points.

In figure~\ref{figure20} we have plotted the successive iterates
worked out numerically for the three archetypical
actions~\citep{Barriuso:2003}.  In the elliptic case, the points $z_N$
revolve in the orbit centered at the fixed point and the system never
reaches the unit circle.  On the contrary, for the hyperbolic and
parabolic actions the iterates converge to one of the fixed points on
the unit circle, although with different laws, which correspond to the
stop band and edges, respectively~\citep{Lekner:2000}.

The $N$th iterate can be easily computed for the canonical forms in
equation~(\ref{canonical}) and then, conjugating as in (\ref{conjC}).
For a hyperbolic action one has
\begin{equation}
  z_N =  \frac{1- \xi^N }{1 - \xi^N  (z_{f +}/z_{f -})}z_{f +}  \,  ,
\end{equation}
where $ \xi = (\alpha + \beta z_{f -})/(\alpha + \beta z_{f +})$ is a
complex number satisfying $| \xi | < 1$ and $z_{f \pm}$ are the fixed
points. Analogously, for the parabolic case we have
\begin{equation}
  z_N = \frac{N  \beta z_f^2}{ N \beta z_f - 1}   \,  ,
\end{equation}
$z_f$ being the (double) fixed point. In both cases, $z_N$ converges
to one of the fixed points on the unit circle, so $| z_N | \rightarrow
1$ when $N$ increases, a typical behavior of a stop band (or, in other
terms, a perfect mirror). In the mathematical literature this limit
point is referred to as the Denjoy-Wolff point of the
map~\citep{Kapeluszny:1999}.

\subsection{Bandgaps in the half-plane}

The previous formalism can be translated to the hyperbolic
half-plane $\mathbb{H}$ by  the unitary transformation
(\ref{Cayley}), as explained in section 4.2. However, to give a
physical feeling, we prefer to illustrate this point with the simple
yet interesting case of optical beams~\citep{Barriuso:2005fv}.

In paraxial-wave optics, axially symmetric (monochromatic scalar)
beams are specified in the Hilbert space of complex-valued
square-integrable wave-amplitude functions $\Psi
(x)$~\citep{Gloge:1969}, with $x$ labeling the axis. To deal with
partially coherent beams we specify the field not by its amplitude,
but by its cross-spectral density. The latter is defined in terms of
the former as
\begin{equation}
  \label{Gammaesp}
  \Gamma (x_1, x_2) = \langle \Psi^\ast (x_1) \Psi (x_2) \rangle \, ,
\end{equation}
where the angular brackets denote ensemble averages.

There is a wide family of beams, the Schell-model fields
\citep{Wolf:1978rr,Foley:1978nx,Starikov:1982ai,Friberg:1982hc,
  Gori:1983,Gori:1984,Friberg:1988,Ambrosini:1994}, for which the
cross-spectral density (\ref{Gammaesp}) factors in the form
\begin{equation}
  \Gamma (x_1, x_2) =  \sqrt{I(x_1) I (x_2)} \,  \mu (x_1 - x_2)   \, .
\end{equation}
Here $I$ is the intensity distribution and $\mu$ is the normalized
degree of coherence, which is translationally invariant. When these
two fundamental quantities are Gaussians
\begin{equation}
  I(x) =  \frac{\mathcal{I}}{\sqrt{2 \pi} \sigma_I}
  \exp \left ( - \frac{x^2}{2 \sigma_I^2} \right )  \, ,
  \qquad \qquad
  \mu (x) =   \exp \left ( - \frac{x^2}{2 \sigma_\mu^2} \right )  \,  ,
\end{equation}
the beam is said to be a Gaussian Schell model (GSM). Here,
$\mathcal{I}$ is a constant independent of $x$ that can be identified
with the total irradiance and $\sigma_I$ and $\sigma_\mu$ are,
respectively, the effective beam width and the transverse coherence
length. Other well-known families of Gaussian fields are special cases
of these GSM fields: when $\sigma_\mu \ll \sigma_I$ we have the
Gaussian quasihomogeneous field, and when $\sigma_\mu \rightarrow
\infty$ we have the coherent Gaussian field. 

Anyhow, the crucial point for our purposes is that for GSM fields one
can assign a complex parameter $Q$~\citep{Simon:1984,Simon:1985,
  Simon:1988,Simon:1993,Dragoman:1996,Baskal:2002}
\begin{equation}
  \label{Qpar}
  Q =  \frac{1}{R} + i \frac{1} {k \, \sigma_I \; \delta}  \,  ,
\end{equation}
where
\begin{equation}
  \frac{1}{\delta} = \sqrt{\frac{1}{\sigma_\mu^2}
    + \frac{1}{4 \sigma_I^2}}  \, ,
\end{equation}
and $R$ is the wave front curvature radius.  This parameter fully
characterizes the beam and satisfies the Kogelnik $abcd$ law; namely,
after propagation through a first-order optical system described by
the matrix $\Matriz{M}$ as in (\ref{defMgeo}), the parameter $Q$
changes to $Q^\prime$ via
\begin{equation}
  \label{MOBQ}
  Q^\prime =  \Phi [\Matriz{M}, Q ] =   \frac{\mathfrak{d}  \,   Q  + \mathfrak{c}}
  {\mathfrak{b}  \,  Q + \mathfrak{a}} \, .
\end{equation}
 On account of $\im Q > 0$ by the definition (\ref{Qpar}), one immediately
checks that $\im Q^\prime > 0$ and we can thus view the action of the
system as a bilinear transformation on the upper complex
half-plane. When we use the metric $ds = |dQ|/\im Q$ to measure
distances, what we get again is the standard model of the hyperbolic
plane $\mathbb{H}$~\citep{Stahl:1993}.

The whole real axis, which is the boundary of $\mathbb{H}$, is also
invariant under (\ref{MOBQ}) and represents wave fields with unlimited
transverse irradiance (contrary to the notion of a beam). On the other
hand, for the points in the imaginary axis we have an infinite wave
front radius, which defines the corresponding beam waists. The origin
renders a plane wave.

\begin{figure}
  \centering
  \includegraphics[height=6.5cm]{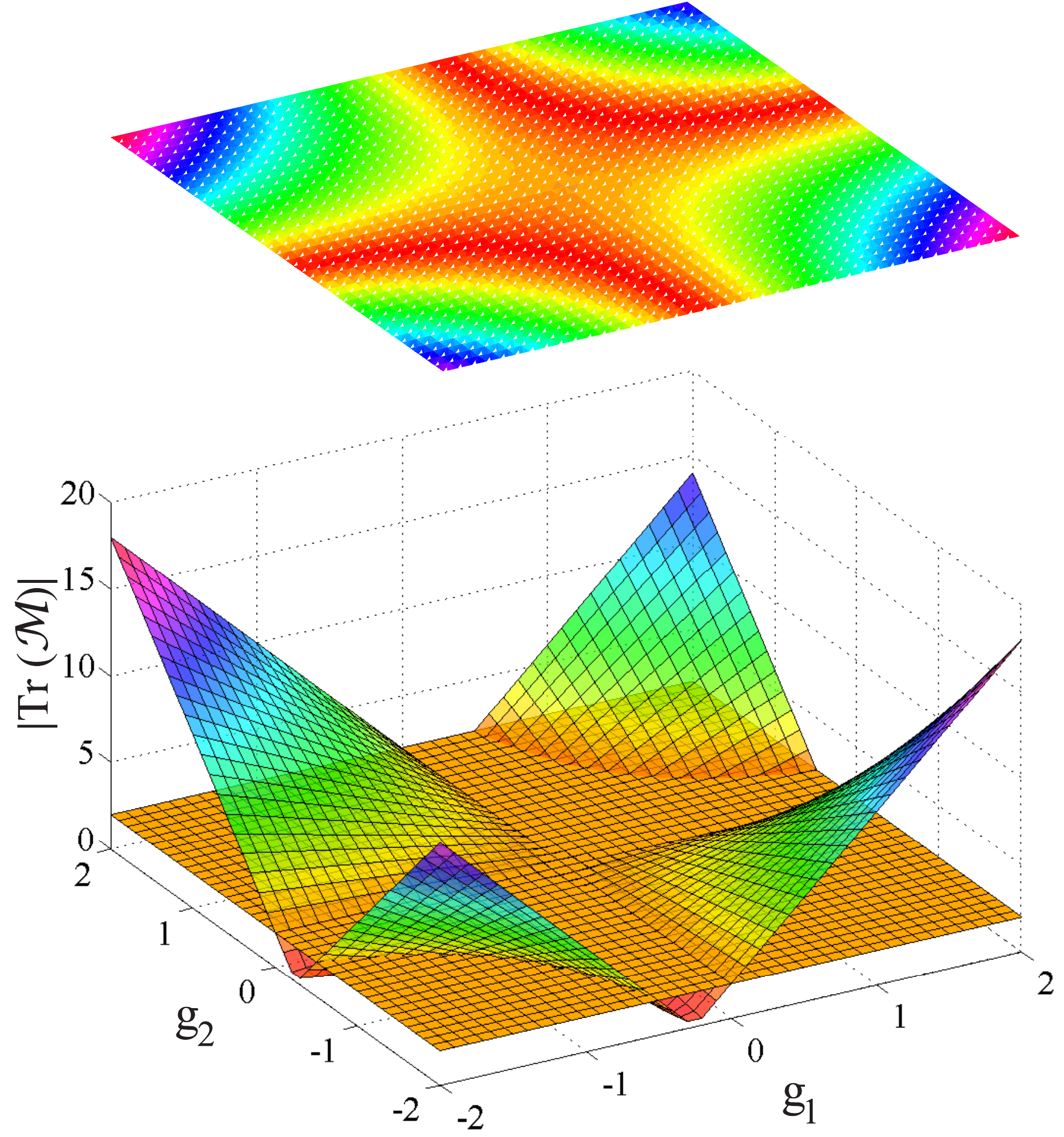}
  \caption{Plot of $|\Tr(\Matriz{M})|$ in terms of the parameters
    $g_1$ and $g_2$ of the optical resonator. The plane
    $|\Tr(\Matriz{M})|=2 $ is also shown. The density plot of the
    three-dimensional figure appears at the top.}
  \label{figure21}
\end{figure}

The geometrical scenario presented before allows one to describe the
evolution of a GSM beam by means of the associated orbits. Let us go
back to the example of the optical cavity treated in section 3.3. The
associated transfer matrix $\Matriz{M}$ fulfills
\begin{equation}
  \label{trest}
  \Tr \Matriz{M}  = 2 (2g_1 g_2-1) \,  .
\end{equation}
Since the trace determines the fixed point and the orbits of the
system, the $g$ parameters establish uniquely the geometrical action
of the cavity.  To illustrate this point, in figure~\ref{figure21} we
have plotted the value of $|\Tr(\Matriz{M})|$ in terms of $g_1$ and
$g_2$. The plane $|\Tr( \Matriz{M} )|= 2$, which delimits the
boundary between elliptic and hyperbolic action, is also shown. At the
top of the figure, a density plot is presented, with the
characteristic hyperbolic contours~\citep{Kogelnik:1966}.

Assume now that the light bounces $N$ times through this system. The
overall transfer matrix is $\Matriz{M}^N$ and the transformed
beam is represented by 
\begin{equation}
  \label{iterat1}
  Q_N = \Phi[\Matriz{M}, Q_{N-1}] =
  \Phi [\Matriz{M}^N, Q_0]  \,  ,
\end{equation}
where $Q_0$ is the initial point.

Note that all the $Q_N$ lie in the orbit associated to $Q_0$ by the
single round trip, which is determined by its fixed points.  By
varying the parameters $g$ of the cavity we can choose to work in the
elliptic, the hyperbolic, or the parabolic case~\citep{Baskal:2002}.

\begin{figure}[t]
  \centering
  \includegraphics[height=4.25cm]{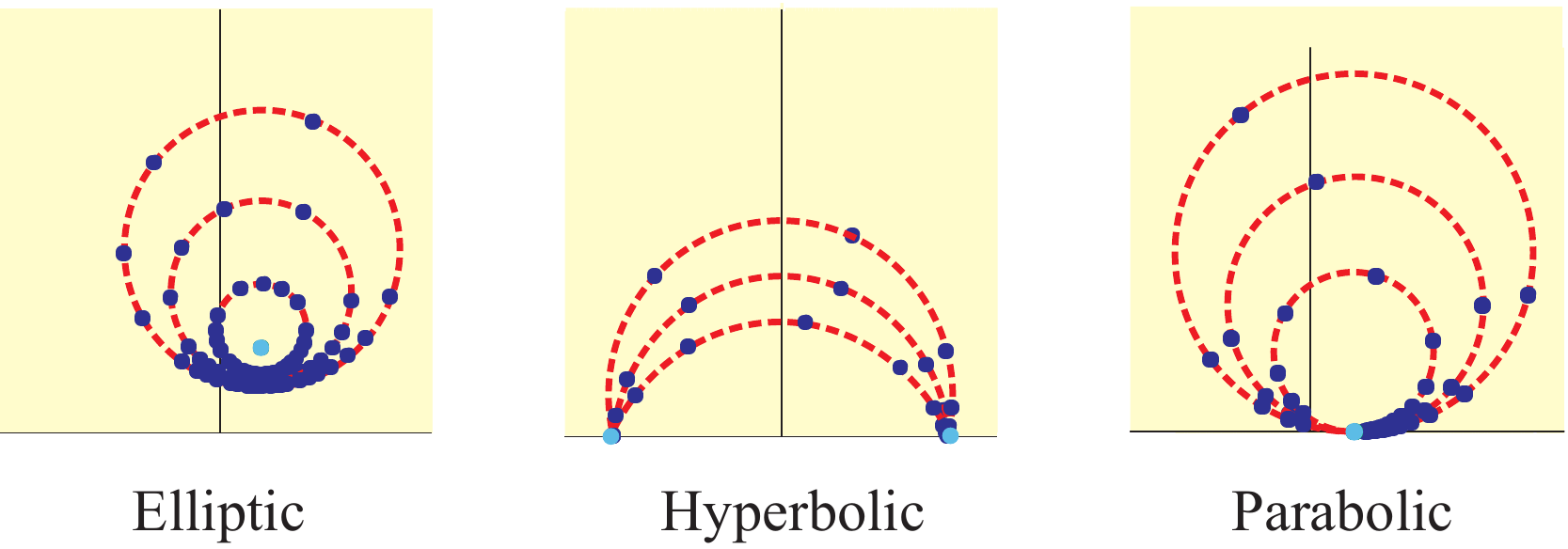}
  \caption{Successive iterates in the half plane $\mathbb{H}$ for
    typical elliptic, hyperbolic, and parabolic actions. For
    hyperbolic and parabolic actions, the iterates tend to the real
    axis. The fixed points are marked in cyan.}
  \label{figure22}
\end{figure}

In figure~\ref{figure22} we have plotted the sequence of successive
iterates  for different kind of ray-transfer matrices,
according to our previous classification.  In the elliptic case, it is
clear that the points $Q_N$ revolve in the orbit centered at the fixed
point and the system never reaches the real axis.  On the contrary,
for the hyperbolic and parabolic cases the iterates converge to one of
the fixed points on the real axis, although with different
laws~\citep{Barriuso:2005fv}.

The iterates of hyperbolic and parabolic actions produce solutions
fully unlimited, which are incompatible with our ideas of a beam. The
only beam solutions are thus generated by elliptic actions and,
according with equation~(\ref{trest}), the stability criterion is
\begin{equation}
  \label{stco}
  0 \le | 2g_1 g_2-1 | = | \cos (\phi/2) | \le 1  \,   ,
\end{equation}
where $\phi$ is the parameter in the canonical form $\hat{\Matriz{K}}
$ in equation~(\ref{Iwasa2}).  Such a condition is usually worked out
in terms of algebraic arguments using ray-transfer matrices \citep{Siegman:1986}.

\subsection{Quasiperiodic sequences}

The most relevant property of periodic systems is the possibility of
generating bandgaps for wave propagation.  Spatial periodicity,
however, is not necessarily a requirement for that: the presence of
large bandgaps has been reported in aperiodic structures. Moreover,
they share distinctive physical properties with both periodic (e.g.,
the formation of gaps) and disordered random media (e. g., the
presence of highly localized states characterized by high-field
enhancement and anomalous transport properties).

In few words, aperiodic structures are made up of two (or more)
incomplete periods stacked together to form an overall system that is
neither fully periodic nor random, but somewhere in
between~\citep{Macia:2009}.  These structures can be roughly
sorted into two classes, those that are quasiperiodic and those
that are not.  The former present a discrete Fourier spectrum
characterized by self-similar Bragg peaks (quite reminiscent of
periodicity), whereas the latter usually exhibit complex diffraction
spectra with singular scattering peaks, and multifractal scaling
properties.

From a mathematical viewpoint quasiperiodic functions belong to the
class of almost periodic functions \citep{Bohr:1951ij}. These
functions can be uniformly approximated by Fourier series containing a
countable infinity of pairwise incommensurate frequencies. When the
set of frequencies required can be generated from a finite-dimensional
basis, the resulting function is referred to as a quasiperiodic
one. The simplest example of a quasiperiodic function is
\begin{equation}
  \label{eq:7}
  V(x) = \cos x + \cos (\lambda x) \, ,
\end{equation}
where $\lambda$ is an irrational number.

Additionally, quasiperiodic structures lack translational invariance
but possess a high degree of rotational symmetry, while the not
quasiperiodic lack both translational and rotational symmetry but
display remarkable self-similarity (scale invariance symmetry) in
their structural and spectral features~\citep{Macia:2006}.

The interest in these sequences was originally fueled by the
theoretical predictions that they should manifest peculiar electron
and phonon critical states~\citep{Ostlund:1984,Kohmoto:1987},
associated with highly fragmented fractal energy
spectra~\citep{Kohmoto:1983,Luck:1989,Suto:1989,Bellissard:1989,
Oh:1993kl,Chakrabarti:1995,Liu:1997,Monsoriu:2006wd}.
On the other hand, the practical fabrication of
Fibonacci~\citep{Merlin:1985} and Thue-Morse~\citep{Merlin:1987}
superlattices triggered a number of experimental achievements that
have provided new insights into the capabilities of quasiperiodic
structures~\citep{Zarate:2001dk,Velasco:2003,Albuquerque:2003il}.  In
particular, possible optical applications have deserved major
attention and some intriguing properties have been
demonstrated~\citep{Tamura:1989,Avishai:1990ye,Kolar:1991qf,Hattori:1994,
Vasconcelos:1999,Hollingworth:2001la,Lusk:2001,Barriuso:2005,Mugassabi:2009dz}.
Underlying all these theoretical and experimental efforts a crucial
fundamental question remains concerning whether quasiperiodic devices
would achieve better performance than usual periodic ones for some
specific applications~\citep{Macia:2001,Barriuso:2003}.

A simple understanding of a well ordered but aperiodic arrangement of
numbers can be grasped by thinking of the Fibonacci numbers sequence
$F_{n} = \{ 1, 1, 2, 3, 5, 8, 13, 21, \ldots \}$.  The terms in this
sequence are generated from the recursive equation $F_{n+1} = F_{n} +
F_{n-1}$, starting with $F_0 =1$ and $ F_1= 1$. Hence, each number in
the sequence is just the sum of the preceding two. The sequence is
then perfectly ordered, but the rule used to generate it has nothing
to do with periodicity. The symbolical analog of the Fibonacci
sequence, constructed using two types of building blocks, say $A$ and
$B$, can be obtained from the substitution rule $A \mapsto AB$ and $B
\mapsto A$, whose successive application generates the sequence of
letters $A, AB, AB A, AB AAB, AB AAB AB A, \ldots$ and so on.  In this
way, we get a perfectly ordered word which is not periodic at all.

The standard method of constructing aperiodic structures is thus
through a substitution rule operating on a finite alphabet $\{A, B,
\ldots \}$. which consists of certain number of letters. In actual
realizations each letter will correspond to a different type of
building block in the structure.  In particular, the substitutional
sequences that act upon a two-letter alphabet $\{A, B \}$ are
especially important: in this case the algorithm reduces to
\begin{equation}
  \label{substi}
  A \mapsto \sigma_{1} (A, B )  \, ,
  \qquad  \qquad   
  B \mapsto  \sigma_{2} (A, B) \, ,
\end{equation}	
where $\sigma_{1}$ and $\sigma_{2}$ can be any string of $A$ and $B$.
In table~\ref{tabla1} we list some representatives among the plethora
of aperiodic structures grown during the last two decades.

\begin{table}
  \caption{List of the substitution rules determining the sequences
    usually considered in the study of self-similar aperiodic systems.}
  \begin{tabular}{lll}
    \hline
    Sequence  &  Alphabet  &  Substitution rule \\
    \hline
    Fibonacci  & $\{A, B \}$  & $ \sigma_{1} =AB$, $\sigma_{2} =A $ \\
    Thue-Morse  & $\{A, B\}$  & $ \sigma_{1} =AB$, $\sigma_{2} =BA$ \\
    Period doubling &$\{A,B\} $& $ \sigma_{1} =AB$, $\sigma_{2} =AA$ \\
    Rudin-Shapiro  & $\{A,B,C,D\}$ & $\sigma_{1} =AC$, 
    $ \sigma_{2} =DC$, $\sigma_{3} =AB$, $\sigma_{4} =DB$\\
    Circular &  $\{A, B, C \}$ & $ \sigma_{1}  =CAC$, $ \sigma_{2} =ACCAC$, $ \sigma_{3}=ABCAC$\\
    \hline
  \end{tabular}
  \label{tabla1}
\end{table}

To each rule we can associate a substitution matrix $\matriz{T}$,
whose columns give the number the letters $A$ and $B$ which occur in
the substitutions $\sigma_{1} $ and $\sigma_{2} $
\begin{equation}
  \matriz{T}  =
  \left (
    \begin{array}{cc}
      n_{A} [ \sigma_{1} (A, B)] &  n_{A} [ \sigma_{2}  (A, B) ] \\
      & \\
      n_{B} [ \sigma_{1} (A, B) ] &  n_{B} [ \sigma_{2} (A, B) ]    
    \end{array}
  \right )  \, .
\end{equation}
This matrix does not depend on the precise form of the substitutions
(the order of the letters), only on the number of letters $A$ or $B$.

The eigenvalues of the substitution matrix $\matriz{T}$ contain a lot
of information. In fact, according to a theorem by
\citet{Bombieri:1986}, if the spectrum of $\matriz{T}$ contains a
Pisot number, the structure is quasiperiodic; otherwise it is not.  A
Pisot number is a positive algebraic number (i.e., a number that is a
solution of an algebraic equation) greater than one, all of whose
conjugate elements (the other solutions of the algebraic equation)
have modulus less than unity~\citep{Bertin:1992,Godreche:1992}.  For
example, let us consider the sequence $\sigma_{1} = AAAB$, $
\sigma_{2} =BBA$, whose characteristic matrix is:
\begin{equation}
  \matriz{T}  =
  \left (
    \begin{array}{cc}
      3 & 1 \\
      1 & 2
    \end{array}
  \right ) \, .
\end{equation}\\
The eigenvalues are $\lambda_{1} = (5 + \sqrt{5}/2)$ and
$\lambda_{2}=(5-\sqrt{5}/2)$. Since $\lambda_{1}>\lambda_{2}>1$,
the sequence does posses the Pisot property.  In
Table~\ref{tabla2} we give a sketch of the properties of the
substitutional sequences considered thus far.

\begin{table}
  \caption{Substitution matrices and related eigenvalues for the
    sequences listed in Table~\ref{tabla1}.}
  \begin{tabular}{lll}
    \hline
    Sequence  &  Substitution matrix  &  Eigenvalues \\
    \hline
    Fibonacci  & $ \left (
      \begin{array}{cc}
        1& 1 \\
        \displaystyle
        1 &0
      \end{array}
    \right )  $  & $ \lambda_1=(1+\sqrt{5})/2$,   $\lambda_2=(1-\sqrt{5})/2 $ \\
    Thue-Morse  & $ \left (
      \begin{array}{cc}
        1& 1 \\
        \displaystyle
        1 &1
      \end{array}
    \right )  $   & $  \lambda_1=2$,   $\lambda_2=0$ \\
    Period doubling &$ \left (
      \begin{array}{cc}
        1& 1 \\
        \displaystyle
        2 &0
      \end{array}
    \right )  $ & $  \lambda_1=2$,   $\lambda_2=-1$ \\
    Rudin-Shapiro  &$ \left (
      \begin{array}{cccc}
        1& 0&1&0 \\
        \displaystyle
        0& 0&1&1 \\
        \displaystyle
        1& 1&0&0 \\
        \displaystyle
        0 &1&0&1
      \end{array}
    \right )  $   &  $  \lambda_1=0$,   $\lambda_2=2$,  $\lambda_3=\sqrt{2}$,  $\lambda_4=-\sqrt{2}$\\
    Circular & $ \left (
      \begin{array}{ccc}
        1& 0&2 \\
        \displaystyle
        2 &0&3\\
        2&1&2
      \end{array}
    \right )  $  & $  \lambda_1=-1$, $\lambda_2=2+\sqrt(5)$,  $\lambda_3=2-\sqrt{5}$\\
    \hline
  \end{tabular}
  \label{tabla2}
\end{table}

The sequences can be also characterized by the nature of their Fourier
spectrum \citep{Severin:1989}.  The Fourier spectrum corresponding to
a perfect infinite periodic system contains delta functions centered
in wave numbers associated to the reciprocal lattice (this is the
origin of the Bragg peaks). On the contrary, a disordered structure
has a very flat spectrum.  Aperiodic heterostructures following a
deterministic sequence display characteristic spectral properties
absent in either of these extreme cases.
 
For a specific sequence of length $N$, the discrete Fourier transform
is 
\begin{equation}
  W_{N} (k) =   \frac{1}{\sqrt{N}} \sum_{j=1}^{N-1} 
  w(j) \exp \left( \frac{-2\pi i  k}{N} \right ) \, ,
\end{equation}
where $w(j)$ is a numerical sequence obtained by assigning to each
letter of the alphabet a fixed amplitude. This assignment is otherwise
arbitrary and does not change any conclusion.  In
consequence, one could, e.g., use $A \mapsto -1$ and $B \mapsto 1$.
The structure factor (or power spectrum) is \citep{Cheng:1990}
\begin{equation}
  S_{N}(k) = |W_{N}(k)|^{2} \, .
\end{equation}

From a rigorous viewpoint, the only well-established concept attached to
the Fourier spectrum is its spectral measure. If we define
\begin{equation}
  d\nu_N(k) = S_N(k) \, dk \, ,
\end{equation}
we will be concerned with the nature of the limit
\begin{equation}
  \label{specmeas}
  d\nu(k)= \lim_{N \rightarrow \infty} d\nu_{N} (k) \, ,
\end{equation}
which corresponds to an infinite structure and a continuous variable
$k$. Just as any positive measure, (\ref{specmeas}) has a unique
decomposition \citep{Reed:1980}
\begin{equation}
  d\nu(k)=d \nu_{\mathrm{pp}}(k) + d \nu_{\mathrm{ac}}(k) + d \nu_{\mathrm{sc}}(k)
\end{equation}
into its pure point, absolutely continuous and singular continuous
parts.

The pure point part refers to the presence of Bragg peaks; the
absolute continuous part is a differentiable function (diffuse
scattering), while the singular continuous part it is neither
continuous nor does it have Bragg peaks. It shows broad peaks, which
are never isolated and, with increasing resolution, split again into
further broad.

The Fibonacci sequence has a pure point spectrum; the Thue-Morse
sequence has a singular continuous Fourier spectrum, while the
Rudin-Shapiro sequence shows an absolute continuous one.  For a very
detailed and up-to-date discussion of these issues, the reader is
referred to comprehensive book by \citet{Macia:2009}.

\subsection{Hyperbolic tilings}

The rich properties of these aperiodic structures suggest the utility
of studying of systems based on more general sequences
\citep{Spinadel:1999th}. Periodicity is intimately connected with
tessellations, i.e., tilings by identical replicas of a unit cell (or
fundamental domain) that fill the plane with no overlaps and no
gaps. Of special interest is the case when the primitive cell is a
regular polygon with a finite area~\citep{Zieschang:1980}. In the
Euclidean plane, the associated regular tessellation is generically
noted $\{p, q\}$, where $p$ is the number of polygon edges and $q$ is
the number of polygons that meet at a vertex. Geometrical constraints
limit the possible regular tilings $\{p, q\}$ to those verifying
\begin{equation}
  \label{eq:restcris}
  (p-2)(q -2) = 4 \, .
\end{equation}
This includes the classical tilings $\{4, 4\}$ (tiling by squares) and
$\{6, 3\}$ (tiling by hexagons), plus a third one, the tiling $\{3,
6\}$ by triangles (which is dual to the $\{6, 3\}$).

On the contrary, in the hyperbolic disk regular tilings exist provided
$(p - 2)(q - 2) > 4$, which now leads to an infinite number of
possibilities~\citep{Magnus:1974}.  The fundamental polygons are
connected to the discrete subgroups of isometries (or congruent
mappings); they are Fuchsian groups~\citep{Ford:1972} and play
for the hyperbolic geometry a role similar to that of crystallographic
groups for the Euclidean geometry~\citep{Beardon:1983}.

A tessellation of the hyperbolic plane by regular polygons has a
symmetry group that is generated by reflections in geodesics, which
are inversions across circles in the unit disc. These geodesics
correspond to edges or axes of symmetry of the polygons.  Therefore,
to construct a tessellation of the unit disk one just has to built one
tile and to duplicate it by using reflections in the edges.

\begin{figure}[t]
  \centering
  \includegraphics[height=6cm]{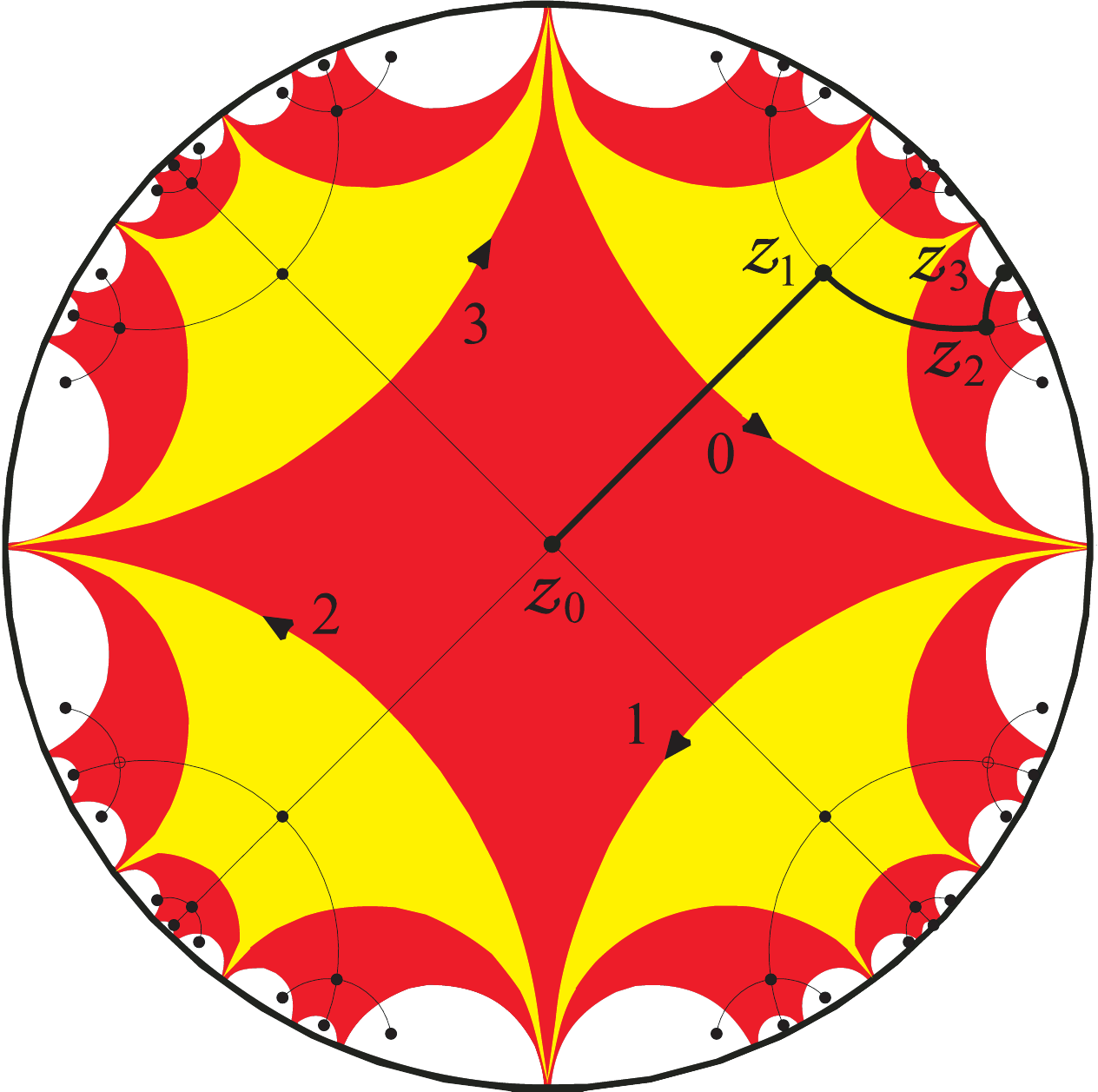}
  \caption{Tessellation of the unit disk with the matrices
    (\ref{matcuad}).  The marked points are the centers of the
    squares. All of them are the transformed of the origin by a matrix
    that have as reflection coefficient the complex number that links
    the origin with the center of the square.}
  \label{figure23}
\end{figure}

To go straight to the point let us consider the following parabolic
transformations
\begin{equation}
  \label{matcuad}
  \matriz{A}  =
  \left (
    \begin{array}{cc}
      1-i & 1 \\
      \displaystyle
      1 &1+i
    \end{array}
  \right ) \, , 
  \qquad \qquad 
  \matriz{B}  =
  \left (
    \begin{array}{cc}
      1+i & 1 \\
      \displaystyle
      1 &1-i
    \end{array}
  \right ) \, ,
\end{equation}\\
with fixed points  $+i$ and $-i$, respectively. A possible implementation of these matrices (and 
their inverses) in terms of two commonly employed materials can be found in
\citet{Barriuso:2009}.  In figure~\ref{figure23}, we have plotted the 
tessellation obtained  by transforming the fundamental square with the 
Fuchsian group generated by the powers of $\{A, B\}$ (and the inverses).

To give an explicit construction rule for the admissible words, we
proceed as follows. First, we arbitrarily choose a side of the
fundamental square and assign to it the value $0$. Then the other
three sides are numbered clockwise as $1$, $2$ and $3$. It is easy to
convince oneself that this assignment fixes once and for all the
numbering for the sides of all the other squares in the
tessellation. However, these squares can be distinguished by their
orientation (as seen from the corresponding center): the clockwise
oriented ones are filled in red, while the counterclockwise ones are
filled in yellow. In short, we have determined a fundamental coloring
of the tessellation~\citep{Grunbaum:1987}.

\begin{table}
  \caption{Explicit rules to obtain the center $z_{n+1}$ from  $z_{n}$ in the 
  tiling by hyperbolic squares. We have indicated the corresponding transformations,
  which depend on the color jump and the sides crossed by going from $z_{n}$ to $z_{n+1}$.
  \label{unica}}
\bigskip
\centering
\begin{tabular}{llcclcc}
  \hline
  & \multicolumn{3}{c}{red $\rightarrow $ yellow} &
  \multicolumn{3}{c}{yellow $\rightarrow $ red} \\
  Side & $T$ & $ \matriz{A}_{n+1}$ & $\matriz{B}_{n+1}$ &
  $T$ & $\matriz{A}_{n+1}$ & $\matriz{B}_{n+1}$ \\ 
  \hline
  0 & $\matriz{A}_{n}$ & $\matriz{B}_{n} $ &
  $\matriz{A}_{n} \matriz{B}_{n}\matriz{A}_{n}^{-1}$ & $\matriz{A}_{n}^{-1}$ &
  $\matriz{A}_{n} $ & $\matriz{A}_{n}^{-1} \matriz{B}_{n}\matriz{A}_{n} $  \\
  1 & $\matriz{B}_{n}$ & $\matriz{B}_{n}\matriz{A}_{n}$ $\matriz{B}_{n}^{-1}$ &
  $\matriz{B}_{n} $ & $\matriz{B}_{n}^{-1}$ &
  $\matriz{B}_{n}^{-1}\matriz{A}_{n}$$\matriz{B}_{n}$ & $ \matriz{B}_{n} $   \\
  2 &  $\matriz{B}_{n}^{-1}$ & $ \matriz{B}_{n}^{-1} \matriz{A}_{n} \matriz{B}_{n}$ &
  $\matriz{B}_{n}$ &  $\matriz{B}_{n}$ &
  $ \matriz{B}_{n} \matriz{A}_{n} \matriz{B}_{n}^{-1}$ & $  \matriz{B}_{n}$ \\
  3 &  $\matriz{A}_{n}^{-1}$ & $  \matriz{A}_{n} $ &
  $\matriz{A}_{n}^{-1} \matriz{B}_{n} \matriz{A}_{n}$ &  $\matriz{A}_{n}$ &
  $ \matriz{A}_{n} $ & $ \matriz{A}_{n} \matriz{B}_{n} \matriz{A}_{n}^{-1}$ \\
  \hline
\end{tabular}
\label{tabla3}
\end{table}

To derive a center $z_{n+1}$ from a previous one $z_{n}$, one looks
first at the corresponding color jump. Next, the matrix that takes
$z_{n}$ into $z_{n+1}$ depends on the numbering of the side ($0$, $1$,
$2$, or $3$) one must cross, and appears in the column labeled $T$ in
table~\ref{tabla3}.  The next generation is obtained in much the same
way, except for the fact that $\matriz{A}_{n}$ and $\matriz{B}_{n}$
must be replaced by $\matriz{A}_{n+1}$ and $\matriz{B}_{n+1}$ ,
respectively, as indicated in the table. In creating recursively any
word, the origin is denoted as $z_{0}$ and the matrices
$\matriz{A}_{0}$ and $\matriz{B}_{0}$ coincide with $\matriz{A}$ and
$\matriz{B}$.  One can then construct any word step by step. For
example, the word that transforms $z_{0}$ into $z_{6}$ in the zig-zag
path sketched in figure~\ref{figure23} results:
\begin{equation}
  \label{eq:1}
  \begin{array}{lcl}
    z_{0} \rightarrow z_{1} \quad : \quad \matriz{A} \, ,\\
    z_{1} \rightarrow z_{2} \quad : \quad
    \matriz{A} \matriz{B}^{-1} \matriz{A}^{-1}  \, , \\
    z_{2} \rightarrow z_{3} \quad : \quad
    \matriz{A}  \matriz{B}^{-1} \matriz{A}^{-1}  \, ,\\
    z_{3} \rightarrow z_{4} \quad : \quad
    \matriz{A}  \matriz{B}^{-1}  \matriz{B}^{-1} \matriz{A}
    \matriz{B}\matriz{B}  \matriz{A}^{-1}  \, , \\
    z_{4} \rightarrow z_{5} \quad : \quad
    \matriz{A} \matriz{B}^{-1} \matriz{B}^{-1}  \matriz{A} 
    \matriz{B} \matriz{B}   \matriz{A}^{-1} \, , \\
    z_{5} \rightarrow z_{6} \quad : \quad
    \matriz{A} \matriz{B}^{-1} \matriz{B}^{-1}  \matriz{A} 
    \matriz{A} \matriz{B}^{-1} \matriz{A}^{-1} \matriz{A}^{-1} 
    \matriz{B}\matriz{B} \matriz{A}^{-1}  \, .
  \end{array}
\end{equation}

The characteristic matrices of these substitution rules are
\begin{equation}
  \matriz{T}_{\mathrm{odd}}  =
  \left (
    \begin{array}{cccc}
      1& 0 & 1 & 1 \\
      2& 1 & 3 & 1\\
      1& 1 & 1 & 0\\
      3& 1 & 2 & 1
    \end{array}
  \right ) , 
  \quad  \qquad
  \matriz{T}_{\mathrm{even}}  =
  \left (
    \begin{array}{cccc}
      1& 0 & 0 & 0 \\
      0& 1 & 0 & 0\\
      0& 0 & 1 & 0\\
      0& 0 & 0 & 1
    \end{array}
  \right )
\end{equation}\\
with eigenvalues $2+\sqrt{5}$,  $2-\sqrt{5}$, $1$ and $-1$ for $\matriz{T}_{\mathrm{odd}}$ 
and $1$ for $\matriz{T}_{\mathrm{even}}$ (that produces a trivial effect and will no
be taken into account). Since one of the eigenvalues is greater than 1 and the 
others have an absolute value less than or equal to unity (with at least one 
of modulus 1), the substitution possesses the Salem property~\citep{Salem:1963},  
property weaker than the Pisot property.

\begin{figure}
  \centering
  \includegraphics[height=5cm]{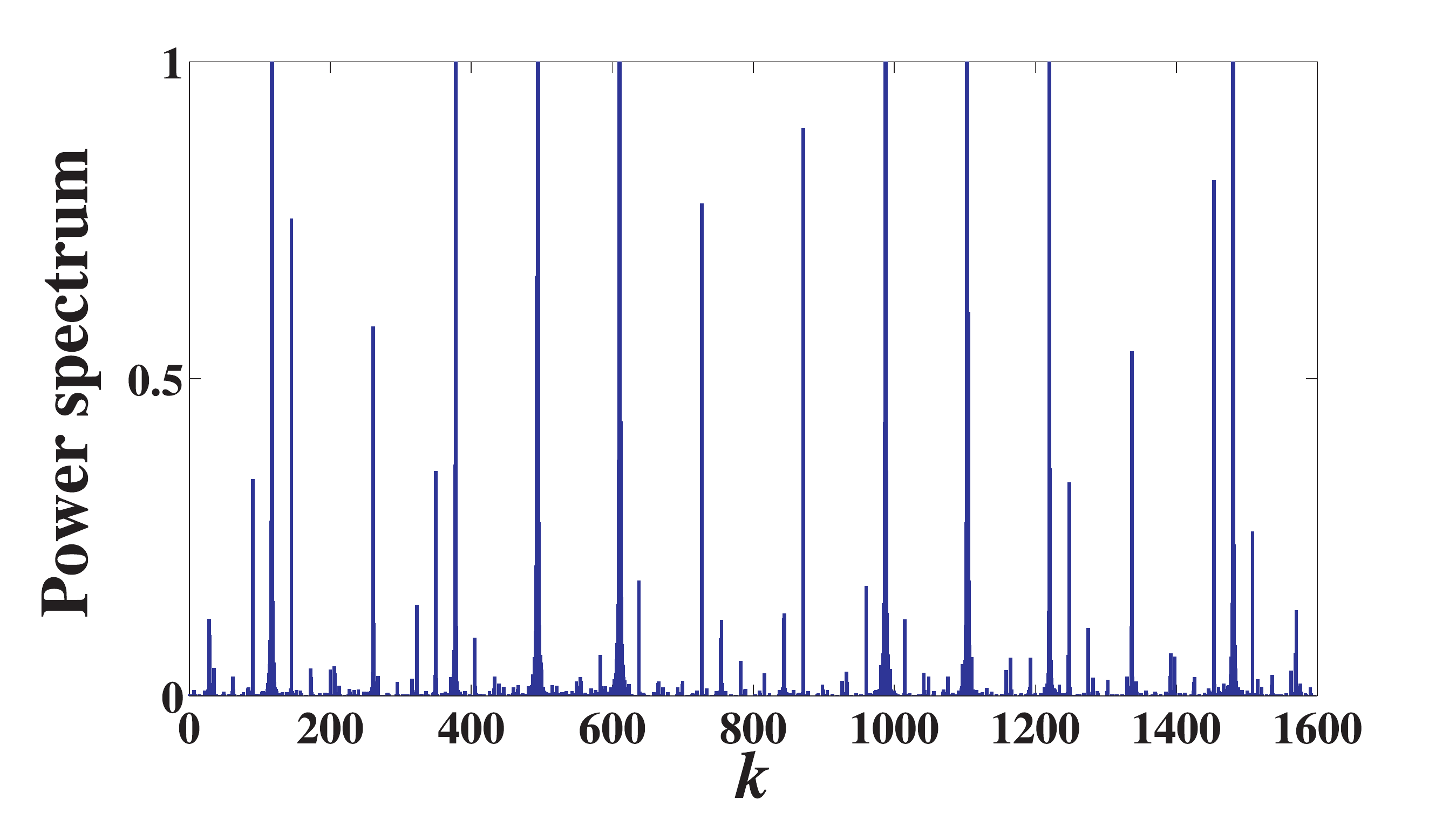}
  \caption{Normalized power spectrum for a word in the zig-zag path
    shown in figure~\ref{figure23} but with 1600 letters.}
  \label{figure24}
\end{figure}

In figure~\ref{figure24} we have plotted the structure factor of this
sequence for a word of 1600 letters that starts at the
origin. Previously, we assigned to each letter in the alphabet the
quartic roots of the identity, namely $ \matriz{A} \mapsto i$,
$\matriz{B} \mapsto + 1$, $\matriz{A}^{-1} \mapsto -i$,
$\matriz{B}^{-1} \mapsto -1$. The gross features of the spectrum are
seen to be humps separated by almost empty regions. Inside these
humps, there is a blurred structure built up of packed delta-spikes.
The dominant peaks tend to be isolated and larger.

To describe the continuous part of the measure in an analytical way
seems to be a hard task. Still we can introduce the
integrals~\citep{Aubry:1987,Godreche:1990}
\begin{equation}
  \label{IntGod}
  \mathcal{I}_{N}=  \int_{0}^{k_{\mathrm{max}}} \sqrt{S_{N}(k)} \, dk
  \, ,
\end{equation}
where $k_{\mathrm{max}}$ is some arbitrary spectral cutoff. The nature
of the limit measure is partly coded in the behavior of these
quantities for large $N$. In fact, for an absolutely continuous
measure they tend to be constant, while they scale as $\mathcal{I}_{N}
\sim N ^{-1/2}$ for a pure point measure~\citep{Godreche:1992}. A
somehow intermediate behavior can be expected for the singular
continuous case. For our system, we have evaluated numerically
(\ref{IntGod}), finding the law
\begin{equation}
  \mathcal{I}_{N} \sim N^{-\gamma}, \qquad  \qquad \gamma=0.36 \pm 0.02.
\end{equation}	
This rules out the existence of an absolutely continuous component and
suggests that only a singular spectrum is present. This exponent can
also be related to the theory of multifractals: if the measure $d\nu
(k)$ has a generalized dimension function $D_{k}$, then $\gamma$
should be linearly related to $D_{k}$~\citep{Hentschel:1983}.

\section{Concluding remarks}

This review is concerned with the transfer matrix, a powerful tool
that relies only on the linearity of a system with two input and two
output channels. Therefore, it is not surprising the variety of
domains in which this object has been successfully employed.

Instead of embarking in a detailed description of a particular model,
our thread has been to put geometry to work. This provides a useful
and, at the same time, simple language in which numerous physical
ideas may be clearly formulated and effectively treated.

In a first step, we have transplanted the transfer matrix into
space-time phenomena. This gateway works in both directions: here, it
has allowed us to establish a relativistic presentation of the
transfer matrix, but specific models can be also used as an instrument
for visualizing special relativity. This is more than an academic
curiosity: in fact, some intricate relativistic effects, such as,
e.g., the Wigner angle, can be measured (and not merely inferred) by
optical setups.

By resorting to elementary notions of hyperbolic geometry, we have
interpreted in a natural way the action of the transfer matrix as a
mapping on the unit disk and on the upper half-plane. The trace 
turns out to classify and characterize the basic
geometrical actions, which has physical relevance. In fact, in
this arena, nontrivial phenomena can be
understood in terms of analogues of vectors in Euclidean
geometry.

We have applied this perspective to periodic systems,
explaining the existence of bandgaps in appealing terms. In
addition, we have presented schemes to generate quasiperiodic 
sequences based on tessellations of the unit disc. 

Nothing of the material presented here is applicable \textit{per se}, but
everything can be relevant for researchers in other fields. This is
the beauty of the approach. To our mind, the manuscript is better than
ever. May each reader benefit and enjoy!

\section*{Acknowledgments}

Our efforts towards understanding the problems posed in this paper
were fueled in part, and were made much more interesting, by the
interaction with a number of colleagues and friends. We warmly 
thank Gunnar Bj\"{o}rk, Luis Joaqu\'{\i}n Boya, Antonio F. Costa,
Angel Felipe, Alberto Galindo, Hubert de Guise, Andrei Klimov, Gerd
Leuchs, Carlos L\'opez Lacasta, Enrique Maci\'a, Jos\'e Mar\'{\i}a
Montesinos, Ma\-ria\-no Santander, and Teresa Yonte.

Financial support from the Spanish Research Agency (Grants
FIS2005-06714, FIS2008-04356, FIS2011-26786) and the UCM-BCSH program
(Grant GR-920992) is gratefully acknowledged.


\end{document}